\documentclass[11pt]{article}
\usepackage[top=25truemm,bottom=25truemm,left=20truemm,right=20truemm]{geometry}
\usepackage{cite}
\usepackage{graphicx}
\usepackage{amsfonts}
\usepackage{amstext}
\usepackage{amsmath}
\usepackage{amssymb}
\usepackage{amsthm}
\usepackage[mathscr]{eucal}
\makeatletter
 
 \@addtoreset{equation}{section}
\newtheorem{theorem}{Theorem}
\newtheorem{proposition}[theorem]{Proposition}
\newtheorem{lemma}[theorem]{Lemma}
\newtheorem{definition}[theorem]{Definition}
\newtheorem{corollary}[theorem]{Corollary}
  \@addtoreset{theorem}{section}

 \makeatother
\def\a{\alpha}
\def\A{\mathcal{A}}
\def\Ai{{\rm Ai}}
\def\b{\beta}

\def\bv{\bar{v}}
\def\B{\mathcal{B}}
\def\C{\mathbb{C}}
\def\d{\delta}

\def\e{\epsilon}

\def\g{\gamma}
\def\G{\mathbb{G}}

\def\i{\infty}

\def\l{\lambda}

\def\N{\mathbb{N}}
\def\o{\omega}

\def\vp{\varphi}
\def\P{\mathbb{P}}
\def\Pe{\mathcal{P}}
\def\1{\bf{1}}
\def\R{\mathbb{R}}

\def\s{\sigma}

\def\S{\mathcal{S}}

\def\t{\tau}
\def\th{\theta}
\def\T{\mathbb{T}}
\def\th{\theta}

\def\Z{\mathbb{Z}}

\def\z{\zeta}

\def\k{\kappa}

\begin{document}
\title{Fluctuations for stationary $q$-TASEP}
\author{Takashi Imamura
\footnote { Department of Mathematics and Informatics, 
Chiba University,~E-mail:imamura@math.chiba-u.ac.jp}
, Tomohiro Sasamoto
\footnote { Department of Physics, 
Tokyo Institute of Technology,~E-mail: sasamoto@phys.titech.ac.jp}}
\maketitle
%
We consider the $q$-totally asymmetric simple exclusion process ($q$-TASEP)
in the stationary regime and study the fluctuation of the position of a particle. 
We first observe that the problem can be studied as a limiting case of an $N$-particle 
$q$-TASEP with a random initial condition and with particle dependent hopping rate.  
Then we explain how this $N$-particle $q$-TASEP can be encoded in a dynamics on a two-sided 
Gelfand-Tsetlin cone described by a two-sided $q$-Whittaker process and present a 
Fredholm determinant formula for the $q$-Laplace transform of the position of a particle. 
Two main ingredients in its derivation is the Ramanujan's bilateral summation formula and the 
Cauchy determinant identity for the theta function with an extra parameter. Based on this we 
establish that the position of a particle obeys the universal stationary KPZ distribution 
(the Baik-Rains distribution) in the long time limit.

\section{Introduction}
Large time fluctuations of certain class of interface growth exhibit 
universal critical behaviors, characterized by the Kardar-Parisi-Zhang (KPZ) 
universality class. For systems in the one dimensional KPZ class, the 
fluctuations of the height of an interface at a position grow as 
$O(t^{1/3})$ for large time $t$. The exponent $1/3$ are observed 
for example in various simulation models such as the ballistic deposition 
model and the Eden model\cite{BS1995}. 
Theoretically it was predicted by an application 
of a dynamical version of the renormalization group method 
\cite{FNS1977, KPZ1986} to 
the celebrated KPZ equation, 
\begin{equation}
\partial_t h = \tfrac12\partial_x^2 h + \tfrac12(\partial_x h)^2 + \eta , 
\label{KPZeq}
\end{equation} 
where $h=h(x,t)$ represents the height at position $x\in\R$ and at time $t\geq 0$ and 
$\eta=\eta(x,t)$ is the space-time Gaussian white noise with mean zero and covariance 
$\langle \eta(x,t)\eta(x',t')\rangle = \delta(x-x')\delta(t-t')$. 
Later the exponent was also 
confirmed by exactly solvable models. 

One of the standard models in this one dimensional KPZ class is 
the totally asymmetric exclusion process (TASEP), in which each particle 
on one dimensional lattice hops to the right neighboring site with rate one 
as long as the target site is empty. By the interpretation of the occupied site
(resp. empty site) as upward (resp. downward) slope, the TASEP can be 
mapped to a surface growth model, called the single step model. 
For TASEP the 1/3 exponent was found exactly by looking at the spectral gap 
of the generator for the process on a ring \cite{GS1992}. 
For TASEP with step initial condition, in which all sites are occupied 
(resp. empty) to the left (resp. right) of the origin, this exponent was proved 
and even the limiting distribution of the height fluctuation was determined in 
\cite{Johansson2000}. 
In this case, the limiting distribution turned out to be the Tracy-Widom(TW) 
distribution of GUE(Gaussian unitary ensemble) type, which describes 
the largest eigenvalue distribution of the GUE random matrix ensemble
in the limit of large matrix dimension\cite{TW1994}.  

The limiting distributions, though universal in the sense that the same 
distribution appears in many other models and are even observed in 
experiments \cite{TS2010,TS2012}, turned out to have certain boundary/initial condition dependence. 
For instance for the flat interface, the limiting distribution is given by the TW distribution of
the GOE (Gaussian orthogonal emsenble) type\cite{TW1996}, 
as found in \cite{BR2000,PS2002a,Sasamoto2005}. 
The stationary case was studied in \cite{FS2006} based on a previous work by 
\cite{IS2004}, in which the limiting distribution 
is given by the Baik-Rains distribution $F_0$ \cite{BR2000}.
The wedge, flat and stationary are the most typical three initial conditions 
in the whole space. 

More recently,  exact solutions for the height distribution have been 
obtained for other growth models, notably the KPZ equation (\ref{KPZeq}) itself. 
For the case of wedge initial condition this was established in 
\cite{SS2010a,SS2010b,SS2010c,ACQ2011} based on 
previous works \cite{BG1997, TW2009a} on ASEP. 
See also \cite{Dotsenko2010a, CLDR2010} for a different approach at almost the same time by replica 
method. Similar results have 
been obtained for the O'Connell-Yor(OY) polymer model \cite{OConnell2012,BC2014}
and a few other models.  
Most of the results have been obtained for the wedge (or step in TASEP 
language) case. Exceptions are the KPZ equation for which the stationary 
case was already studied in \cite{IS2012,IS2013,BCFV2015}.  
(See the remark at the end of this introduction.) 

The $q$-TASEP, which is also in this class and will be of our main interest in this paper, 
is a version of TASEP in which each particle tries to hop to the 
right neighboring site with rate $1-q^{\text{gap}}$ where $0<q<1$ and "gap" means the number 
of empty sites between the particle in question and the first particle ahead.  
In the limit $q\to 0$, this becomes the usual TASEP. The $q$-TASEP was introduced 
in \cite{BC2014}, though the dynamics of the gaps among particles described 
as the totally asymmetric zero range process (TAZRP) 
was introduced earlier in \cite{SW1998}. The KPZ exponent for this $q$-TAZRP 
was confirmed in \cite{Po2004} by using Bethe ansatz. 
In the stationary measure of $q$-TASEP, all the gaps among the particles are independent 
and each gap is distributed as the $q$-Poisson random variable (\ref{q-geo}) with a 
parameter $\a\in [0,1)$ by which the average density of particles in the system can be controlled.
In this article, we study the $q$-TASEP for this stationary situation with the condition 
that there is a particle at the origin initially  and establish that the 
distribution of the position of a particle is given by the Baik-Rains distribution, 
$F_0$ \cite{BR2000}, in a suitably scaled large time limit. 

Due to a version of the Burke theorem \cite{Burke1956,FerrariFontes1994}, which allows us to 
replace the infinite particles to the right by a single Poisson random walker with rate $\a$, and 
the fact that in TASEP a particle can not affect the dynamics of particles ahead, our studies on 
stationary $q$-TASEP are reduced to an $N$ particle problem. 
Let us denote the position of the $i$-th particle from the right at time $t$ as $X_i(t), 1\leq i\leq N$,  
and consider a generalized setting in which the hopping rate is particle dependent: 
the hopping rate of the $i$-th particle is $a_i(1-q^{x_{i-1}-x_i+1}), 1\leq i\leq N$ with the 
convention $x_0=+\infty$ and the assumption $0<a_i\leq 1, 1\leq i\leq N$. 
We will study the $q$-TASEP with a random initial condition, 
\begin{align}
-1-X_1(0)=Y_1,~X_{i-1}(0)-X_{i}(0)-1=Y_i \text{~for}~i=2, 3, \cdots ,N,
\label{h-st-ic}
\end{align}
where $Y_1, Y_2, \cdots$ are independent $q$-Poisson random variables with 
parameter $\a/a_i$ with $0\leq \a/a_i < 1,1\leq i\leq N$. 
This does not really correspond to our stationary $q$-TASEP in two respects. 
One is that the initial position of the first particle is taken to be random, not fixed 
at the origin. This can be handled as follows. 
Let us denote by $X_N^{(0)}(t)$ the position of he $i$th particle at time $t$  
when the first particle is at the origin initially,  i.e., the initial condition now reads
\begin{align}
X_1^{(0)}(0)=0,~X_{i-1}^{(0)}(0)-X_{i}^{(0)}(0)-1=Y_i \text{~for}~i=2, 3, \cdots ,N,
\label{h-st-ic0}
\end{align}
where $Y_2, Y_3, \cdots$ are independent $q$-Poisson random variables with 
parameter $\a/a_i$ with $0\leq \a/a_i < 1,2\leq i\leq N$.
Then the difference between 
the two cases comes only from the randomness of $Y_1$ and a simple 
relation $X_i(t) = X_i^{(0)}(t) -Y_1, 1\leq i\leq N$ holds. Since the distribution 
of $Y_1$ is independent of $X_i^{(0)}(t)$ and $Y_1$ is just $q$-Poisson distributed, 
one can study $X_i^{(0)}(t)$ by a combination of information on $X_i(t)$ and $Y_1$.  
The other difference of (\ref{h-st-ic}) from the original problem we want to study 
is that the hopping rates depend on particles.   
The stationary case can be accessed by specializing to  
$a_1=a, a_i=1,2\leq i\leq N$ and then taking the $a\to \a$ limit.
In the $a\to\a$ limit, the distribution of $Y_1$ becomes singular 
(since the parameter of the $q$-Poisson distribution becomes one)
but this difficulty can be handled by analytic continuation from $\a<a$ case.  

\bigskip 
The fact that the largest eigenvalue of certain random matrix ensembles 
and the height of an interface growth share the same limiting distribution 
is staggering. It is still a challenging problem to understand 
the connection in full generality, but some part of it may be understood 
by a common determinatal structure which appear in the studies of 
a few concrete models. 
For GUE, the joint eigenvalue distribution is written as the square of a 
product of differences \cite{Mehta2004, Forrester2010}. 
As is well-known in linear algebra, a product of differences 
can be written as a Vandelmonde determinant. Hence GUE is associated 
with a measure in the form of a product of two determinants. Once this 
structure is found, one can show that all correlation functions are written 
as determinants in terms of the same correlation kernel \cite{TW1998}. 
A determinantal point process is a point process whose correlation functions can 
be written as determinants in terms of the same correlation kernel
\cite{Soshnikov2000,Borodin2009}. 
Hence GUE is an example of a determinantal point process. 

The TASEP with step initial condition is, via the RSK mapping, 
associated with Schur measure, which is in the form of a product of two Schur 
functions \cite{Johansson2000}. It is well known that a Schur function can be written as a single 
determinant (Jacobi-Trudi identity) \cite{Mac1995, Stanley1999}. Hence TASEP with step initial condition 
is associated with a measure in the form of a product of two determinants and the 
above machinery of the determinantal processes can be applied. 
Basically the same determinatal structure can be utilized to study the stationary case
\cite{IS2004,FS2006}. For the flat case, a modified signed determinatal structure exists \cite{Sasamoto2005,BFPS2007}. 
 
 Situation is different for the KPZ equation, the OY polymer, $q$-TASEP, ASEP and so on. 
There the final results are written as Fredholm determinants after long calculations but not as
a consequence of an underlying determinantal process (see the remark at the end of this 
introduction for recent related developments). 
In \cite{BC2014} the $q$-TASEP appears as a special case of the Macdonald process, 
which can be written as a product of two Macdonald polynomials \cite{Mac1995}. 
For the Macdonald polynomials, a single determinat formula has not been found but   
the authors of \cite{BC2014} could use other properties of the Macdonald polynomials to 
study $q$-TASEP.
There is another approach using duality \cite{Schuetz1997a,IS2011,BCS2014,CGRS2016}.  
In either method, one first finds a multiple integral formula for the $q$-deformed 
moments of the form $\langle q^{kX} \rangle$ and sees that its certain generating function 
can be rewritten as a Fredholm determinant. 
 
For the OY polymer, which is a finite temperature semi-discrete directed 
polymer model,  some determinantal structures have been discussed.
In  \cite{OConnell2012} a connection to a GUE random matrix with a source 
is found 
though the resulting formula was not suited for asymptotics. In \cite{BCR2013}, 
this formula was transformed to the one in \cite{BC2014} for which the aysmptotic 
analysis is possible. Another determinantal structure was also discussed in \cite{IS2016}. 
But all these discussions have been made after the application of the 
 Bump-Stade identity \cite{Bump1989,Stade2002} which allows us to rewrite a 
 certain quantity of the original model in an integral from with a better product structure 
 in the integrand in a magical way. 
 
\bigskip
In this paper we analyze the $N$-particle $q$-TASEP for the initial condition (\ref{h-st-ic}). 
First we show that the process can be described as a marginal of a dynamics on an 
enlarged state space, which is a two-sided version of the Gelfand-Tsetlin cone $\G_N$.
Here "two-sided" means that each component of the element takes a value from 
$\Z$ rather than $\N$. The element of $\G_N$ is in the form, 
$\underline{\l}_N=(\l^{(1)},\ldots, \l^{(N)})$ where $\l^{(i)}=(\l_1^{(i)},\ldots ,\l_i^{(i)}),1\leq i\leq N$ 
is a signature (two-sided partition), meaning $\l_j^{(i)}\in\Z$, with the interlacing condition
$\l^{(i+1)}_{j+1}\le \l^{(i)}_{j}\le \l^{(i+1)}_{j}, 1\le j \le i \le N-1$. 
We consider a Markov dynamics on $\G_N$, s.t. that probability 
$\P[\underline{\l}_N(t)=\underline{\l}]$ is given by a measure in the form, 
\begin{align}
P_t (\underline{\l}_N)
 := 
 \frac{\prod_{j=1}^N P_{\l^{(j)}/\l^{(j-1)}} \left( a_j\right) \cdot Q_{\l^{(N)}} \left(\a,t\right)}{\Pi(a;\a,t)} .
\label{i2qWh-pr}
\end{align}
Here $P_{\l/\mu}(a)$ is the $q$-Whittaker function of one variable, with the label 
$\l$ being a signature.  
$Q_{\l}(\a,t)$ is a generalization of a version of the Macdonald polynomial $Q_\l(x)$ 
to our situation with a finite density of particles to the left. 
(Note that our case is not included as the non-negative specialization treated in 
\cite{BC2014} because $\l$ is not restricted to partitions but is associated with a 
positive measure.) $\Pi(a;\a,t)$ is the normalization. 
We call this the two-sided $q$-Whittaker process. 
We will see that the dynamics of the $N$ particle $q$-TASEP with (\ref{h-st-ic}) is encoded as a 
marginal dynamics of $\l_j^{(j)}(t), 1\leq j\leq N$ in this process as $X_j(t)+j = \l_j^{(j)}(t),1\leq j\leq N$. 

To study the position of the $N$th particle in our $q$-TASEP, one can instead focus on
the marginal for $\l^{(N)}(t)$, which turns out to be written as
\begin{equation}
 \P[\l^{(N)}(t)=\l] = \frac{1}{\Pi(a;\a,t)} P_{\l}(a_1,\ldots, a_N) Q_\l(\a,t). 
 \label{qWh-meas}
\end{equation} 
Here $P_{\l}(a_1,\ldots, a_N)$ is the two-sided $q$-Whittaker function and 
we call this marginal measure the two-sided $q$-Whittaker measure.  
We will calculate the $q$-Laplace transform, 
$\bigl\langle \frac{1}{(\zeta q^{\l_N};q)_\i} \bigr\rangle$ with respect to this measure. 
Here $(a;q)_\i=\prod_{i=1}^\i (1-aq^i)$ is the $q$-Pochhammer symbol. 
A difficulty to study a random initial condition for $q$-TASEP is that the $q$-moment 
$\langle q^{k \l_N}\rangle$ diverges for large $k$ \cite{BCS2014}. 
So one can not expand the above $q$-Laplace transform as the generating 
function of these $q$-moments using the $q$-binomial theorem (\ref{qbinom}). 
Instead we compute the $q$-Laplace transform directly. To do so, 
we rewrite the Cauchy identity for Macdonald polynomials by separating the factor only on 
$N$th row of a Young diagram, see (\ref{hci}), and for the remaining summation of 
$\l_N$ over $\Z$, we use the Ramanujan's birateral summation formula, see (\ref{ram}). 
Then, after some calculation, we notice a determinantal structure for our quantity in terms of 
the theta function. We can use the Cauchy determinant identity for the theta function
with an extra parameter, (\ref{cdet}), 
to arrive at the following Fredholm determinant formula,
which is our first main result (Corollary \ref{main1}). 
\begin{theorem}\label{ql-det}
For $q$-TASEP with the initial condition (\ref{h-st-ic}) with $0\leq \a < a_j \leq 1,1\leq i,j\leq N$ and 
with $\zeta\neq q^n,n\in\Z$, 
\begin{equation}
 \Big\langle \frac{1}{(\z q^{X_N(t)+N};q)_\i} \Big\rangle
 =
 \det(1-f K)_{L^2(\Z)} 
 \label{det_formula_i}
\end{equation}
where $\langle \cdots \rangle$ means the average, 
$ f(n)  =  \frac{1}{1-q^n/\z}$ and  the kernel in the Fredholm determinant is written as   
\begin{align}
 K(n,m) &= \sum_{l=0}^{N-1} \phi_l(m) \psi_l(n).
 \end{align}
Here the functions $\phi_n,\psi_n$ are explicitly given in the form of contour integrals  
 (\ref{phi}),(\ref{psi}) with $\a_1=\a,\a_i=0,2\leq i\leq N$. 
\end{theorem} 
\noindent
Some part of the procedure to find this Fredholm determinant formula is basically the 
same as that in \cite{BCR2013} for the Log-Gamma (and OY) polymer case in which 
the trigonometric Cauchy determinant was used after the Bump-State identity. 
Novelty in our calculation is that 
we do not  rely on the Bump-Stade identity and that we use the Cauchy identity 
at the elliptic level with an extra parameter \cite{KN2003}. 

Then we consider the stationary case by specializing to  $a_1=a,a_i=1,i\geq 2$ and
taking the limit $a\to \a$. The formula for the $q$-Laplace transform for the 
stationary $q$-TASEP is given in (\ref{fl31}). We also perform the asymptotic analysis 
to our expression and establish that the limiting distribution is given by the Baik-Rains 
distribution, $F_0$. 
This is our second main result in the paper. 
\begin{theorem}
For the stationary $q$-TASEP, with the parameter $\a=q^{\th},\th>0$ determining the 
average density through (\ref{rhoal}), we have, for $\forall s\in\R$, 
\begin{align}
\lim_{N\rightarrow\infty}\P(X^{(0)}_N(\kappa N) >(\eta-1) N-\g N^{1/3}s)
=
F_0(s)
\label{introF0}
\end{align}
where $\k=\k(\th,q),\eta=\eta(\th,q),\g=\g(\th,q)$ are given by (\ref{f32}),(\ref{f33}),(\ref{f34}).
$F_0$ is a special case of the Baik-Rains distribution (\ref{fp51}). 
\end{theorem}
\noindent
This is a special ($\o=0$) case of Theorem \ref{main2}. 
The coefficients $\k,\eta,\g$ in the theorem can be understood from the KPZ scaling 
theory \cite{PS2002a, Spohn2012,FerrariVeto2015}. 
In the main text we will study the case where $\a$ is also rescaled and obtain a 
parameter family of limiting distribution $F_\o$ \cite{BR2000, FS2006, IS2013}. 
Our result is only for the one point distribution. As a conjecture it is expected that 
joint distributions of scaled height would be described by the process identified 
in the studies of TASEP \cite{BFP2010}. 
The scaled two point correlation function would also show universal behaviors. 
Such analysis have been done for the PNG model, TASEP and the KPZ equation
\cite{PS2004,FS2006,IS2013}. 

In the proof of the theorem we show that some part of the formula tends to the GUE Tracy-Widom 
distribution. In Appendix \ref{TWlimit}, we provide two lemmas to establish this limit when 
the kernel is written in a specific form. This would be useful for other situations as well.

Our approach can be applied to other models. 
As an example, we discuss two limiting cases, the OY polymer and the KPZ equation, 
in a companion paper \cite{IS2017p2}. 
For the OY model, the analysis for the case with multi parameters was already studied in 
\cite{BCFV2015}. In \cite{IS2017p2}, we will provide a formula for the real stationary case.
One can also consider the limit to the stationary KPZ equation. 

\bigskip

The paper is organized as follows. 

In section 2, we explain basic properties of $q$-TASEP and the corresponding 
$q$-boson TAZRP. We discuss their stationary measures, average density and 
current, hydrodynamic description, Burke theorem, and the KPZ scaling theory. 

In section 3, we introduce a two-sided version of the Gelfand-Tsetlin (GT) cone, 
$q$-Whittaker function and $q$-Whittaker process. For the Schur case this type 
of generalization was already studied by \cite{Borodin2011}. The $q$-version 
was also mentioned in \cite{CP2015,MatveevPetrov2017}.  
Next we see that the dynamics of the $q$-TASEP is encoded in the marginal 
$\l_j^{(j)}, 1\leq j\leq N$ on one edge of GT cone. For the step case, corresponding 
relation is stated in \cite{BC2014} as a limit from discrete time dynamics but we 
show this more directly for the continuous time model. At the end of the section 
we discuss the two-sided $q$-Whittaker measure, as the marginal for $\l^{(N)}$
of the two-sided $q$-Whittaker process. In particular an expression for the 
distribution of $\l_N^{(N)}$ is given, which is essential for the subsequent 
discussions. 

In section 4, we present a Fredholm determinant expression for the $q$-Laplace 
transform for the $N$ particle position. We will use in an essential way the 
Ramanujan's bilateral summation formula and the Cauchy determinant 
identity for theta function. 

In section 5, we first discuss in more detail the relationships between the two 
cases in which the initial position of the first particle is random and is fixed to 
be at the origin. Then we present a formula for the position of the particle 
for the stationary measure. Finally we perform asymptotic analysis and 
establish the Baik-Rains distribution for the position of the particle.  

In section 6, we discuss the $q\to 0$ limit corresponding to the usual  TASEP 
case studied in \cite{Johansson2000,FS2006}.   

In Appendix A, some $q$ notation and functions are summarized. 
In Appendix B, we explain and discuss a few basic properties of the ordinary (not two-sided)
$q$-Whittaker function and process. 
In Appendix C, we discuss the Tracy-Widom limit when the kernel is written in a 
specific form. 
In Appendix D,  we discuss the inverse of the $q$-Laplace transform. 

\medskip 
\noindent
{\bf Remark.} 
After the main contents of this paper have been obtained and while the manuscripi is 
prepared, a few papers in a similar direction have appeared. 
In \cite{AB2016p}, the authors consider "generalized step Bernoulli initial data" and 
discuss the BBP transition. 
In \cite{Aggarwal2016p}, the stationary ASEP is studied. 
In \cite{OrrPetrov2016p}, the authors study some generalized initial condition for 
discrete $q$-TASEP. They are based on a new 
 interesting observation on the relationship 
between certain observables of the Macdonald measure and the Schur measure \cite{Borodin2016p}. 
Their approach also give a better understanding on the determinantal structures
of the models. Our approach is somewhat different and 
it would be certainly an 
interesting question to have a better undersanding about the relationship between 
their methods and ours. 

\smallskip
\noindent
{\bf Acknowledgements.}
Part of this work was performed during the stay at KITP in Santa Barbara, USA. 
This research was supported in part by the National Science Foundation under 
Grant No. NSF PHY11-25915. The work of T.I. and T.S. is also supported by JSPS 
KAKENHI Grant Numbers JP25800215, JP16K05192 and JP25103004, JP14510499, 
JP15K05203, JP16H06338 respectively.

\section{Stationary $q$-TASEP and $q$-TAZRP}
\label{StqTASEP}
\begin{figure}[h]
\begin{center}
\includegraphics[scale=0.6]{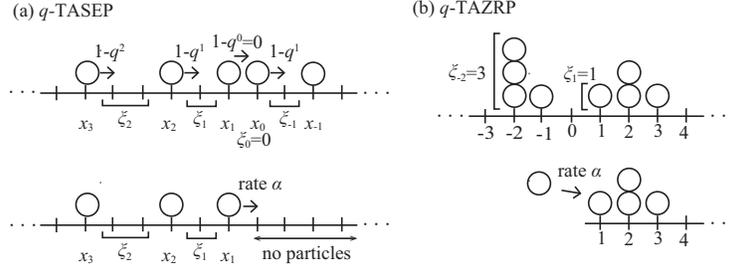}
\caption{\label{figqtaseptazrp}
$q$-TASEP and $q$-TAZRP (top), 
with equivalent half systems after an application of the Burke theorem (bottom). 
}
\end{center}
\end{figure}

The $q$-TASEP on $\Z$, which was introduced rather informally in Introduction,
is constructed in a standard manner \cite{Liggett1985} (cf \cite{BC2014} Sec 3.3.3), 
based on its generator, 
\begin{equation}
 L_{q-{\rm \!TASEP}} f(\eta) = \sum_{x\in\Z} (1-q^{g_x(\eta)})\eta(x) [f(\eta^{x,x+1})-f(\eta)]
\end{equation}
where $\eta\in\{0,1\}^\Z$ is a configuration of particles in $q$-TASEP, 
$g_x(\eta)+1$ is the distance between $x$ and the first particle to the right
in configuration $\eta$, $\eta^{x,x+1}$ is the configuration in which a particle at 
$x$ (if it exists) moves to $x+1$, and $f$ is a local function on $\{0,1\}^\Z$. 
The dynamics of the "gaps", $\xi_k=x_k-x_{k+1}-1$, where $x_k$ is the position of 
the $k$th particle (labelled from right to left, with the 0th particle starting at the 
position which is positive and closest to the origin), is a version of zero range 
process whose generator is given by 
\begin{equation}
 L_{q-{\rm \!TAZRP}}  f(\xi) = \sum_{k\in\Z} c(\xi_k) f[(\xi^{k,k+1})-f(\xi)]
 \label{genZRP}
\end{equation}
with 
\begin{equation}
 c(l) = 1-q^l, ~ l\in\N
 \label{gqT}
\end{equation}
where $\xi\in\N^{\Z}$ is a configuration of particles in a zero range process, 
and $f$ is a local function on $\N^{\Z}$. See Fig.  \ref{figqtaseptazrp}.
It is known that, when the function 
$c$ in (\ref{genZRP}) satisfies the condition, 
\begin{equation}
 \sup_l |c(l+1)-c(l)| < \i, 
\end{equation} 
the process can be constructed (for a large class of initial conditions). The 
function $c$ (\ref{gqT}) correspondinding to our $q$-TAZRP certainly satisfies 
this condition when $0<q<1$. 
Because particles hop only in one direction and the (adjoint) generator is written 
in terms of $q$-boson \cite{SW1998},  we call this the $q$-boson TAZRP 
(or simply $q$-TAZRP). 
We first explain some basic facts for this $q$-TAZRP. 

\subsection{$q$-TAZRP} 
\begin{figure}[h]
\begin{center}
\includegraphics[scale=0.6]{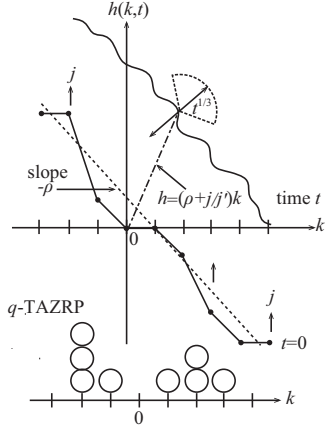}
\caption{\label{figheight}
Height picture corresponding to $q$-TAZRP. The characteristic line starting at the 
origin is also indicated which has a parametric representation $(k,h)=(j't,(j-\rho j')t)$. 
}
\end{center}
\end{figure}

For a zero range process (with certain mild conditions), 
the stationary measures (which are translationary invariant and extremal) 
are known \cite{Andjel1982}.
For our $q$-TAZRP, the stationary measure is given by the one in which  
the number of particles at sites are independent and at each site the number is 
distributed by the $q$-Poisson distribution, 
\begin{align}
\P[\xi_x=n]=(\a;q)_{\infty}\frac{\a^n}{(q;q)_n}, ~n\in\N,
\label{q-geo}
\end{align}
with $\a\in [0,1)$ a parameter. 
Here $(a;q)_n,(a;q)_\i$ are the $q$-Pochhammer symbols, see Appendix \ref{q}.  
One observes that this is normalized due to the $q$-binomial theorem (\ref{qbinom}) 
and that it becomes singular as $\a\uparrow 1$. 
We call this the $q$-Poisson distribution because if we scale $\a$ to $(1-q)\a$ and 
take the $q\to 1$ limit, this tends to the Poisson distribution (In \cite{BC2014}, 
it was called the $q$-geometric distribution since it tends to the geometric 
distribution when $q=0$).  
We sometimes write (\ref{q-geo}) as $\xi_x \sim q{\rm Po}(\a)$.
The average density for a given $\a$ is 
\begin{equation}
 \rho(\a) = \langle \xi_x \rangle = (\a;q)_\i \sum_{n=0}^{\i} \frac{n \a^n}{(q;q)_\i}
 =(\a;q)_\i \a\frac{d}{d\a}\frac{1}{(\a;q)_\i}=
 \a \sum_{k=0}^\i \frac{q^k}{1-\a q^k},
 \label{rhoal}
\end{equation}
where $\langle \cdots \rangle$ means the expectation value. 
We denote by $\a(\rho)$ the inverse function. 
The average current (to the right) is 
\begin{equation}
 j(\rho) = \langle 1-q^{\xi_x} \rangle =\a(\rho). 
 \label{jtazrp}
\end{equation}
Let us consider a quantity $h(k,t)$ defined by 
\begin{equation}
 h(k,t) 
 = 
 N_0(t) 
 +
 \begin{cases} - \sum_{l=1}^{k} \xi_l(t), & k\geq 1, \\
                       0, & k=0, \\
                       + \sum_{l=k+1}^{0} \xi_l(t), & k<0, 
 \end{cases}
\end{equation}
where $N_0(t)$ is the number of particles which 
hop from $0$ to $1$ between the time period $[0,t]$. This can be interpreted as 
a height function. 
Since $h(k,t)-h(k-1,t)=-\xi_k, k\in\Z$, the average density $\rho$ is the 
(minus) average slope and the average current $j$ is the average growth speed in the 
surface picture.  
See Fig.  \ref{figheight}.

Macroscopically the evolution of the density profile $\rho(x,t)$ (at a macroscopic 
space-time position $x$ and $t$) is expected to 
be described by the hydrodynamic equation, 
\begin{equation}
 \frac{\partial}{\partial t} \rho(x,t) + \frac{\partial}{\partial x} j(\rho(x,t))= 0.
\end{equation}
See for instance \cite{Spohn1991, KipnisLandim1999, Ferrari2016p, Gon2014} for more information. 
In the stationary case, the macroscopic density is just a constant $\rho(x,t)=\rho$. 
In the surface picture, macroscopic height profile is $h(x,t) =  j(\rho) t -\rho x$. 
We will be interested in the fluctuation around the macroscopic shape. Since we 
are considering the stationary case, the nontrivial fluctuation is expected to 
be observed near the characteristic line from the origin $x=j'(\rho) t$,
along which the effect of fixing the height initially (i.e. $h(0,0)=0$) is propagated
(cf e.g. \cite{Ferrari2016p, CFP2010}).

A totally asymmetric zero range process can be considered as a system of 
queues in tandem. In the correspondence each site $k$ is a service counter 
and the number of particles $\xi_k$  is the number of customers waiting for a 
service at the counter at site $k$. The service at each site is provided with  
rate $1-q^{\xi_k}$.  Queues are in tandem, which means that once a customer 
at a site $k$ gets a service he/she moves to the next queue at site $k+1$. $N_0(t)$ 
represents the number of customers who get the service at site 0 and move
to the site 1 between time period $[0,t]$. 
This is called the output process (at site 0) in the queue language. Likewise the input 
process for each service counter is the number of customers who join its 
queue between time period $[0,t]$. 
In the stationary situation, the number of customers at each site is distributed 
as the $q$-Poisson distribution (\ref{q-geo}). For this situation, we have 
\begin{proposition}
\label{Burke}
For the queues in tandem corresponding to the stationary $q$-TAZRP with parameter $\a$, 
the output process at the origin is the Poisson process with rate $\a$. 
\end{proposition}
\smallskip
\noindent
{\bf Proof.}
This is basically due to the Burke theorem \cite{Burke1956}, which says that 
when a queue having an exponential waiting time for a service and 
the input process is the Poisson process, then the output process 
is also the Poisson process with the same rate. 

The case of infinite queues in tandem corresponding to ordinary TASEP 
($q=0$) was proved in \cite{FerrariFontes1994}. The arguments there
can be applied to our case ($q\neq 0$) with $\xi_x$ dependent 
hopping rate as well. One first notices that the process is reversible 
with respect to the stationary measure with parameter $\a$. The 
output process at site 0 of the original process is in distribution the same as 
the input process at site $0$ in the reversed process, which is by construction
a Poisson process with rate calculated as the ratio of the stationary 
probability of having $n$ and $n+1$ particles at site 1 times the 
particle hopping rate when $n+1$ particles are at site 1, i.e., 
$\frac{\a^{n+1}}{(q;q)_{n+1}}\frac{(q;q)_n}{\a^n} (1-q^{n+1}) = \alpha$. 
\qed

\medskip\noindent
This proposition implies that, when we are only interested in the 
right half of the TAZRP, one can replace the whole left half by 
the Poisson input process at site 1 with rate $\a$. We can 
suppose that there are infinite particles at the origin which acts 
as a source of particles with rate $\a$, see the bottom-right figure 
in Fig. \ref{figqtaseptazrp}. 

\medskip
Next we explain the KPZ scaling theory \cite{PS2002a, Spohn2012} 
for our stationary $q$-TASEP.  
According to this conjectural theory, for a given density $\rho$, 
it is expected that the scaled height around the characteristic line, 
\begin{equation}
 \xi_t = (h(k,t) - (j(\rho) t-\rho k))/c(\rho) t^{1/3}, ~ \text{with}~k=j'(\rho)t, 
 \label{scalingZRP}
\end{equation}
exhibits universal behaviors, and the coefficient $c(\rho)$ can be determined 
by the knowledge of the average current $j(\rho)$ and the growth rate 
of the variance of height in space direction in the stationary measure, 
\begin{equation}
 A(\rho) = \lim_{k\to\i} \frac{1}{k} \langle (h(k,0)-h(0,0))^2\rangle_C ,
\end{equation}
where the bracket is taken with respect to the stationary measure with 
a density $\rho$ and $C$ means the cumulant. Then one should take 
\begin{equation}
 c(\rho) = -(-\frac12 j''(\rho) A(\rho)^2)^{1/3}.  
\label{KPZc}
\end{equation}
In our case since $h(k,0)=-\sum_{l=1}^{k} \xi_l(0),k>0$ is a sum of $k$ independent 
$q$-Poisson variables with parameter $\a$, $A(\rho)$ can be easily calculated as 
\begin{equation}
 A(\rho) = \sum_{k=0}^\i \frac{\a(\rho) q^k}{(1-\a(\rho) q^k)^2}.
\label{Arho}
\end{equation}
From (\ref{jtazrp}), we have
\begin{align}
 j'(\rho) &= \a'(\rho) = \frac{1}{\rho'(\a)}, \\
 j''(\rho) &= \a'(\rho) \frac{d}{d\a} \left( \frac{1}{\rho'(\a)}\right) 
               = -\frac{\rho''(\a)}{(\rho'(\a))^3}, \label{j2prime}
\end{align}
where by (\ref{rhoal}) 
\begin{align}
 \rho'(\a) = \sum_{k=0}^\i \frac{q^k}{(1-\a q^k)^2}, \quad
 \rho''(\a) = 2\sum_{k=0}^\i \frac{q^{2k}}{(1-\a q^k)^3}. \label{rhop}
\end{align}
Substituting (\ref{j2prime}) with (\ref{rhop}) and (\ref{Arho})(note $A(\rho)=\a \rho'(\a)$) into (\ref{KPZc}), 
the constant $c(\rho)$ (or $c(\rho)^3$) in our case is 
\begin{equation}
 c(\rho)^3 
 =
 -\frac{\a^2 \rho''(\a)}{2\rho'(\a)} 
 =
 -\frac{\sum_{k=0}^\i \frac{\a^2 q^{2k}}{(1-\a q^k)^3}}{\sum_{k=0}^\i \frac{q^{2k}}{(1-\a q^k)^2}}
\end{equation}
where $\a$ on rhs should be understood as $\a(\rho)$ defined below (\ref{rhoal}).

\subsection{$q$-TASEP} 
Next we discuss corresponding and similar statements for $q$-TASEP. 
As already explained in the Introduction, in the stationary measure with parameter $\a$, 
all the gaps among the particles are independent and each gap is distributed as the 
$q$-Poisson random variable (\ref{q-geo}). 
First, since the average distance between two consecutive particles is 
$1+\langle {\rm gap}\rangle$, the average density for a given $\a$ is 
\begin{equation}
 \rho_0(\a) = \frac{1}{1+\a \sum_{k=0}^\i \frac{q^k}{1-\a q^k}} = \frac{1}{1+\rho(\a)}. 
 \label{rho0al}
\end{equation}
The inverse function is denoted by $\a_0(\rho_0)$. 
For $q$-TASEP, the average current for a given density $\rho$ 
(or for a given $\a$ through $\rho=\rho_0(\a)$) is given by 
\begin{equation}
 j_0(\rho)|_{\rho=\rho_0(\a)}
 = 
 \a \rho_0(\a). 
\end{equation} 
This is understood by noting that the average speed of each particle is the 
same as the average hopping probability at each site, which is equal to 
the average current (\ref{jtazrp}) for $q$-TAZRP
and that the average current is given by this 
times the average density $\rho$. 

\medskip
Next we explain how one can reduce the problem of the fluctuation of the position of a 
particle in stationary regime, with the conditioning that there is a particle at the origin initially, 
to the one of the $N$-particle $q$-TASEP with the initial condition (\ref{h-st-ic}). 
First, for $q$-TASEP,  the 
proposition \ref{Burke} implies that, in the stationary measure with a parameter $\a$, 
the marginal dynamics of each particle is a Poisson random walk with rate $\a$. 
Therefore, if one is only interested in particles starting from the left of a specific particle,  
one can replace the whole dynamics of particles to the right of that particles by postulating 
that the particle is performing the Poisson random walk with parameter $\a$,
see the bottom-left figure of Fig. \ref{figqtaseptazrp}. 
Because of the conditioning of the presence of a particle at the origin at $t=0$, 
we put this particle at the origin initially. 
Next since in TASEP particles can not give any influence on the particles ahead, 
when we are interested in the fluctuation of the $N$th particle, it is enough to study 
the first $N$ particles. Let us denote the position of these $N$ particles as 
$X_i^{(0)}(t), 1\leq i\leq N$. The initial condition is in the form (\ref{h-st-ic0}) in which 
$Y_i\sim q{\rm Po}(\a),  2\leq i\leq N$ are i.i.d. random variables. 
Let us consider a generalized setting in which the hopping rate is particle dependent: 
the hopping rate of the $i$-th particle is $a_i(1-q^{x_{i-1}-x_i+1}), 1\leq i\leq N$ with the 
convention $x_0=+\infty$.  
We also generalize the distribution of $Y_i$'s to $Y_i \sim q{\rm Po}(\a/a_i), 2\leq i\leq N$.
For the $q$-Poisson distributions to make sense, one should take $a_i>\a, 2\leq i\leq N$.
From the discussions above, the stationary case corresponds to $a_1=\a, a_i=1,2\leq i\leq N$. 

To have better algebraic structures (as we will see in the next section), 
it is useful to consider a situation in which the position 
of the first particle is also random, i.e., we consider the initial condition (\ref{h-st-ic}) where
$Y_i \sim q{\rm Po}(\a/a_i), a_i>\a, 1\leq i\leq N$ are i.i.d. random variables. If we denote 
by $X_i(t),1\leq i\leq N$ the position of the $i$th particle with this initial condition, the 
difference between $X_i(t)$ and $X_i^{(0)}(t)$ is only $Y_1$. Since the law of $Y_1$ 
is known explicitly, one can study $X_i^{(0)}(t)$ from the information on $X_i(t)$ and $Y_1$. 
More precise connections will be given in section 5.1 for our quantity of interest. 
A tricky point here is that the distribution of $Y_1$ becomes singular in the stationary 
limit $a_1\downarrow\a, a_i=1,2\leq i\leq N$. But one can take this limit after performing 
an analytic continuation of the formula for $X_N^{(0)}(t)$. In this way one can study the 
fluctuation of the position of a particle by considering $N$-particle $q$-TASEP 
with the initial condition (\ref{h-st-ic}). 

\medskip
To see the nontrivial fluctuation 
for the stationary case,  one needs to look at the fluctuations along the 
characteristic line from the origin, which is now given by $x(t) = j_0'(\rho) t$.
(The reasonings here are exactly the same as for the $q$-TAZRP, but the 
current functions of $q$-TAZRP and $q$-TASEP, and hence also their 
characteristic lines, are different.) 
We will study the fluctuation of the position of the $N$th particle when a nontrivial fluctuation 
is expected, that is when it comes around the characteristic line. 
For a given density $\rho$ (or for a given $\alpha$ through $\rho=\rho_0(\a)$), 
\begin{equation}
 j_0'(\rho)|_{\rho=\rho_0(\a)} = \a + \a_0'(\rho_0)\rho_0(\a) = \a + \rho_0(\a)/\rho_0'(\a)
 \label{jprime}
\end{equation} 
determines the characteristic line. 
If we parametrize $\a=q^{\th},0<\th<\i$, 
for a given large $N$, the time and the position that the $N$-th particle is around this 
characteristic line are given by $t=\k N, x = (\eta-1)N$ with 
\begin{align}
\k
&=
\frac{\Phi_q\rq{}(\theta)}{(\log q)^2q^{\theta}}
=\sum_{n=0}^{\infty}
\frac{q^n}{(1-q^{\theta+n})^2},
\label{f32}
\\
\eta
&=
\frac{\Phi_q\rq{}(\theta)}{(\log q)^2}
-
\frac{\Phi_q(\theta)}{\log q}
-
\frac{\log (1-q)}{\log q}
=
\sum_{n=0}^{\infty}
\frac{q^{2\theta+2n}}{(1-q^{\theta+n})^2},
\label{f33}
\end{align}
where we used the $q$-digamma function $\Phi_q(z)=\partial_z\log\Gamma_q(z)$,
see Appendix \ref{q}. 
This is confirmed by checking that the average distance traveled by the 
particle is equal to the speed times time, 
\begin{equation}
 \a \times \k N = (\eta-1)N+\frac{1}{\rho} N,
\end{equation}
and that $t=\k N, x=(\eta-1) N$ in fact is on the characteristic line, i.e. 
\begin{equation}
 j_0'(\rho) \k N = (\eta-1) N
\end{equation}
with $j_0'(\rho)$ given by (\ref{jprime}). 

We are interested in the fluctuation of the particle position around this 
macroscopic value. If one applies again the KPZ scaling theory 
\cite{Spohn2012, FerrariVeto2015} to our $q$-TASEP 
(here again the reasonings are the same as for $q$-TAZRP but 
note the differences of the current functions and the stationary 
measures), it is expected we should scale the position as  
\begin{equation}
 \frac{X_N^{(0)}(\k N) -(\eta-1)N}{\g N^{1/3}} 
 \label{scalingQT}
\end{equation} 
where  
\begin{align}
\gamma
&=
-\frac{1}{\log q}\left(\frac
{\Phi_q\rq{}(\theta)\log q-\Phi_q\rq{}\rq{}(\theta)}
{2}\right)^{1/3}
=\left(\sum_{n=0}^{\infty}
\frac{q^{2\theta+2n}}{(1-q^{\theta+n})^3}\right)^{1/3}. 
\label{f34}
\end{align}
In fact one can check that this scaling for the $N$th particle position 
of $q$-TASEP is equivalent to (\ref{scalingZRP}) for $q$-TAZRP.  
First note  $h(N,t)=X_N^{(0)}(t)+N$. By changing the large parameter 
from $t$ to $N$ by $N=j'(\rho)t$ and observing
\begin{align}
 \frac{1}{j'(\rho)} = \kappa, \quad
 \frac{j(\rho)}{j'(\rho)}-\rho = \eta, \quad
 -c(\rho)^3/j'(\rho) = \g^3,
\end{align}
one sees that the rhs of (\ref{scalingZRP}) is written as (\ref{scalingQT}). 
The fact that the coefficients $\k,\eta,\g$ are consistent with the KPZ scaling 
theory was already confirmed in \cite{FerrariVeto2015} (though with different 
character sets $\k,f,\chi$) but a remark here is that the calculations of the 
coefficients are somewhat easier in $q$-TAZRP language because the 
stationary measure of $q$-TAZRP is written in a product form. 

So far our discussions have been for a fixed $\a$, but it is useful to 
also consider the scaling of $\a$  as 
\begin{equation}
\a=q^{\th}\left(1+\frac{\o}{\gamma N^{1/3}}\right),
\label{f31}
\end{equation}
for a fixed $\th>0$. 
This scaling limit will be discussed in subsection \ref{LTL}.

\section{Two-sided $q$-Whittaker process and $q$-TASEP}
\label{qWhT}

\subsection{Definitions}
For the case of the step initial condition, the dynamics of particles in 
$q$-TASEP can be described in terms of partitions, which are 
$n$-tuples of non-increasing non-negative integers \cite{BC2014}. 
To study the $q$-TASEP for a random initial condition, it is useful to 
consider $n$-tuples of non-increasing integers, each of which 
can take a value from $\Z$,  
\begin{align}
\S_n := 
\{\l=
(\l_1,\cdots,\l_n)\in\Z^n
|\l_1\ge \l_2\ge\cdots\ge\l_n
\}. 
\label{signature}
\end{align}
Note that a signature $\l\in\S_n$ becomes an ordinary partition when $\l_n\geq 0$. 
The set $\S_n$ can be seen as a discrete version of the Weyl chamber but here 
we call an element $\l\in\S_n$ a signature of length $n$. In this section we discuss this 
type of generalizations of partitions, $q$-Whittaker functions and so on. 
The Schur ($q=0$)  case is discussed in \cite{Borodin2011}. Some definitions 
and properties of the ordinary partitions and the $q$-Whittaker functions will be explained 
and discussed in Appendix \ref{stdqWh}. 

\begin{figure}[t]
\begin{center}
\includegraphics[scale=1.0]{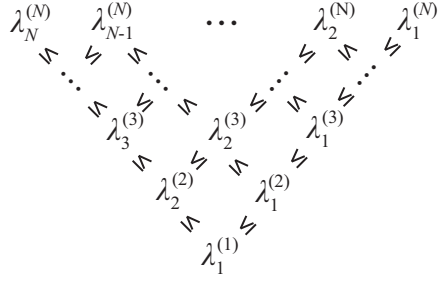}
\caption{\label{figGT}
The Gelfand-Tsetlin cone as a triangular array. 
}
\end{center}
\end{figure}

We also consider a set of $N(N+1)/2$-tuples of integers with interlacing conditions, 
\begin{align}
\mathbb{G}_N
:=
&\{ (\l^{(1)},\l^{(2)},\cdots,\l^{(N)}), \l^{(n)}\in\S_n,1\leq n\leq N
| \notag\\
&\quad \l^{(m+1)}_{\ell+1}\le \l^{(m)}_{\ell}\le \l^{(m+1)}_{\ell}, 
1\le \ell \le m \le N-1
\}.
\label{GT}
\end{align}
See Fig.  \ref{figGT}.
Note that an element of $\underline{\l}_N\in\G_N$ can also be regarded as a
point in $\Z^{N(N+1)/2}$, 
with the above interlacing conditions. 
We call $\G_N,N\in\N$ the Gelfand-Tsetlin (GT) cone 
for signatures. 
Recall that the ordinary GT cone $\G^{(0)}$ is given by (\ref{GN0}), in which 
the sum is over the partitions rather than the signatures. 
The (ordinary) $q$-Whittaker process, introduced  in~\cite{BC2014},
is defined as a process on $\G^{(0)}_N$ and a marginal of the process 
gives  the $q$-TASEP with the step initial condition. 
As a generalization, we will see in the following that a marginal of a process 
on our generalized GT cone $\G_N$  gives the $q$-TASEP with the 
initial condition~\eqref{h-st-ic} with~\eqref{q-geo}.

First we introduce the (skew) $q$-Whittaker function labeled by signatures. 
\begin{definition}{\label{d1}}
Let 
$\l\in\S_n,\mu\in\S_{n-1}$ be two signatures of length $n$ and $n-1$ respectively 
and $a$ an indeterminate. 
The skew $q$-Whittaker function (with 1 variable) is defined as 
\begin{align}
 P_{\l/\mu}\left(a \right)
 =\prod_{i=1}^n a^{\l_i}\cdot
	\prod_{i=1}^{n-1}\frac{ a^{-\mu_i} (q;q)_{\l_i-\l_{i+1}}}
        {(q;q)_{\l_i-\mu_i}(q;q)_{\mu_i-\l_{i+1}}} .
\label{skewWh}
\end{align} 
Using this, for a signature $\l\in\S_N$ and $N$ indeterminates $a=(a_1,\cdots,a_N)$, 
we define the $q$-Whittaker function with $N$ variables  as  
\begin{equation}
P_\l\left(a\right)
=
\sum_{\substack{\l_i^{(k)}, 1\leq i\leq k\leq N-1\\
            \l_{i+1}^{(k+1)} \leq \l_i^{(k)} \leq \l_i^{(k+1)}}}
	\prod_{j=1}^N P_{\l^{(j)}/\l^{(j-1)}}\left(a_j\right).
	\label{Wh}
\end{equation}
Here the sum is over the Gelfand-Tsetlin cone $\G_N$ with the condition $\l^{(N)}=\l$.
\end{definition}
\noindent
Note that the notation $a$ is used for both 1 variable and $N$ variable cases. 
In this paper the skew functions always have only one variable.   
When $\l$ and $\mu$ are ordinary partitions, these definitions 
reduce to the combinatorial definitions of
the skew $q$-Whittaker function (with one variable) (\ref{skewWhp}) and the 
$q$-Whittaker function (\ref{Whp}) respectively~\cite{Mac1995,BC2014}.

Next we define another function labeled by a signature. 
\begin{definition}{\label{d2}}
For a signature $\l$ of length $N$~\eqref{signature}, 
$t>0$
and 
$\a=(\a_1,\cdots,\a_N)\in [0,1)^N$, we define
\begin{align}
Q_\l\left(\alpha,t\right)
 =\prod_{i=1}^{N-1}(q^{\l_i-\l_{i+1}+1};q)_{\infty}
\int_{\T^N}\prod_{i=1}^N\frac{dz_i}{z_i}\cdot P_\l \left(1/z\right)
 \Pi\left(z;\a,t\right)m_N^q\left(z\right), 
\label{defQ}
\end{align}
where $z=(z_1,\cdots,z_N)$, 
\begin{equation}
m_N^q(z)=\frac{1}{(2\pi i)^NN!}\prod_{1\le i<j\le N}(z_i/z_j;q)_{\infty}(z_j/z_i;q)_{\infty}
\label{Sk}
\end{equation} 
is the $q$-Sklyanin measure, 
\begin{align}
\Pi\left(z;\a,t\right)=\prod_{i,j=1}^N\frac{1}{(\a_i/z_j;q)_{\infty}}
\cdot
\prod_{j=1}^Ne^{z_j t} ,
\label{d14}
\end{align}
and $1/z$ in $P_\l$ is a shorthand notation for $(1/z_1,\cdots,1/z_N)$.  
\end{definition}
\noindent 
When $\a_j=0,j=1,\cdots, N$, this function $Q_\l(\a,t)$ becomes zero unless 
$\l\in\G_N^{(0)}$ and on $\G_N^{(0)}$ reduces to 
a representation of the $q$-Whittaker function using the torus scalar product
with the Plancherel specialization, see (\ref{Qtorus}) with the remark below it. 
The case with nonzero $\alpha_j$ is not included as a special case of 
the nonnegative specialization defined for the ordinary $q$-Whittaker functions and 
treated in Sec.2.2.1 of~\cite{BC2014}. 
But one can see that (\ref{defQ}) is in fact nonnegative. 
The notation $\a$ here is for $N$ variables, but when we discuss the $q$-TASEP
we consider the case where only one of them is nonzero, whose value is denoted by
$\alpha\in [0,1)$ corresponding to the parameter in the last section. 

Let us discuss a few properties of the $q$-Whittaker 
functions labeled by signatures. When two signatures $\l,\nu\in\S_n$ 
are related by 
\begin{align}
 \nu_j  = \l_j+k, ~ k\in\Z, ~ 1\leq j\leq n, 
\label{shift}
\end{align}
let us call $\nu$ to be the shift of $\l$ by $k$. 
Suppose that $\nu\in\S_n,\tau\in\S_{n-1}$ are the shifts of 
$\l\in\S_n,\mu\in\S_{n-1}$  respectively by the same $k\in\Z$. 
Then the skew $q$-Whittaker function (\ref{skewWh}) of them are related by 
\begin{equation}
 P_{\l/\mu}(a) = a^{-k} P_{\nu/\tau}(a). 
 \label{shiftP2}
\end{equation}
Accordingly, when $\nu\in\S_N$ is the shift of $\l\in\S_N$ by $k\in\Z$, 
the $q$-Whittaker function (\ref{Wh}) satisfies
\begin{align}
P_\l(a)
=
\prod_{\ell=1}^N a_{\ell}^{-k} \cdot 
P_\nu(a).
\label{shiftP}
\end{align}

For a given signature $\l=(\l_1,\ldots ,\l_n)$ of length $n$, 
one can define another signature of the same length, 
which we call the negation of $\l$ and denote by $-\l$, by
\begin{equation}
 -\l=(-\l_n,-\l_{n-1},\cdots,-\l_1). 
 \label{negl}
\end{equation}  
Also, for a given element $\underline{\l}_N=(\l^{(1)},\ldots ,\l^{(N)})$ 
in the Gelfand-Tsetlin cone $\G_N$, 
we define another element $-\underline{\l}_N\in\G_N$  by
\begin{equation}
-\underline{\l}_N=(-\l^{(1)},\cdots,-\l^{(N)}). 
\label{negGT}
\end{equation}

The (skew) $q$-Whittaker functions introduced above has simple symmetry properties 
with respect to this negation. 
\begin{lemma}
Let $\l\in\S_n,\mu\in\S_{n-1}$ be two signatures of length $n$ and $n-1$ respectively.
We have
\begin{align}
P_{\l/\mu}(a)=P_{(-\l)/(-\mu)}\left(1/a\right).
\label{negP2}
\end{align}
For a signature $\l$ of length $N$ and 
$a=(a_1,\ldots, a_N),\a=(\a_1,\ldots ,\a_N)$, we have 
\begin{align}
&P_\l\left(a\right)=P_{-\l}\left(1/a\right),
\label{negP} \\
&Q_\l(\a,0)
=
\begin{cases}
Q_{-\l}(\a), & \text{when}~~\l^{(N)}_N\le \cdots\le \l^{(N)}_1\le 0,
\\
0, & {\text{~otherwise}},
\end{cases}
\label{negQ}
\end{align}
where $Q_{\l}(x),\l\in\Pe_N$ is defined in (\ref{Q}). 
For the $q$-Sklyanin measure (\ref{Sk}) there is a related symmetry, 
\begin{equation}
m_q(z)=m_q\left(1/z\right).
\label{negSk}
\end{equation}
\end{lemma}

\smallskip
\noindent
{\bf Proof.}
The relations (\ref{negP2}), (\ref{negP})  and (\ref{negSk}) are obvious from the 
definitions~\eqref{skewWh},  ~\eqref{Wh} and~\eqref{Sk}.

For (\ref{negQ}), 
we first note that $\Pi(e^{-i\th};\a,0)=
\prod_{i,j=1}^N\frac{1}{(\a_i e^{i\th_j};q)_\i}$ is a function of $\theta_1,\cdots,\theta_N$
and is periodic with the period $2\pi$ for each $\th_j$. Since $0\leq \a_i <1,1\leq i\leq N$ 
it is expanded uniquely as the Fourier series with some coefficients $d_m$, 
\begin{align}
\Pi\left(e^{-i\theta};\a,0\right)
&=
\sum_{m\in \Z^N}d_m \prod_{j=1}^Ne^{im_j\th_j}.
\label{nl410}
\end{align}
In addition, noting $\Pi(e^{-i\th};\a,0)$ is symmetric in $\th_1,\cdots,\th_N$,
we further write
\begin{align}
\Pi\left(e^{-i\theta};\a,0\right)
=
\sum_{m\in \S_N}\tilde{d}_{m}\sum_{\s\in \Pe_N}\prod_{j=1}^Ne^{i m_{\s(j)}\th_j}
=
\sum_{m\in \S_N}\tilde{d}_{m}
e^{im_N\sum_{j=1}^N\th_j}
\sum_{\s\in \Pe_N}\prod_{j=1}^Ne^{i(m_{\s(j)}-m_N)\th_j},
\label{nl41}
\end{align}
where $\Pe_N$ denotes the set of all permutations of $\{1,\cdots,N\}$.
Since $\sum_{\s\in \Pe_N}\prod_{j=1}^Ne^{i(m_{\s(j)}-m_N)\th_j}$
is symmetric in $\th_1,\cdots, \th_N$ and $m_{\s(j)}-m_N\ge 0$, we
see that it is uniquely expanded by the $q$-Whittaker function with 
some coefficients $\tilde{\gamma}_\nu$ as 
\begin{align}
\sum_{\s\in S_N}\prod_{j=1}^Ne^{i(m_{\s(j)}-m_N)\th_j}
=
\sum_{\nu\in\Pe_N}\tilde{\gamma}_{\nu}P_{\nu}(e^{i\th}).
\label{nl42}
\end{align}
Combining~\eqref{nl41} with~\eqref{nl42} and noting~\eqref{shiftP},
we see that $\Pi\left(e^{-i\theta};\a,0\right)$ can be expanded 
by $P_{\l}(e^{i\th})$ with $\l\in\S_N$,
\begin{align}
\Pi\left(e^{-i\theta};\a,0\right)
=
\sum_{m\in \S_N}\tilde{d}_{m} e^{im_N\sum_{j=1}^N \th_j}
\sum_{\nu\in\Pe_N}\tilde{\gamma}_{\nu}P_{\nu}(e^{i\th})
=
\sum_{\l\in\S_N}\gamma_{\l}P_{\l}(e^{i\th}),
\label{nl43}
\end{align}
with some coefficient $\g_{\l}$. 

Next we discuss the orthogonality of $P_{\l}(z)$. When $\l$ is a partition, 
it is well-known that $P_{\l}(z)$'s are orthogonal with respect to 
the torus scalar product (\ref{torus}), see 
(\ref{torus-ortho}).  
But in fact we can see that the same orthogonality relation holds even for our generalized 
case, in which $\l$ is a signature. 
Note that the relation \eqref{nl46} and thus the first equality in~\eqref{nl47} hold
also for the sigunatures. 
Furthermore, one sees in~\eqref{nl46}, $m\in \Z_{\ge 0}^{N-1}$
even in the case $\mu\in\S_N$.
Thus we can apply~\eqref{nl48} to the first relation in~\eqref{nl47} and 
find that the orthogonal relation~\eqref{torus-ortho} holds also for the case $\mu,\l\in\S_N$. 

Using this, we find that the coefficient $\gamma_{\l}$ in~\eqref{nl43} can be expressed as
\begin{align}
\gamma_{\l}
&=
\prod_{i=1}^{N-1}(q^{\l_i-\l_{i+1}+1};q)_{\infty}
\left\langle
\Pi(e^{-i\th};\a,0),P_{\l}(e^{i\th})
\right\rangle' 
\notag\\
&=
\prod_{i=1}^{N-1}(q^{\l_i-\l_{i+1}+1};q)_{\infty}
\int_{[0,2\pi)^N}\prod_{j=1}^Nd\th_j 
\cdot
P_{\l}(e^{-i\th})\Pi(e^{-i\th};\a,0)m_N^q(e^{i\th}). 
\label{nl49}
\end{align}
By ~\eqref{negP}, ~\eqref{negSk} and (\ref{defQ}), this is written as 
\begin{equation}
\gamma_\l
=
\prod_{i=1}^{N-1}(q^{\l_i-\l_{i+1}+1};q)_{\infty}
\int_{[0,2\pi)^N}\prod_{j=1}^Nd\th_j 
\cdot
P_{-\l}(e^{-i\th})\Pi(e^{i\th};\a,0)m_N^q(e^{i\th})
=Q_{-\l}(\a,0). 
\label{gamma1}
\end{equation} 
On the other hand, using the Cauchy identity (\ref{CI}) in (\ref{nl49}), we find 
\begin{equation}
\gamma_\l 
=
\begin{cases}
Q_\l(\a), & \l~ \text{is a partition}, \\
0, & \text{otherwise}.   
\end{cases}
\label{gamma2}
\end{equation} 
Comparing these two expressions (\ref{gamma1}) and (\ref{gamma2}), 
we get~\eqref{negQ}.
\qed

\subsection{Two-sided $q$-Whittaker process}
Using Definitions~\ref{d1} and~\ref{d2}, we introduce a measure on $\G_N$. 
\begin{definition}
 For $\underline{\l}_N\in\G_N$, we define
\begin{align}
P_t (\underline{\l}_N)
 := 
 \frac{\prod_{j=1}^N P_{\l^{(j)}/\l^{(j-1)}} \left( a_j\right) \cdot Q_{\l^{(N)}} \left(\a,t\right)}{\Pi(a;\a,t)} .
\label{2qWh-pr}
\end{align}
\end{definition}
\noindent
We call this the two-sided $q$-Whittaker process.  At this point it is not obvious that 
the rhs of (\ref{2qWh-pr}) indeed gives  a probability measure but we will see it is 
indeed the case in the next subsection.  

When $\a_j=0$ for all $j=1,2,\cdots,N$, 
this reduces to the ascending $q$-Whittaker
process with the Plancherel specialization~\cite{BC2014}, 
for which it has been shown that the marginal with respect to 
$(\l^{(1)}_1,\cdots,\l^{(N)}_N)$ describes the probability density function
of the particle positions in the $q$-TASEP with step initial condition 
at time $t$.

We will show that the $P_t(\underline{\l}_N)$ is a pdf for a 
dynamics of interacting random walkers on  $\G_N$ at time $t$.
Furthermore we show that its marginal density of
$(\l^{(1)}_1,\cdots,\l^{(N)}_N)$ in the special 
case where only one of $\a_j$ remains finite and all the other ones
are zero,
describes the $q$-TASEP with the initial condition (\ref{h-st-ic})  
defined in the previous section. 

First we show that at $t=0$ the measure (\ref{2qWh-pr}) is the 
pure $\a$ specialization~\cite{BC2014} for $-\l$. 
Substituting (\ref{negP2}) and~\eqref{negQ} into~\eqref{2qWh-pr}, we have 
\begin{proposition}
\label{p2} 
At $t=0$, $P_0(\underline{\l}_N)$ 
has a support on $-\G^{(0)}_N$~\eqref{GN0} and
on this support it is written as the $q$-Whittaker process
with pure $\a$ specialization, 
i.e., with $\underline{\mu}_N = -\underline{\lambda}_N$
defined by  ~\eqref{negGT}, 
\begin{align}
P_0(\underline{\l}_N)
=
\begin{cases}
 \frac{\prod_{j=1}^NP_{\mu^{(j)}/\mu^{(j-1)}}\left(1/a_j\right)
\cdot Q_{\mu^{(N)}}\left(\a\right)}{\Pi(a;\a;0)},
& \underline{\mu}_N\in\G^{(0)}_N,\\
0,
&\underline{\mu}_N\notin\G^{(0)}_N ,
\end{cases}
\label{p20}
\end{align}
where $P_{\mu^{(j)}/\mu^{(j-1)}}$ is the skew $q$-Whittaker function (\ref{skewWhp}) 
and $Q_{\mu^{(N)}}(\a)$ is given by (\ref{Q}). 
In particular $P_{0}(\underline{\l}_N)$ is a pdf on $-\G_N^{(0)}$. 
\end{proposition}

\noindent
The $q$-TASEP with the initial condition (\ref{h-st-ic}) corresponds to  
the case in which one of $\a_j=\a,1\leq j\leq N$ and $\a_k=0,k\neq j$.
But since $Q_{\mu^{(N)}}(\a)$ is a symmetric function, the value of 
$P_0(\underline{\l}_N)$ in~\eqref{p20} does not depend on which $\a_j$
is nonzero. The Proposition~\ref{p2} restricted to this case is the following:
\begin{proposition}
In the case where one of $\a_j=\a$ and $\a_k=0,k\neq j,1\leq j,k\leq N$, 
we have, for $\underline{\l}_N\in\G_N$, 
\begin{align}
P_0(\underline{\l}_N)
=
\prod_{j=1}^N
\left(\frac{\a}{a_j};q\right)_{\infty}
\frac{\left(\a/a_j\right)^{\l^{(j-1)}_{j-1}-\l^{(j)}_j}}
{(q;q)_{\l^{(j-1)}_{j-1}-\l^{(j)}_j}}\cdot
\prod_{k=1}^N
\prod_{j=1}^{k-1}\delta_{\lambda^{(k)}_{j},0}.
\label{p0}
\end{align}
\end{proposition}

\smallskip
\noindent 
This vanishes unless $\l_j^{(k)}=0, 1\leq j<k\leq N$ and 
$\l^{(N)}_N\le\l^{(N-1)}_{N-1}\le\cdots\le\l^{(1)}_1\le 0$. 
Focusing on $\l_j^{(j)},1\leq j\leq N$ and comparing with  (\ref{h-st-ic}), 
we see that the marginal density at time $t=0$
is exactly the initial condition (\ref{h-st-ic})  for the $q$-TASEP. 

\smallskip
\noindent
{\bf Proof.}
As we mentioned above,~\eqref{p20} is independent of which $\a_j$ remains finite.   
We adopt the choice $j=N$, i.e. $\a_1=\cdots=\a_{N-1}=0,~\a_N=\a$,
where we can give the simplest proof.

First we consider the factor 
\begin{align}
Q_{\l^{(N)}}(\a,t=0)
&=
Q_{\mu^{(N)}}(0,\cdots,0,\a) 
=
\frac{P_{\mu^{(N)}}(0,\cdots,0,\a)}
{(q;q)_{\mu^{(N)}_N}\prod_{j=1}^{N-1}(q;q)_{\mu^{(N)}_{j}-\mu^{(N)}_{j+1}}},
\label{i-2e}
\end{align}
where in the first equality we used (\ref{negQ}) and in the second the definition~\eqref{Q} 
with (\ref{Pnorm}).
By the definition~\eqref{Whp}, $P_{\mu^{(N)}}(0,\cdots,0,\a)$ is given as
\begin{align}
~P_{\mu^{(N)}}(0,\cdots,0,\a)
&=
\sum_{\mu^{(1)},\cdots,\mu^{(N-1)}}
	\prod_{j=1}^{N-1} P_{\mu^{(j)}/\mu^{(j-1)}}\left(0\right)
\cdot P_{\mu^{(N)}/\mu^{(N-1)}}\left(\a\right)
\notag
\\
&=
\sum_{\mu^{(1)},\cdots,\mu^{(N-1)}}
	\prod_{j=1}^{N-1} \delta_{\mu^{(j)},\varnothing}
\prod_{i=1}^N\a^{\mu^{(N)}_i}\cdot\prod_{i=1}^{N-1}
\frac
{(q;q)_{\mu^{(N)}_i-\mu^{(N)}_{i+1}}}
{(q;q)_{\mu^{(N)}_{i}}(q;q)_{-\mu^{(N)}_{i+1}}}
\notag\\
&=
\prod_{i=1}^N\a^{\mu^{(N)}_i}\cdot\prod_{i=1}^{N-1}
\frac
{(q;q)_{\mu^{(N)}_i-\mu^{(N)}_{i+1}}}
{(q;q)_{\mu^{(N)}_{i}}(q;q)_{-\mu^{(N)}_{i+1}}}
=\a^{\mu^{(N)}_1}\prod_{j=1}^{N-1}\delta_{\mu^{(N)}_{j+1},0},
\label{i-2d}
\end{align}
where we wrote the summation in~\eqref{Whp} as $\sum_{\mu^{(1)},\cdots,\mu^{(N-1)}}$ and in the second equality we used the fact 
$P_{\mu^{(j)}/\mu^{(j-1)}}\left(0\right)
=\delta_{\mu^{(j)},\mu^{(j-1)}}$ while in the last equality we used another
fact: for $n\in\N$, 
$
1/(q;q)_{-n}
=
(q^{n+1};q)_{\infty}/
(q;q)_{\infty}
=\delta_{n,0}
$.
Thus from~\eqref{i-2e} and~\eqref{i-2d}, we see that 
\begin{align}
Q_{\mu^{(N)}}(0,\cdots,0,\a)
=\frac{\a^{\mu^{(N)}_1}}{(q;q)_{\mu^{(N)}_1}}
   \prod_{j=1}^{N-1}\delta_{\mu^{(N)}_{j+1},0}
\label{i-2}
\end{align}
and that $P_{0}(\underline{\l}_N)$ can be written as
\begin{align}
&P_{t=0}(\underline{\l}_N)
=
\prod_{j=1}^N\left(\frac{\a}{a_j};q\right)_\infty\cdot\prod_{j=1}^{N-1}
P_{\mu^{(j)}/\mu^{(j-1)}}\left(\frac{1}{a_j}\right)
\cdot
P_{\mu^{(N)}/\mu^{(N-1)}}\left(\frac{1}{a_N}\right)
\frac{\a^{\mu^{(N)}_1}}{(q;q)_{\mu^{(N)}_1}}
   \prod_{j=1}^{N-1}\delta_{\mu^{(N)}_{j+1},0}.
\label{i-10}
\end{align}
Here we find that the last factor in this equation can be calculated as 
\begin{align}
&P_{\mu^{(N)}/\mu^{(N-1)}}\left(\frac{1}{a_N}\right)
\frac
{\a^{\mu^{(N)}_1}}
{(q;q)_{\mu^{(N)}_1}}
\prod_{j=1}^{N-1}\delta_{\mu^{(N)}_{j+1},0}
\notag
\\
&=
\left(\frac{1}{a_N}\right)^{\mu^{(N)}_1-\mu^{(N-1)}_1}
\frac
{(q;q)_{\mu^{(N)}_1}}
{(q;q)_{\mu^{(N)}_{1}-\mu^{(N-1)}_1}(q;q)_{\mu^{(N-1)}_{1}}}
\prod_{i=2}^{N-1}
\frac{a_N^{\mu^{(N-1)}_i}}
{(q;q)_{-\mu^{(N-1)}_i}(q;q)_{\mu^{(N-1)}_i}}
\cdot
\frac
{\a^{\mu^{(N)}_1}}
{(q;q)_{\mu^{(N)}_1}}
\prod_{j=1}^{N-1}\delta_{\mu^{(N)}_{j+1},0}
\notag\\
&=
\frac
{\a^{\mu^{(N-1)}_1}}
{(q;q)_{\mu^{(N-1)}_1}}
\prod_{j=1}^{N-2}\delta_{\mu^{(N-1)}_{j+1},0}
\cdot
\frac
{(\a/a_N)^{\mu^{(N)}_{1}-\mu^{(N-1)}_1}}
{(q;q)_{\mu^{(N)}_{1}-\mu^{(N-1)}_1}}
\prod_{j=1}^{N-1}\delta_{\mu^{(N)}_{j+1},0}.
\label{i-3}
\end{align}
Substituting this into (\ref{i-10}), 
we get
\begin{align}
P_{t=0}(\underline{\l}_N)&=
\prod_{j=1}^N\left(\frac{\a}{a_j};q\right)_\infty
\cdot
\frac
{\left(\frac{\a}{a_N}\right)^{\mu^{(N)}_{1}-\mu^{(N-1)}_1}}
{(q;q)_{\mu^{(N)}_{1}-\mu^{(N-1)}_1}}
\prod_{j=1}^{N-1}\delta_{\mu^{(N)}_{j+1},0}
\cdot\prod_{j=1}^{N-2}P_{\mu^{(j)}/\mu^{(j-1)}}
\left(\frac{1}{a_{j}}\right)
\notag
\\
&~~~~\times
P_{\mu^{(N-1)}/\mu^{(N-2)}}\left(\frac{1}{a_{N-1}}\right)
\frac{\a^{\mu^{(N-1)}_1}}{(q;q)_{\mu^{(N-1)}_1}}
   \prod_{j=1}^{N-2}\delta_{\mu^{(N-1)}_{j+1},0}.
\label{i-11}
\end{align}
Note that the last factor in this equation is the same as the one in~\eqref{i-10} with $N$ replaced by $N-1$ and~\eqref{i-3} holds for any $N=1,2,\cdots$. 
Applying~\eqref{i-3} with $N-1,~N-2\cdots,~1$ repeatedly
to~\eqref{i-11} and
remembering $\underline{\mu}_N=-\underline{\l}_N$, we obtain~\eqref{p0}.
\qed
\smallskip

\begin{figure}[t]
\begin{center}
\includegraphics[scale=0.6]{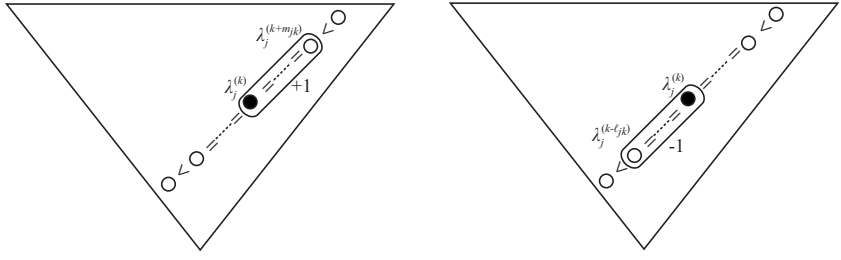}
\caption{\label{figGTpm}
Definition of $m_{j,k}$ and $\ell_{j,k}$ in GT cone. The change in 
$\l^{jk+}$ and $\l^{jk-}$ are also indicated. 
}
\end{center}
\end{figure}

Next we consider a time evolution property of $P_t(\underline{\l}_N)$~\eqref{2qWh-pr}.
For this purpose,
for a $\underline{\lambda}_N\in \G_N$, we define $\ell_{j,k}=\ell_{j,k}(\underline{\l}_N)$ and 
$m_{j,k}=m_{j,k}(\underline{\l}_N)$ by the relations, 
\begin{align}
\l^{(k-\ell_{j,k}-1)}_j<\l^{(k-\ell_{j,k})}_j =\cdots=\l^{(k-1)}_j=\l^{(k)}_j
=\l^{(k+1)}_j=\cdots=\l^{(k+m_{j,k})}_j< \l^{(k+m_{j,k}+1)}_j.
\label{t21}
\end{align} 
Namely, for a given $\underline{\l}_N$, $\ell_{j,k}$ (resp. $m_{j,k}$) denotes the number of particles 
on the GT cone which share the position and the lower indices with $\l_j^{(k)}$ and whose upper indices are less 
(resp. bigger) than $\l_j^{(k)}$. 
We also define $\underline{\l}^{jk\pm}_{N}$ as
\begin{align}
\left(\l^{jk+}\right)^{(a)}_b
&=
\begin{cases}
\l^{(a)}_b+1, & a\in \{k, k+1,\cdots, k+m_{jk}\} \text{~and~} b=j,\\
\l^{(a)}_b,& \text{otherwise},
\end{cases}
\label{t22}
\\
\left(\l^{jk-}\right)^{(a)}_b&=
\begin{cases}
\l^{(a)}_b-1, & a\in \{k-\ell_{jk},k-\ell_{jk}+1,\cdots, k\} \text{~and~} b=j,\\
\l^{(a)}_b, &\text{otherwise}.
\end{cases}
\label{t23}
\end{align}
In other words, $\underline{\l}^{jk+}_{N}$ (resp. $\underline{\l}^{jk-}_{N}$) is 
the configuration of particles in $\G_N$ in which the whole $m_{j,k}+1$
(resp. $\ell_{j,k}+1$) particles in $\underline{\l}_N$ which share the positions with $\l_j^{(k)}$ and whose lower indices are $j$ and the upper one is 
larger (resp. smaller) than $k$ move one step to the right (resp. left). 
See Fig.  \ref{figGTpm}.

Then we have the following Kolmogorov forward equation.
\begin{theorem}\label{thme}
For $\underline{\l}_N\in\G_N$, we have
\begin{align}
\frac{dP_t(\underline{\lambda}_N)}{dt}
=&
\sum_{\underline{\mu}_N\in\G_N}
P_t(\underline{\mu}_N)L(\underline{\mu}_N,\underline{\l}_N)
\label{me}
\end{align}
where the generator $L$ is expressed as
\begin{align}
&L(\underline{\mu}_N,\underline{\l}_N)
=
\sum_{1\le j\le k\le N}
r_{jk}
\left(
\delta_{\underline{\l}_N,\underline{\mu}^{jk+}_N}
-\delta_{\underline{\l}_N,\underline{\mu}_N}
\right).
\label{genL}
\end{align}
Here the rate $r_{j,k}=r_{jk}(\mu_{j-1}^{(k-1)},\mu_j^{(k-1)},\mu_j^{(k)},\mu_{j+1}^{(k)})$ is given by 
\begin{align}
r_{jk}
=
a_k
\frac
{(1-q^{\mu^{(k-1)}_{j-1}-\mu^{(k)}_j})(1-q^{\mu^{(k)}_{j}-\mu^{(k)}_{j+1}+1})}
{1-q^{\mu^{(k)}_{j}-\mu^{(k-1)}_j+1}}, 
\label{defr}
\end{align}
with the convention $\mu_{0}^{(j)}=\infty,~\mu_{j+1}^{(j)}=-\infty$.
\end{theorem}
\smallskip
We find $L$ in~\eqref{genL} indeed satisfies the conditions of a generator of a
Markov chain, see for instance the chapter 2 of \cite{Liggett2010}. It is easy to see
$L(\underline{\mu}_N,\underline{\l}_N)\geq 0$ 
for $\underline{\mu}_N\neq \underline{\l}_N$ since $r_{jk}\geq 0$ 
for $\underline{\mu}_N\in \G_N$.
One also finds $\sum_{\underline{\l}_N}L(\underline{\mu}_N,\underline{\l}_N)=0$,
by noting that if $\underline{\mu}_N\in\G_N$, 
\begin{align}
r_{jk}=0
\Leftrightarrow 
\mu^{(k-1)}_{j-1}=\mu^{(k)}_j
\Leftrightarrow
\underline{\mu}_N^{jk+}\notin\G_N.
\label{ll82}
\end{align}
This theorem tells us that $P_t(\underline{\l}_N)$ represents
a (discrete) probability density of an interacting particle system on $\G_N$~\eqref{GT} 
described by the following rules:
Suppose for $j=1,\cdots,k,~k=1,\cdots,N$ each particle labeled $(j,k)$ 
has the position $\l^{(k)}_j$ at time $t$ where $(\l^{(1)},\cdots,\l^{(N)})\in\G_N$.
Then the particle $(j,k)$ hops from $\l^{(k)}_j$ to $\l^{(k)}_j+1$ with rate
$r_{jk}$. 
Furthermore when the particle $(j,k)$ hops, the other particles labeled 
$(j,k+1),\cdots,(j,k+m_{jk})$ are pushed to proceed forward one step, to keep the interlacing 
condition of the GT cone.
Note that from~\eqref{ll82}, $r_{jk}=0$ when $\l^{(k)}_j=\l^{(k-1)}_{j-1}$, which means that
the particle $(j,k)$ is blocked by the particle $(j-1,k-1)$ when they have the same position, again to 
maintain the interlacing condition of the GT cone. 
In addition, one notices that the rules for $\l_j^{(j)}(t),1\leq j\leq N$ are independent from the rest and 
each $\l_j^{(j)}$ hops to the right with rate $a_j(1-q^{\l_{j-1}^{(j-1)}-\l_j^{(j)}})$ unless 
$\l_{j-1}^{(j-1)}=\l_j^{(j)}$.  

The above rules of the dynamics are the same as the $q$-Whittaker process 
with Plancherel specialization introduced in \cite{BC2014}. 
In particular, if we set $X_j(t)=\l_j^{(j)}(t)-j+1$, the rules of the dynamics of $X_j(t)$ are 
exactly the ones of $q$-TASEP. In \cite{BC2014} this dynamics is shown to appear as 
a limiting case of a discrete time dynamics of the Macdonald process for the step initial 
condition (the last part of section 2.3 in \cite{BC2014}. See also Proposition~\ref{p26}
for a little reformulated proof for the discrete dynamics). 
Actually the generator~\eqref{genL} can be obtained
by the expansion of 
the transition matrix $G_M(\underline{\mu}_N,\underline{\l}_N)$
of the discrete time case~\eqref{a4} as,
\begin{align}
G_M(\underline{\mu}_N,\underline{\l}_N)=\delta_{\underline{\mu}_N,\underline{\l}_N}
+\e
L(\underline{\mu}_N,\underline{\l}_N)+O(\e^2)
\label{p923}
\end{align}
where in~\eqref{a4} we set $M=t \e$ and expand $G_M(\underline{\mu}_N,\underline{\l}_N)$
in powers of $\e$. 

In the rest of this subsection, we give a proof of Theorem~\ref{thme}. Namely 
we will show that $P_t(\underline{\l})$ satisfies the Kolmogorov forward equation (\ref{me}) directly,
without detouring through a discrete time dynamics, for our random initial condition, 
using a few properties of the $q$-Whittaker functions.  
Below we will also use notation 
$P(\mu,\l;a)=P_{\l/\mu}(a),P(\l;a)=P_{\l}(a),Q(\l;\a,t)=Q_{\l}(\a,t)$ 
for the (skew) $q$-Whittaker 
functions to avoid complicated subscripts. 
We also use a notation $\l\pm 1_j$ to denote the Young diagram with one box 
added to (or removed from) $\l$. 
For the proof, we prepare three lemmas. 
\begin{lemma}
\label{ll8}
The rate $r_{jk}$~\eqref{defr} can be expressed as
\begin{align}
r_{jk}
=
(1-q^{\mu^{(k)}_{j-1}-\mu^{(k)}_j})
\frac
{P(\mu^{(k-1)},\mu^{(k)}+1_j;a_k)}
{P(\mu^{(k-1)},\mu^{(k)};a_k)}.
\label{ll81}
\end{align}
\end{lemma}

\smallskip
\noindent
{\bf Proof.}
It follows immediately from the relation
for the $q$-Whittaker functions,
\begin{align}
&
P(\mu^{(k-1)},\mu^{(k)}+1_j;a_k)
=
a_k
\frac
{1-q^{\mu^{(k-1)}_{j-1}-\mu^{(k)}_j}}
{1-q^{\mu^{(k)}_{j}-\mu^{(k-1)}_{j}+1}}
\frac
{1-q^{\mu^{(k)}_{j}-\mu^{(k)}_{j+1}+1}}
{1-q^{\mu^{(k)}_{j-1}-\mu^{(k)}_j}}
P(\mu^{(k-1)},\mu^{(k)};a_k),
\label{l61}
\end{align}
which is obtained by the definition~\eqref{skewWhp}.
\qed

\smallskip

The next lemma is a direct consequence of the 
properties (\ref{l413}),(\ref{l415}) of the $q$-Whittaker functions. 
\begin{lemma}\label{lem3}
For $\l\in\S_k$ and $\mu\in\S_{k-1}$,~$k=1,2,\cdots$, we have
\begin{align}
&\sum_{l=1}^k(1-q^{{\l}_{l-1}-\l_{l}})P(\mu,\l+1_l;a_k)
\notag
\\
&~~~~=a_kP(\mu,\l ;a_k)+
\sum_{l=1}^{k-1}(1-q^{\mu_{l-1}-\mu_{l}+1})P(\mu-1_l,\l ;a_k),
\label{l411}
\\
&
\frac{\partial}{\partial t} Q(\l; \a,t)
=
\sum_{l=1}^k (1-q^{\l_{l-1}-\l_{l}+1}) Q(\l-1_l;\a,t)
\label{l412}
\end{align}
with the convention $\l_0 = +\infty$. 
\end{lemma}
\noindent
The following special case of (\ref{l411}) will be useful in the following. 
\begin{corollary}
\label{corPP}
For $\l\in\S_k$ and $\mu\in\S_{k-1}$, suppose $\mu=\l+1_j$ holds for some
$j, 1\le j\le k-1$. Then we have
\begin{align}
(1-q^{\l_{j-1}-\l_j})P(\mu,\l+1_j;a_k)
=
(1-q^{\mu_{j-1}-\mu_j+1})P(\mu-1_j,\l;a_k).
\end{align}
\end{corollary}
\noindent
{\bf Proof.}
Recall the skew $q$-Whittaker function $P(\mu,\l;a)=P_{\l/\mu}(a)$ vanishes 
if $\l/\mu$ is not a skew diagram. With the condition $\mu=\l+1_j$, only the 
$j$th terms on both sides (\ref{l411}) remain. 
\qed

\medskip
\noindent
{\bf Proof of Lemma \ref{lem3}.}
Remembering~\eqref{shiftP2} and \eqref{shiftP}, and noting
$\nu$ in~\eqref{shift} becomes an ordinary partition with
positive integer values for sufficiently large $k$, 
one sees it is enough to show~\eqref{l411} and~\eqref{l412}
for ordinary partitions.

First~\eqref{l411} comes from a property (\ref{l413}) of the $q$-Whittaker functions.
Noting $Q^{(\b)}_{\l/\mu}(r)$ with one variable (\ref{a1}) can be expanded as 
\begin{align}
Q_{\l/\mu}^{(\beta)}(r)
=
\delta_{\l,\mu}+r\sum_{j=1}^{\ell(\mu)}(1-q^{\mu_{j-1}-\mu_{j}})\delta_{\l,\mu+1_j}
+O(r^2)
\label{l414}
\end{align}
for small $r$ and comparing the coefficients of the $O(r^1)$ terms 
in both sides of~\eqref{l413} with $a=a_k$ and $\l,\mu$ exchanged, 
we obtain~\eqref{l411}.

Next, for~\eqref{l412}, we use the relation (\ref{l415}). 
In a similar way to the previous case, using~\eqref{l414}
and the representation with the torus scalar product (cf (\ref{Qtorus}) with (\ref{nnsp})), 
\begin{align}
Q_{\l}^{(\beta)}(r_1,\cdots,r_M)
=
\prod_{i=1}^{k-1}(q^{\l_i-\l_{i+1}+1};q)_{\infty}
\int_{\T^k}\prod_{j=1}^k\frac{dz_j}{z_j}\cdot P\left(\l;\frac{1}{z}\right)
 \prod_{i=1}^M\prod_{j=1}^k\left(1+r_i z_j\right)m_N^q\left(z\right),
\label{l416}
\end{align}
we compare the coefficients of $O(r_{M+1}^1)$ term in both hand sides of~\eqref{l415},
which leads to the relation for $P_{\l}(z)$,
\begin{align}
\sum_{j=1}^k (1-q^{\l_{j-1}-\l_j+1})P(\l-1_j;1/z)
=\sum_{j=1}^k z_jP(\l;1/z).
\label{l417}
\end{align}
Differentiating (\ref{defQ}) this respect to $t$ and using (\ref{l417}), we get~\eqref{l412}.
\qed

\smallskip

The last lemma provides a decomposition of the generator 
$L(\underline{\mu}_N,\underline{\l}_N)$~\eqref{genL}.
\begin{lemma}
\label{p9}
The generator $L$ in (\ref{genL}) can be written as 
\begin{align}
L(\underline{\mu}_N,\underline{\l}_N)
=
\sum_{r=1}^N A_r(\underline{\mu}_N,\underline{\l}_N)+
B(\underline{\mu}_N,\underline{\l}_N),
\label{p94}
\end{align}
where 
\begin{align}
A_r(\underline{\mu}_N,\underline{\l}_N)
&=\sum_{1\le j\le k\le r-1}
(1-q^{\mu_{j-1}^{(r-1)}-\mu_{j}^{(r-1)}})
\prod_{m=k}^{r-1}
\frac
{P(\l^{(m-1)}, \l^{(m)};a_m)}
{P(\mu^{(m-1)}, \mu^{(m)};a_m)}\delta_{m_{jk},r-1-k}
~\delta_{\underline{\l}_N,\underline{\mu}^{jk+}_N}
\notag
\\
&~~~~-
\sum_{j=1}^{r-1}
(1-q^{\mu_{j-1}^{(r-1)}-\mu_{j}^{(r-1)}+1})
\frac
{P(\mu^{(r-1)}-1_j,\mu^{(r)};a_r)}
{P(\mu^{(r-1)},\mu^{(r)};a_r)}
\delta_{\underline{\l}_N,\underline{\mu}_N}
-
a_r
\delta_{\underline{\l}_N,\underline{\mu}_N},
\label{p98}
\\
B(\underline{\mu}_N,\underline{\l}_N)
&=
\sum_{1\le j\le k\le N}
(1-q^{\mu_{j-1}^{(N)}-\mu_{j}^{(N)}})
\prod_{m=k}^{N}
\frac
{P(\l^{(m-1)}, \l^{(m)};a_m)}
{P(\mu^{(m-1)}, \mu^{(m)};a_m)}
\cdot
\delta_{m_{jk},N-k}
\delta_{\underline{\l}_N,\underline{\mu}^{jk+}_N}.
\label{p914}
\end{align}
\end{lemma}
\smallskip
\noindent
{\bf Proof.} 
By~\eqref{ll81}, $L(\underline{\mu}_N,\underline{\l}_N)$~\eqref{genL}
is written as
\begin{align}
L(\underline{\mu}_N,\underline{\l}_N)
=
\sum_{1\le j\le k\le N}
(1-q^{\mu^{(k)}_{j-1}-\mu^{(k)}_j})
\frac
{P(\mu^{(k-1)},\mu^{(k)}+1_j;a_k)}
{P(\mu^{(k-1)},\mu^{(k)};a_k)}
\left(
\delta_{\underline{\l}_N,\underline{\mu}^{jk+}_N}
-\delta_{\underline{\l}_N,\underline{\mu}_N}
\right).
\label{p91}
\end{align}
Using  $1=\sum_{r=k}^{N}\delta_{m_{jk},r-k}$,
we further rewrite $L(\underline{\l}_N,\underline{\mu}_N)$~\eqref{p91} as
\begin{align}
L(\underline{\mu}_N,\underline{\l}_N)
&=
\sum_{r=1}^N
\sum_{1\le j\le k\le r}
(1-q^{\mu_{j-1}^{(k)}-\mu_{j}^{(k)}})
\frac
{P(\mu^{(k-1)},\mu^{(k)}+1_j;a_k)}
{P(\mu^{(k-1)},\mu^{(k)};a_k)}
\left(
\delta_{m_{jk},r-k}
\delta_{\underline{\l}_N,\underline{\mu}^{jk+}_N}
-
\delta_{r,k}\delta_{\underline{\l}_N,\underline{\mu}_N}
\right),
\label{p93}
\end{align}
from which we find a decomposition of the generator (\ref{genL}) of the 
form (\ref{p94}) with $A_r,B$ replaced by 
\begin{align}
\tilde{A}_r(\underline{\mu}_N,\underline{\l}_N)
&=
\sum_{1\le j\le k\le r-1}
(1-q^{\mu_{j-1}^{(k)}-\mu_{j}^{(k)}})
\frac
{P(\mu^{(k-1)},\mu^{(k)}+1_j;a_k)}
{P(\mu^{(k-1)},\mu^{(k)};a_k)}
\delta_{m_{jk},r-1-k}
\delta_{\underline{\l}_N,\underline{\mu}^{jk+}_N}
\notag
\\
&~~~~-\sum_{j=1}^r
(1-q^{\mu_{j-1}^{(r)}-\mu_{j}^{(r)}})
\frac
{P(\mu^{(r-1)},\mu^{(r)}+1_j;a_r)}
{P(\mu^{(r-1)},\mu^{(r)};a_r)}
\delta_{\underline{\l}_N,\underline{\mu}_N},
\label{p95}
\\
\tilde{B}(\underline{\mu}_N,\underline{\l}_N)
&=
\sum_{1\le j\le k\le N}
(1-q^{\mu_{j-1}^{(k)}-\mu_{j}^{(k)}})
\frac
{P(\mu^{(k-1)},\mu^{(k)}+1_j;a_k)}
{P(\mu^{(k-1)},\mu^{(k)};a_k)}
\delta_{m_{jk},N-k}
\delta_{\underline{\l}_N,\underline{\mu}^{jk+}_N}.
\label{p96}
\end{align}
In fact in (\ref{p93}), the  $(r+1)$th summand, $0\leq r\leq N-1$, 
of the first term and the $r$th summand of the second term after the sum over $k$ is 
taken gives~\eqref{p95} whereas the $N$-th summand of the first term gives $\tilde{B}$. 
Combining these two, we get the decomposition~\eqref{p94} with 
(\ref{p95}),(\ref{p96}). 
Below we show $\tilde{A}_r=A_r, \tilde{B}=B$. 

First we see $\tilde{A}_r=A_r$. 
Setting $\mu=\mu^{(m-1)}+1_j$ and $\l=\mu^{(m)}$ for $m=k+1,\cdots,r-1$
in~\eqref{l411}, we see
\begin{align}
1
=
\frac
{1-q^{\mu^{(m)}_{j-1}-\mu^{(m)}_j}}
{1-q^{\mu^{(m-1)}_{j-1}-\mu^{(m-1)}_j}}
\frac
{P(\mu^{(m-1)}+1_j, \mu^{(m)}+1_j,a_m)}
{P(\mu^{(m-1)},\mu^{(m)},a_m)}
\label{p97}
\end{align}
where we used the fact that $\mu^{(k)}_j=\cdots=\mu^{(r)}_j$ (since $m_{jk}=r-1-k$) and
\begin{align}
P(\mu^{(m-1)}+1_j,\mu^{(m)}+1_l,a_m)=P(\mu^{(m-1)}+1_j-1_l,\mu^{(m)},a_m)=
P(\mu^{(m-1)}+1_j,\mu^{(m)},a_m)=0
\label{p915}
\end{align}
for $m\neq j$.
Using~\eqref{p97} in the first term of~\eqref{p95} and~\eqref{l411} with $\mu=\mu^{(r-1)}$
and $\l=\mu^{(r)}$ in the second term, we have
\begin{align}
&\tilde{A}_r(\underline{\mu}_N,\underline{\l}_N)
\notag
\\
&=
\sum_{1\le j\le k\le r-1}
(1-q^{\mu_{j-1}^{(k)}-\mu_{j}^{(k)}})
\frac
{P(\mu^{(k-1)},\mu^{(k)}+1_j;a_k)}
{P(\mu^{(k-1)},\mu^{(k)};a_k)}
\notag
\\
&~~~~~~~~~~~~\times\prod_{m=k+1}^{r-1}
\frac
{1-q^{\mu^{(m)}_{j-1}-\mu^{(m)}_j}}
{1-q^{\mu^{(m-1)}_{j-1}-\mu^{(m-1)}_j}}
\frac
{P(\mu^{(m-1)}+1_j, \mu^{(m)}+1_j;a_m)}
{P(\mu^{(m-1)}, \mu^{(m)};a_m)}
\cdot
\delta_{m_{jk},r-1-k}
\delta_{\underline{\l}_N,\underline{\mu}^{jk+}_N}
\notag
\\
&~~~~-
\sum_{j=1}^{r-1}
(1-q^{\mu_{j-1}^{(r-1)}-\mu_{j}^{(r-1)}+1})
\frac
{P(\mu^{(r-1)}-1_j,\mu^{(r)};a_r)}
{P(\mu^{(r-1)},\mu^{(r)};a_r)}
\delta_{\underline{\l}_N,\underline{\mu}_N}
-
a_r
\delta_{\underline{\l}_N,\underline{\mu}_N}.
\label{p916}
\end{align}
Using 
\begin{align}
&(1-q^{\mu_{j-1}^{(k)}-\mu_{j}^{(k)}})
\prod_{m=k+1}^{r-1}
\frac
{1-q^{\mu^{(m)}_{j-1}-\mu^{(m)}_j}}
{1-q^{\mu^{(m-1)}_{j-1}-\mu^{(m-1)}_j}}
=
1-q^{\mu^{(r-1)}_{j-1}-\mu^{(r-1)}_{j}},
\label{p917}
\\
&\frac
{P(\mu^{(k-1)},\mu^{(k)}+1_j;a_k)}
{P(\mu^{(k-1)},\mu^{(k)};a_k)}
\prod_{m=k+1}^{r-1}
\frac
{P(\mu^{(m-1)}+1_j, \mu^{(m)}+1_j;a_m)}
{P(\mu^{(m-1)}, \mu^{(m)};a_m)}
=
\prod_{m=k}^{r-1}
\frac
{P(\l^{(m-1)}, \l^{(m)};a_m)}
{P(\mu^{(m-1)}, \mu^{(m)};a_m)},
\label{p918}
\end{align}
where $\l^{(k-1)}=\mu^{(k-1)}$, 
$\l^{(m)}=\mu^{(m)}+1_j$ 
for $m=k,\cdots,r-1$,
it is easy to see $\tilde{A}_r=A_r$.

Note that $\tilde{B}$ 
is equal to the first term of $A_{r}$ with $r=N+1$.
Thus we find $\tilde{B}$ is rewritten as~\eqref{p914} in a similar way 
to the first term of~\eqref{p98}.
\qed

\smallskip
\noindent
As we mentioned below Theorem~\ref{thme}, the generator
$L(\underline{\mu}_N,\underline{\l}_N)$ (\ref{genL}) appears in the 
continuous time limit of the transition matrix $G_M(\underline{\mu}_N,\underline{\l}_N)$,
see~\eqref{p923}. 
The decomposition in the proposition is related to the one in~\eqref{a5}
for the discrete-time dynamics, which is introduced
for the proof of Proposition~\ref{p26}. 
Applying the small $\e$ expansion to the representation~\eqref{a5}, one would see
$A_r$ (resp. $B$) in~\eqref{p94} 
corresponds to $A_r^{(\beta)}$ (resp. $B^{(\b)}$). 

In addition, note that $A_r^{(\beta)}$ can be written in two ways as 
\begin{align}
A^{(\beta)}_r(r_{M+1})
=
\frac
{Q^{(\beta)}_{\l^{(r-1)}/\mu^{(r-1)}}(r_{M+1})}
{\sum_{\k}P_{\k/\l^{(r-1)}}(a)Q^{(\beta)}_{\k/\mu^{(r)}}(r_{M+1})}
=
\frac
{Q^{(\beta)}_{\l^{(r-1)}/\mu^{(r-1)}}(r_{M+1})}
{(1+a_r r_{M+1})\sum_{\nu}P_{\mu^{(r)}/\nu}(a)Q^{(\beta)}_{\l^{(r-1)}/\nu}(r_{M+1})}.
\label{p924}
\end{align}
The first one is the definition itself~\eqref{a6} and the second one
is obtained by applying~\eqref{l413} to $\Delta(\l^{(j-1)},\mu^{(j)})$ in~\eqref{a6}.
The two expressions for $A_r$ and $B$ in the decomposition (\ref{p93}) correspond 
to this difference. 
The expressions $\tilde{A}_r$~\eqref{p95} and $\tilde{B}$~\eqref{p96} (resp. the expressions
$A_r$~\eqref{p98} and $B$~\eqref{p914}) are associated with
the first one (resp. second one) in~\eqref{p924}. 
Here we gave a proof of the equivalence between the two expressions
without passing through the discrete time dynamics.
For the proof of Theorem~\ref{thme}, the expressions in the proposition are more useful.  

\smallskip
\noindent
{\bf Proof of Theorem~\ref{thme}.}
First from the definition (\ref{2qWh-pr}) with (\ref{defQ}) one observes 
\begin{align}
\frac{\partial}{\partial t}
P_t(\underline{\l}_N)
&=
-\sum_{r=1}^N a_r\cdot P_t(\underline{\l}_N)
+
\frac
{\prod_{l=1}^N P(\l^{(l-1)},\l^{(l)};a_l)}
{\Pi(a;\a,t)}
\frac{\partial}{\partial t}
Q (\l^{(N)};\a,t).
\label{p922}
\end{align}
Hence, for proving the relation, it is sufficient to show
\begin{align}
\sum_{\underline{\mu}_N}P_t(\underline{\mu}_N)A_r(\underline{\mu}_N,\underline{\l}_N)
&=
-a_rP_t(\underline{\l}_N),
\label{p911}
\\
\sum_{\underline{\mu}_N}P_t(\underline{\mu}_N)B(\underline{\mu}_N,\underline{\l}_N)
&=
\frac
{\prod_{l=1}^N P(\l^{(l-1)},\l^{(l)};a_l)}
{\Pi(a;\a,t)}
\frac{\partial}{\partial t}
Q (\l^{(N)};\a,t),
\label{p913}
\end{align}
where $A_r(\underline{\l}_N,\underline{\mu}_N)$ and $B(\underline{\mu}_N,\underline{\l}_N)$ are given as the second expressions~\eqref{p98},
and~\eqref{p914} respectively. We use Lemma~\ref{p9}.

First we prove~\eqref{p911}. We see that the part of $\sum_{\underline{\mu}_N}P_t(\underline{\mu}_N)A_r(\underline{\mu}_N,\underline{\l}_N)$ coming from the first term of~\eqref{p98} can be written as
\begin{align}
&\sum_{\underline{\mu}_N}P_t(\underline{\mu}_N)
\sum_{1\le j\le k\le r-1}
(1-q^{\mu_{j-1}^{(r-1)}-\mu_{j}^{(r-1)}})
\prod_{m=k}^{r-1}
\frac
{P(\l^{(m-1)}, \l^{(m)};a_m)}
{P(\mu^{(m-1)}, \mu^{(m)};a_m)}\delta_{m_{jk}(\underline{\mu}_N),r-1-k}
~\delta_{\underline{\l}_N,\underline{\mu}^{j,k+}_N}
\notag
\\
&=
\sum_{j=1}^{r-1}
(1-q^{\l^{(r-1)}_{j-1}-\l^{(r-1)}_j+1})
\prod_{\substack{l=1\\l\neq r}}^N P(\l^{(l-1)},\l^{(l)};a_l)
\cdot
P(\l^{(r-1)}-1_j,\l^{(r)};a_r)
\frac
{Q(\l^{(N)};\a,t)}
{\Pi(a;\a,t)}.
\label{p919}
\end{align}
Here we used the definition (\ref{2qWh-pr}) and the fact that each factor of 
$\prod_{l=1}^N P(\mu^{(l-1)},\mu^{(l)};a_l)$ from $P_t(\underline{\mu}_N)$ can be 
replaced by $P(\l^{(l-1)},\l^{(l)};a_l)$ except $l=r$ since 
$\mu^{(l)}=\l^{(l)}, 1\leq l\leq k-1,r+1\leq l\leq N$ while the factor 
for $l=r$ changes to $P(\l^{(r-1)}-1_j,\l^{(r)};a_r)$ since 
 $\mu^{(r-1)}=\l^{(r-1)}-1_j$. 
We also rewrite the Kronecker's deltas as 
\begin{align}
\delta_{m_{jk}(\underline{\mu}_N),r-1-k}
\delta_{\underline{\l}_N,\underline{\mu}^{jk+}_N}
=
\delta_{\ell_{j,r-1}(\underline{\l}_N),r-1-k}
\delta_{\underline{\mu}_N,\underline{\l}^{j,(r-1)-}_N}
\label{p919-2}
\end{align}
which follows from the definitions~\eqref{t21}--\eqref{t23} (see also Fig. \ref{figGTpm}), and use
$\sum_{k=j}^{r-1}\delta_{\ell_{j,r-1}(\underline{\l}_N),r-1-k}
=\sum_{\underline{\mu}_N}\delta_{\underline{\mu}_N,\underline{\l}^{j,(r-1)-}_N}=1$. 

Clearly the remaining parts of  $\sum_{\underline{\mu}_N}P_t(\underline{\mu}_N)A_r(\underline{\mu}_N,\underline{\l}_N)$, which come from the second and the last terms of~\eqref{p98} become
\begin{align}
\quad
&-\sum_{\underline{\mu}_N}P_t(\underline{\mu}_N)
\sum_{j=1}^{r-1}
(1-q^{\mu_{j-1}^{(r-1)}-\mu_{j}^{(r-1)}+1})
\frac
{P(\mu^{(r-1)}-1_j,\mu^{(r)};a_r)}
{P(\mu^{(r-1)},\mu^{(r)};a_r)}
\delta_{\underline{\l}_N,\underline{\mu}_N},
\notag\\
&\hspace{1cm}=
-\sum_{j=1}^{r-1}
(1-q^{\l^{(r-1)}_{j-1}-\l^{(r)}_j})
\prod_{\substack{l=1\\l\neq r}}^N P(\l^{(l-1)},\l^{(l)};a_l)
\cdot
P(\l^{(r-1)}-1_j,\l^{(r)};a_r)
\frac
{Q(\l^{(N)};\a,t)}
{\Pi(a;\a,t)}
\label{p920}
\\
&-\sum_{\underline{\mu}_N}P_t(\underline{\mu}_N)(a_r\delta_{\underline{\l}_N,\underline{\mu}_N})=-a_rP_t(\underline{\l}_N).
\label{p921}
\end{align}
Noting ~\eqref{p919} and ~\eqref{p920} cancel, only~\eqref{p921} remains and
we obtain~\eqref{p911}.

At last we prove~\eqref{p913}.
Note that $B(\underline{\mu}_N,\underline{\l}_N)$~\eqref{p914}
corresponds to the first term of $A_r(\underline{\mu}_N,\underline{\l}_N)$~\eqref{p98}
with $r=N+1$. Thus as in~\eqref{p919},  we get
\begin{align}
\sum_{\underline{\mu}_N}P_t(\underline{\mu}_N)B(\underline{\mu}_N,\underline{\l}_N)
=
\sum_{j=1}^{N}
(1-q^{\l^{(N-1)}_{j-1}-\l^{(N)}_j})
\prod_{l=1}^N P(\l^{(l-1)},\l^{(l)};a_l)
\cdot
\frac
{Q(\l^{(N)}-1_j;\a,t)}
{\Pi(a;\a,t)}.
\label{p912}
\end{align}
Applying~\eqref{l412}, we arrive at~\eqref{p913}.
\qed

\subsection{Marginal distributions}
We will be interested in the distribution of $\l_N^{(N)}(t)$. To study it, it is useful to 
consider the marginal distribution of the two-sided $q$-Whittaker process with respect to the 
particles on the top line $\l^{(N)}(=(\l^{(N)}_1,\cdots,\l^{(N)}_N))$ in the GT cone.  
By recalling the definition of the $q$-Whittaker function~\eqref{Wh}, one easily finds 
\begin{proposition}\label{p-Wh-meas}
For $\l\in\S_N$, 
\begin{align}
\P[\l^{(N)}(t) = \l]
=\frac{P_\l(a)Q_\l(\a,t)}{\Pi (a;\a,t)}.
\label{ptn}
\end{align}
\end{proposition}
\noindent
We call this the two-sided $q$-Whittaker measure. Since this is the probability measure, 
the summation over $\l$ gives a version of the Cauchy identity for our case. 
\begin{corollary}
\begin{equation}
 \sum_{\l} P_\l(a)Q_\l(\a,t) = \Pi (a;\a,t).
\end{equation}
\end{corollary}
\noindent

\noindent
When studying the  distribution of $\l^{(N)}_N(t)$, the following representation will be useful.
\begin{proposition}\label{laNN}
For $l\in\Z$, 
\begin{align}
\P[\l_N^{(N)}(t)=l]= 
(q;q)_{\infty}^{N-1} \int_{\T^N}\prod_{j=1}^N \frac{dz_j}{z_j}
\cdot
\left(
\frac{A}{Z}\right)^l m^q_N(z)
\frac{\Pi(z;\a,t)}{\Pi(a;\a,t)}
\cdot
\frac
{\left(A/Z;q\right)_{\infty}}
{\prod_{i,j=1}^N(a_i/z_j;q)_{\infty}},
\label{p82}
\end{align}
where $A=\prod_{i=1}^N a_i$ and $Z=\prod_{i=1}^N z_i$.
\end{proposition}
\smallskip

\noindent
{\bf Proof of Proposition \ref{laNN}.}
Using (\ref{defQ}) and (\ref{p81}), the two-sided $q$-Whittaker measure 
(\ref{ptn}) is written as 
\begin{equation}
\P[\l^{(N)}(t) = \l]
=
\int_{\T^N}
\prod_{j=1}^{N}\frac{dz_j}{z_j}\cdot \left(\frac{A}{Z}\right)^{\l_N^{(N)}}
m^q_N(z)
\frac{\Pi(z;\a,t)}{\Pi(a;\a,t)}
\prod_{k=1}^{N-1}\frac{(q;q)_{\infty}}{(q;q)_{\ell_j}}
\cdot
R_\ell(a)R_\ell\left(1/z\right) .
\end{equation}
Using the Cauchy identity in the form (\ref{hci}), we find 
\begin{align}
&\P[\l_N^{(N)}(t)=l] 
=
\sum_{\lambda \in\S_N ~\text{s.t.}~ \l_N = l}
\P[\lambda^{(N)}(t)=\l]
\notag\\
&=
\int_{\T^N}
\prod_{j=1}^{N}\frac{dz_j}{z_j}\cdot \left(\frac{A}{Z}\right)^{\l_N^{(N)}}
m^q_N(z)
\frac{\Pi(z;\a,t)}{\Pi(a;\a,t)}
\sum_{
\substack{\ell_j=0\\
			j=1,\cdots,N-1}
}^{\infty}
\prod_{k=1}^{N-1}\frac{(q;q)_{\infty}}{(q;q)_{\ell_j}}
\cdot
R_\ell(a)R_\ell\left(1/z\right)
\notag\\
&=
(q;q)_{\infty}^{N-1}
\int_{\T^N}\prod_{j=1}^N \frac{dz_j}{z_j}
\cdot
\left(
\frac{A}{Z}
\right)^l
m^q_N(z)
\frac{\Pi(z;\a,t)}{\Pi(a;\a,t)}
\cdot
\frac
{\left(A/Z;q\right)_{\infty}}
{\prod_{i,j=1}^N(a_i/z_j;q)_{\infty}}.
\end{align}
\qed

\section{Fredholm determinant formula for the $q$-Laplace transform}
\label{qLap}
In this section, we derive a Fredholm determinant formula for the $q$-Laplace 
transform of the position of the $N$-th particle in $q$-TASEP with the initial condition
(\ref{h-st-ic}).  
When taking the sum over the position of the particle and getting a determinantal expression, 
we need two essential ingredients. One is the Ramanujan's summation formula and the other 
is the Cauchy determinant formula for the theta function. The former played an important role when 
studying the limit from ASEP to the KPZ equation in \cite{SS2010a} but does not seem to 
have been utilized in the analysis of related models since then. Some part of the calculations in this 
section can be regarded as a generalization of those in \cite{BCR2013}. 

Let us first recall the Ramanujan's summation formula (\ref{Ram}). 
\begin{theorem}\label{rama}
For $|q|<1, |b/a|<|z|<1$, $a\notin q^n, n\in\Z$, 
\begin{equation}
\sum_{n\in\Z} \frac{(bq^n;q)_\i}{(aq^n;q)_\i} z^n 
=
\frac{(az;q)_\i (\frac{q}{az};q)_\i (q;q)_\i (\frac{b}{a};q)_\i}{(a;q)_\i (\frac{q}{a};q)_\i (z;q)_\i (\frac{b}{az};q)_\i} . 
\label{ram}
\end{equation}
\end{theorem}

Next we introduce a modified Jacobi theta function (\cite{GR2004} p303), 
\begin{equation}
 \th(z) = (z;q)_\i (q/z;q)_\i , ~ |q|<1, z\neq 0, 
 \label{th}
\end{equation}
which is related to the ordinary theta function as
\begin{equation}
 \th_1(x,e^{\pi i\tau})
 =
 i e^{-ix+\pi i \tau/4} (q;q)_\i \th(e^{2ix};q), \quad q=e^{2\pi i\tau}, {\rm Im}\, \tau>0, x\in\C. 
 \label{thth1} 
\end{equation}
This function has been playing important role in various places. 
In the following we also use 
\begin{equation}
 \tilde{\th}(z) = \frac{1}{\sqrt{z}}\th(z),
 \label{ti-th}
\end{equation}
which has a nicer symmetry property, $\tilde{\th}(1/z)=\tilde{\th}(z)$ where 
the square root is $\sqrt{z}=e^{\frac12 \log z}$ with the standard 
branch cut of logarithm. 

Let $[x]$ be a nonzero holomorphic function which satisfies $[-x]=-[x]$ and the Riemann relation, 
\begin{align}
 &\quad [x+y][x-y][u+v][u-v] \notag\\
 &=[x+u][x-u][y+v][y-v]
 -[x+v][x-v][y+u][y-u] .  
 \label{Riemann} 
\end{align}
It is known that $[x]$ satisfying the above two relations is necessarily in the form 
$e^{a x^2+b}f(cx)$ where $f(x)$ is either $f(x)=x$, $f(x)=\sin \pi x$ or 
$f(x)=\sigma(x)$ 
(cf \cite{WW1927} p451 20$\cdot$53 ex.4 and p461 ex.38, \cite{KN2003}). 
Here the Weierstrass sigma function $\sigma(x)=\sigma(x|\omega_1,\omega_2)$,
with the half periods $\o_1,\o_2$,  
can be written in terms of the ordinary theta function as (cf \cite{WW1927} p473 21$\cdot$43)
\begin{equation}
\sigma(x|\omega_1,\omega_2)
=
\frac{2\omega_1}{\pi \theta_1^{(1)}} \exp\left(-\frac{\pi^2 x^2 \theta_1^{(3)}}{24\o_1^2\theta_1^{(1)}}\right)
\theta_1\bigl(\frac{\pi x}{2\o_1},e^{i\pi\frac{\omega_2}{\omega_1}}\bigr)
\label{sigmath}
\end{equation}
where $\th_1^{(n)} = \frac{d^n }{dx^n} \th_1(x,q)|_{x=0},n\in\N$. 
Combining (\ref{thth1}), (\ref{sigmath}), one sees that our theta function 
$\tilde{\theta}(q^x)$ is written in the form $e^{a x^2+b}\sigma(cx)$ and hence an example of $[x]$. 

\begin{theorem}(cf \cite{KN2003})
\label{cdet}
For $[x]$ as above, the following Cauchy determinant formula holds,
\begin{equation*}
\frac{[\nu+B-C] \prod_{1\leq i<j\leq N} [b_i-b_j][c_j-c_i]}{[\nu] \prod_{i,j=1}^N [b_i-c_j]}
=
\det\left(\frac{[\nu+b_i-c_j]}{[\nu][b_i-c_j]}\right)_{1\leq i,j\leq N} 
\end{equation*}
where $\nu$ is a parameter, $b_i,c_i,1\leq i\leq N$ are $2N$ complex variables and 
$B=\sum_{i=1}^N b_i, C=\sum_{i=1}^N c_i$. 
\end{theorem}
Note that in a certain limit ($\nu\to\infty$ for $f(x)=x$, $\nu\to i\i$ for $f(x)=\sin x$
and for $f(x)=\sigma(x)$ 
using $\sigma(x+2\o_2)= -e^{2\eta_2(x+\o_2)} \sigma(x), 
\eta_2=-\frac{\pi^2\o_2 \th_1^{(3)}}{12\o_1^2 \th_1^{(1)}}-\frac{\pi i}{2\o_2}$,
 cf \cite{WW1927} p448 20$\cdot$412),    
the factors including $\nu$ 
on both sides of the identity cancel, leading to more familiar 
form of the Cauchy identity. This form of Cauchy identity for the theta function 
has already appeared in several contexts in random matrix  theory 
(e.g \cite{Forrester2010} 5.6.3), 
con-colliding diffusions \cite{Katori2015,Katori2016} and so on. 
But in our discussions below, the formula with the 
extra parameter $\nu$ plays an essential role. 

\begin{figure}[t]
\begin{center}
\includegraphics[scale=0.6]{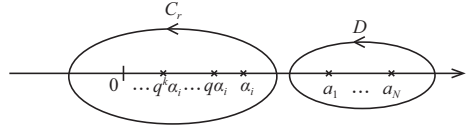}
\caption{\label{figCDcontour}
The contours $C_r$ and $D$. 
}
\end{center}
\end{figure}

Now we come back to the $q$-TASEP. 
By using the above two theorems, we will obtain 
\begin{theorem}\label{ql-det}
For the two-sided $q$-Whittaker measure (\ref{ptn}) with $0\leq \a_i < a_j \leq 1,1\leq i,j\leq N$ 
and with $\zeta\neq q^n,n\in\Z$, 
\begin{equation}
 \Big\langle \frac{1}{(\z q^{\l_N};q)_\i} \Big\rangle
 =
 \det(1-f K)_{\ell^2(\Z)} 
 \label{det_formula}
\end{equation}
where $\langle \cdots \rangle$ means the average and the kernel of the Fredholm detereminant
on rhs is given by   
\begin{align}
 f(n) &=  \frac{1}{1-q^n/\z}, \label{f}\\
 K(n,m) &= \sum_{l=0}^{N-1} \phi_l(m) \psi_l(n), \label{Kernel}\\
 \phi_l(n) 
 &=
  \tau(n)
 \int_D dv \frac{e^{-vt}}{v^{n+N}}\frac{1}{v-a_{l+1}}
 \prod_{j=1}^l \frac{v-\a_j}{v-a_j} \prod_{k=1}^N \frac{(q\a_k/v;q)_\i}{(qv/a_k;q)_\i} , 
 \label{phi}
\\
\psi_l(n)
&=
\frac{a_{l+1}-\a_{l+1}}{\tau(n)} 
 \int_{C_r} dz \frac{e^{zt} z^{n+N}}{z-\a_{l+1}}
 \prod_{j=1}^l \frac{z-a_j}{z-\a_j} \prod_{k=1}^N \frac{(q z/a_k;q)_\i}{(q\a_k/z;q)_\i} . 
 \label{psi}
\end{align}
Here the contour $D$ is around $\{a_i,1\leq i\leq N\}$ 
and the contour $C_r$ is around $\{0,\a_i q^j, 1\leq i\leq N,j\in\N\}$.  See Fig. \ref{figCDcontour}. 
$\tau(n)$ is 
\begin{align}
\tau(n)
=
\begin{cases}
b^n, & n\ge 0,\\
c^n, & n<0.  
\end{cases}
\label{tau}
\end{align}
where $b,c$ are taken to satisfy 
$0<\text{max}\,\a_i<b<\text{min}\,a_i \leq 1<c$.
\end{theorem}

\noindent
By the result in the previous section, we immediately have 
\begin{corollary}
\label{main1}
For $q$-TASEP with the initial condition (\ref{h-st-ic}) with $0\leq \a < a_j \leq 1,1\leq j\leq N$ and 
with $\zeta\neq q^n,n\in\Z$, the $q$-Laplace transform 
$\langle \frac{1}{(\z q^{X_N(t)+N};q)_\i} \rangle$ for the $N$th particle position is 
written as the Fredholm determinant on rhs of (\ref{det_formula}) with 
(\ref{phi}),(\ref{psi}) specialized to $\a_j=\a,\a_k=0,~k\neq j~1\leq j,k\leq N$. 
\end{corollary} 

\noindent
In the rest of this section we will prove Theorem \ref{ql-det}. First we discuss 
a few properties of the functions $\phi_l,\psi_l$ (\ref{phi}), (\ref{psi}) and 
the well-definedness of the Fredholm determinant on rhs. 

\begin{proposition}
The functions $\phi_l(n)$ (\ref{phi}) and $\psi_l(n)$ (\ref{psi}) are biorthonormal, i.e., 
\begin{equation}
 \sum_{n\in\Z} \phi_l(n) \psi_m(n) = \d_{l,m}. 
 \label{on}
\end{equation}
\end{proposition} 
\smallskip
\noindent
{\bf Proof.}
This is easy to show by using their contour integral expressions. 
\qed 

\begin{lemma}
\label{fl2-2}
Fix $t>0, N\in\Z_{>0},l\in\N$ and assume $0< \text{max}\,\a_i < b<\text{min}\,a_i < 1<c$.
When $a_i$'s and $\a_i$'s are all different, 
the functions (\ref{phi}) and (\ref{psi}) have the following simple bounds: 
\begin{align}
|\phi_l(n)|&\le 
\begin{cases}
C_{1 +} b^n \left(\frac{1}{a_{\text{min}}}\right)^n, &\text{for~} n\ge 0,\\
C_{1 -} c^n \left(\frac{1}{a_{\text{max}}}\right)^n, &\text{for~} n\le -1,
\end{cases}
\label{phi_bd}
\\
|\psi_l(n)|&\le 
\begin{cases}
C_{2 +} 
\a_{\text{max}}^n/b^n &\text{for~} n\ge 0,\\
C_{2 -} 
d^n/c^n, &\text{for~} n\le -1.
\end{cases}
\label{psi_bd}
\end{align}
Here $C_{1\pm},C_{2\pm}>0$ are some constants (wrt $n$)
and $d$ is an arbitrary constant which is larger than 1.  
When some of the $a_i$'s or $\a_i$'s are the same, $1/a_{\rm min}$ and 
$\a_{\rm max}$ (resp. $1/a_{\rm max}$) above should be replaced by a 
slightly larger (resp. smaller) value.
Hence $\phi_l$ and $\psi_l$ are in $\ell^2(\Z)$.
\end{lemma}

\smallskip
\noindent
{\bf Proof.}
If we denote by $\hat{\phi}_l,\hat{\psi}_l$ the rhs's of (\ref{phi}),(\ref{psi}) without 
the $\tau$ factors, what we should show are (\ref{phi_bd}),(\ref{psi_bd}) without 
the factors in terms of $b,c$.  
For $\hat{\phi}_l$, when all $a_i$'s are different, by taking the poles in (\ref{phi}), 
it is easy to see,
\begin{align}
\hat{\phi}_{l}(n)&=\sum_{i=1}^N C_i a_i^{-n},
\end{align}
with $C_i, 1\leq i\leq N,$ some constants. When some of $a_i$'s are the same, 
polynomials in $n$ appear in the coefficients but if one does the replacement 
stated in the Lemma, the following discussions would still be valid. 
Large $n$ behaviors of $\hat{\phi}_l(n)$ is dominated by 
the largest or smallest of $a_i$'s depending on the sign of $n$. 
Multiplying $\tau(n)$, we have (\ref{phi_bd}). 

Next we consider $\hat{\psi}_{l}(n)$. For positive $n$, one first shrinks the contour 
in (\ref{f24}) to the one with radius which is slightly larger than $\text{max}\,\a_i$, 
which can be done without passing a singularity. We see the first case of 
(\ref{psi_bd}). For negative $n$, one can enlarge the contour as much as 
one wishes. If one takes the contour with radius $d(>1)$, we have the bound in 
the second case of (\ref{psi_bd}) after division by $\tau(n)$. 

By taking $d>c$, one sees that the new $\phi_l,\psi_l$ go to 
zero exponetially fast and hence they are in $\ell^2(\Z)$.
\qed

\begin{proposition}
\label{trcl}
On rhs of  (\ref{det_formula}) the operator $fK$ is a trace-class operator 
and hence the Fredholm determinant is well-defined. 
\end{proposition} 
\medskip
\noindent
{\bf Proof.} 
If one regards the kernel of $fK$ as a product of two operators with 
kernels,  $\sqrt{f}\phi_l(m),~\sqrt{f}\psi_l(n)$, the above Lemma \ref{fl2-2} 
tells us that each operator is a Hilbert-Schmit operator (Note that the sum over $l$
is finite). Hence $fK$ is a trace-class operator on $\ell^2(\Z)$ 
(Th VI.22 in \cite{RS1980},  see also Prop 2.4 in \cite{Johansson2003} ).  
It is well-known that the Fredholm expansion of a trace-class kernel converges, 
giving the well-defined Fredholm determinant. 
\qed

\medskip
\noindent
Now that a well-definedness of rhs of (\ref{det_formula}) has been checked,  
the remaining is to show how the formula (\ref{det_formula}) is derived.  

\noindent 
{\bf Proof of Theorem \ref{ql-det}.}
In this proof, products such as $\prod_i$ run for $i=1,\ldots , N$ and determinants 
are $N$ dimensional unless otherwise stated. 
By Proposition \ref{laNN} we have,  
with $\{ a_i \}_{1\leq i\leq N}, \{\a_i \}_{1\leq i\leq N}$ where $|a_i|,|\a_i|<1$,  
\begin{equation}
 \Big\langle \frac{1}{(\z q^{\l_N};q)_\i} \Big\rangle
 =
 \sum_{l\in\Z} \frac{1}{(\zeta q^{l};q)_{\infty}} 
 \int_{\T^N} \prod_{i=1}^N \frac{dz_i}{z_i} \left(\frac{A}{Z}\right)^l m^q_N(z) 
 \frac{\Pi(z;\a,t)}{\Pi(a;\a,t)} \frac{(q;q)_\i^{N-1} (A/Z;q)_\i}{\prod_{i,j} (a_i/z_j;q)_\i},
\end{equation}
where $A=\prod_{i=1}^N a_i, Z=\prod_{i=1}^N z_i$. 
The summation over $l$ can be performed by 
using the Ramanujan's formula with $a=\z,b=0,z=A/Z$, 
\begin{equation}
 \sum_{l\in\Z} \frac{1}{(\z q^l;q)_\i}\left(\frac{A}{Z}\right)^l
 =
 \frac{(\frac{\z A}{Z};q)_\i (\frac{qZ}{\z A};q)_\i  (q;q)_\i}
        {(\z,q)_\i (\frac{q}{\z};q)_\i (\frac{A}{Z};q)_\i}
 =
 \frac{\th(\frac{\z A}{Z}) (q;q)_\i}{\th(\z) (\frac{A}{Z};q)_\i}. 
\end{equation}
We find
\begin{equation}
  \Big\langle \frac{1}{(\z q^{\l_N};q)_\i} \Big\rangle
  =
  \frac{(q;q)_\i^N}{N!} \int_{\T^N} \prod_{i=1}^N \frac{dz_i}{2\pi i z_i}
  \frac{\th(\frac{\z A}{Z})}{\th(\z)} \frac{\prod_{i\neq j} (z_i/z_j;q)_\i}{\prod_{i,j} (a_i/z_j;q)_\i}
   \frac{\Pi(z;\a,t)}{\Pi(a;\a,t)}. 
\end{equation}
Next we rewrite the two double products over $i,j$ to apply the Cauchy determinant formula. 
Noting a simple fact 
\begin{equation}
(z/w;q)_\i (w/z;q)_\i 
 = 
 (1-w/z) \th(z/w), 
\end{equation}
we have
\begin{align}
  \prod_{i\neq j} (z_i/z_j;q)_\i
 &=
 \prod_{i<j}(1-z_j/z_i) \prod_{i<j} \th(z_i/z_j) , \\
 \prod_{i,j} (a_i/z_j;q)_\i \prod_{i,j} (z_i/a_j;q)_\i
 &=
 \prod_{i,j}(1-z_j/a_i) \prod_{i,j} \th(a_i/z_j) ,\\
 \prod_{i\neq j} (a_i/a_j;q)_\i
 &=
 \prod_{i<j} (1-a_j/a_i) \prod_{i<j} \th(a_i/a_j) ,
\end{align}
and hence 
\begin{align}
&\quad 
\frac{\prod_{i\neq j}(z_i/z_j;q)_\i}{\prod_{i,j}(a_i/z_j;q)_\i} \notag\\
&=
\frac{\prod_{i<j}(1-z_j/z_i) \prod_{i<j} \th(z_i/z_j) \prod_{i,j}(z_i/a_j;q)_\i \prod_{i<j} (1-a_j/a_i) \prod_{i<j} \th(a_i/a_j)}
       {\prod_{i,j}(1-z_j/a_i)\prod_{i,j}\th(a_i/z_j) \prod_{i\neq j}(a_i/a_j;q)_\i} \notag\\
&=
\frac{\prod_{i<j}(a_i-a_j) \prod_{i<j} (z_i-z_j)}{\prod_{i,j}(a_i-z_j)}
\frac{\prod_{i\geq j} a_i \prod_{i<j} \th(a_i/a_j) \prod_{i<j} \th(z_i/z_j)}{\prod_{i<j}z_i \prod_{i,j}\th(a_i/z_j) }
\frac{\prod_{i,j}(z_i/a_j;q)_\i}{\prod_{i\neq j}(a_i/a_j;q)_\i}   .    
\end{align} 
By rewriting 
\begin{equation}
 \frac{\prod_{i\geq j} a_i }{\prod_{i<j}z_i}
 =
 \frac{\prod_{i<j} \sqrt{a_j/a_i}\sqrt{z_j/z_i}}{\prod_{i,j} \sqrt{z_j/a_i}} \sqrt{Z/A} 
\end{equation} 
and using the definition of $\tilde{\th}(z)$ (\ref{ti-th}), 
we see
\begin{align}
 \Big\langle \frac{1}{(\z q^{\l_N};q)_\i} \Big\rangle
  &=
  \frac{(q;q)_\i^N}{N!} \int_{\T^N} \prod_{i=1}^N \frac{dz_i}{z_i}
  \frac{\prod_{i<j}(a_i-a_j) \prod_{i<j} (z_i-z_j)}{\prod_{i,j}(a_i-z_j)}
  \frac{\prod_{i<j} \tilde{\th}(a_i/a_j)\prod_{i<j} \tilde{\th}(z_i/z_j)}{\prod_{i,j}\tilde{\th}(a_i/z_j) } \notag\\
  &\quad\times
  \frac{\tilde{\th}(\frac{\z A}{Z})}{\tilde{\th(\z)}}
  \prod_i \frac{a_i \prod_k (z_i/a_k;q)_\i g(z_i;\a,t)}{\prod_{k\neq i} (a_i/a_k;q)_\i g(a_i;\a,t)}
\end{align} 
where 
\begin{equation}
 g(z;\a,t) = \frac{e^{zt}}{\prod_j (\a_j/z;q)_\i}.
\end{equation}
By using the Cauchy determinant formula for theta function here, we find 
\begin{equation}
 \Big\langle \frac{1}{(\z q^{\l_N};q)_\i} \Big\rangle
 =
 \frac{1}{N!}
 \int_{\T^N} \prod_{i=1}^N \frac{dz_i}{z_i}
 \det\Big(\frac{a_i}{a_i-z_j}\Big) 
 \det\Big(\frac{\tilde{\th}(\z a_i/z_j)}{\tilde{\th}(\z) \tilde{\th}(a_i/z_j) }\Big)
 \prod_i \frac{\prod_k (z_i/a_k;q)_\i g(z_i;\a,t) (q;q)_\i}
                     {\prod_{k\neq i} (a_i/a_k;q)_\i g(a_i;\a,t)}.
\end{equation}
Note that the three $\tilde{\th}$'s in the above determinant can be replaced by 
$\th$ because the square root factors in (\ref{ti-th}) cancel in this combination. 
Using the Cauchy-Binet identity (or Andreief identity), 
\begin{equation}
 \frac{1}{N!} \int \det(f_i(x_j))_{1\leq i,j\leq N} \det(g_i(x_j))_{1\leq i,j\leq N}  \prod_{i=1}^N d\mu(x_i) 
=
\det\left( \int f_i(x) g_j(x) d\mu(x) \right)_{1\leq i,j\leq N},
\end{equation}
which holds for rather general $f_i,g_i,\mu$ as long as the integrals in the formula 
make sense, we find
\begin{equation}
 \Big\langle \frac{1}{(\z q^{\l_N};q)_\i} \Big\rangle
 =
 \det\left( 
  \int_\T \frac{dz}{z} \frac{a_i}{a_i-z} 
  \frac{\th(\z a_i/z)}{\th(\z) \th(a_i/z) }
  \frac{ (q;q)_\i \prod_k (z/a_k;q)_\i g(z;\a,t)}
         {\prod_{k\neq j} (a_j/a_k;q)_\i g(a_j;\a,t)} \right).
\end{equation}
By making the contour smaller and taking the pole at $z=a_i$, 
\begin{equation}
 \Big\langle \frac{1}{(\z q^{\l_N};q)_\i} \Big\rangle
 =
 \det\left( \delta_{ij}-
  \int_{C_r} \frac{dz}{z} \frac{a_i}{a_i-z} 
   \frac{\th(\z a_i/z)}{\th(\z) \th(a_i/z) }
  \frac{ (q;q)_\i \prod_k (z/a_k;q)_\i g(z;\a,t)}
         {\prod_{k\neq j} (a_j/a_k;q)_\i g(a_j;\a,t)} \right).
\end{equation}
Here using the Ramanujan's formula again with $a=1/\z, b=q/\z, z\to z/a_j$, 
\begin{equation}
\sum_{n\in\Z}\frac{1}{1-q^n/\z} \left( \frac{z}{a_j} \right)^n
=
\frac{(\frac{z}{\z a_j})_\i (\frac{q\z a_j}{z};q)_\i (q;q)_\i^2}
       {(1/\z;q)_\i (q\z;q)_\i (z/a_j;q)_\i (qa_j/z;q)_\i}
=
\frac{\th(\frac{z}{\z a_j})}{\th(1/\z) \th(z/a_j)} (q;q)_\i^2 ,      
\end{equation}
we see 
\begin{align}
 \Big\langle \frac{1}{(\z q^{\l_N};q)_\i} \Big\rangle
 &=
 \det\left( \delta_{ij}- 
  \sum_{n\in\Z}\frac{1}{1-q^n/\z}
  \int_{C_r} \frac{dz}{z} \frac{a_i}{a_i-z} 
  \frac{z^n \prod_k (z/a_k;q)_\i g(z;\a,t)}
         {a_j^n (q;q)_\i \prod_{k\neq j} (a_j/a_k;q)_\i g(a_j;\a,t)} \right) \notag\\
  &=
   \det(\delta_{ij}-\sum_{n\in\Z} A(i,n) B(n,j))_{1\leq i,j \leq N}        
\end{align}
with 
\begin{align}
A(i,n)
&=
\sqrt{f(n)}
\int_{C_r} \frac{dz}{z} \frac{a_i}{a_i-z} z^n \prod_k (z/a_k;q)_\i g(z;\a,t), 
\label{A}\\
B(n,j)
&=
\frac{\sqrt{f(n)}}{a_j^n(q;q)_\i  \prod_{k\neq j} (a_j/a_k;q)_\i g(a_j;\a,t)}.
\label{B}
\end{align}
In the last determinant we state explicitly that this is for an $N$-dimensional matrix. 
Here use $\det(1-A B)_{1\leq i,j\leq N} = \det(1-B A)_{L^2(\Z)}$, which holds for 
arbitrary Hilbert-Schmidt operators $A,B$. 
The fact that $A,B$ in our case 
(\ref{A}),(\ref{B}) are indeed Hilbert-Schmidt operators can be shown in the same 
way as Lemma \ref{fl2-2} for $\phi,\psi$. 
We see that the kernel (or the matrix element of the infinite matrix) is given by 
\begin{align}
(BA)(m,n)
&=
\sum_{i=1}^N B(m,i) A(i,n) \notag\\
&=
\sum_{i=1}^N 
\frac{1}{a_i^m (q;q)_\i \prod_{k\neq i}(a_i/a_k;q)_\i g(a_i;\a,t)}
\frac{1}{1-q^n/\z} \int_{C_r} \frac{dz}{z} \frac{a_i}{a_i-z} z^n \prod_k (z/a_k;q)_\i g(z;\a,t) \notag\\
&=
\frac{-1}{1-q^n/\z} \int_D dv \int_{C_r} \frac{dz}{z} \frac{1}{v-z}
\frac{z^n \prod_k (z/a_k;q)_\i g(z;\a,t)}{v^m \prod_k (v/a_k;q)_\i g(v;\a,t)},
\end{align}
where the contour $D$ is around $\{a_i,1\leq i\leq N\}$.
Rewriting the last product as
\begin{align}
 &\quad 
 \frac{\prod_k (z/a_k;q)_\i g(z;\a,t)}{\prod_k (v/a_k;q)_\i g(v;\a,t)} \notag\\
 &=
 \frac{\prod_k (q z/a_k;q)_\i (q\a_k/v;q)_\i e^{zt}}{\prod_k (q v/a_k;q)_\i (q\a_k/z;q)_\i e^{vt}}
 \frac{(z-a_k)(v-\a_k)}{(v-a_k)(z-\a_k)}
 \left(\frac{z}{v}\right)^N ,\notag\\
\end{align}
we have 
\begin{align}
(BA)(m,n)
  &=
 \frac{1}{1-q^n/\z} \int_D dv \int_{C_r} \frac{dz}{z} 
 \frac{z^{n+N} e^{zt} \prod_k (qz/a_k;q)_\i (q\a_k/v;q)_\i}
        {v^{n+N} e^{vt} \prod_k (qv/a_k;q)_\i (q\a_k/z;q)_\i} \notag\\
 &\quad\times        
 \left( \frac{1}{z-v}\prod_k \frac{(z-a_k)(v-\a_k)}{(v-a_k)(z-\a_k)}-1  \right)         
\end{align}
where $-1$ is inserted since it does not create singularities which change 
the value of the integral. 
By a simple relation, 
\begin{equation}
\frac{1}{z-v}\prod_k \frac{(z-a_k)(v-\a_k)}{(v-a_k)(z-\a_k)}-1  
=
\sum_{l=0}^{N-1}\frac{a_{l+1}-\a_{l+1}}{(z-\a_{l+1})(v-a_{l+1})}
\prod_{j=1}^l \frac{(z-a_j)(v-\a_j)}{(z-\a_j)(v-a_j)},
\end{equation}
we finally arrive at our desired formula. 

This completes the proof of Theorem \ref{ql-det}. 
\qed

\section{Stationary case}
\label{stationarycase}
\subsection{$q$-TASEP with two parameters}
\label{mmodel}
In the last section, we found a formula for the two-sided $q$-Whittaker measure 
(\ref{ptn}) which corresponds to the $q$-TASEP with the initial condition (\ref{h-st-ic})
when $\a_j=\a$ for one $j$ and $\a_k=0, k\neq j$.  
Using this result, in this section, we will study the $q$-TASEP for the stationary situation
with parameter $\a$ under the conditioning that there is a particle at the origin at $t=0$. 
The conditioning can be taken into account by considering the initial condition 
(\ref{h-st-ic0}). 
The stationary situation would be realized if we specialize the parameters as
\begin{align}
\a_j=\a,~\a_k=0, k\neq j , \quad
a_1=a, a_2=\cdots=a_{N}=1, \quad 0<a,\a<1, 
\label{para1}
\end{align}
and take the limit $a\to\a$. 
In this section we suppose $a,\a\in\R$ rather than $a,\a\in\R^N$
as in previous sections.  
Note that the $q$-TASEP with the initial condition (\ref{h-st-ic}) is well-defined 
only when $\a<a$, since the initial position of the first particle $X_1(0)$ obeys 
$q{\rm Po}(\a/a)$ (\ref{q-geo}) which becomes singular when $a\to \a$. 
As in Introduction let us denote by $X_N(t)$ (resp. $X^{(0)}_N(t)$)
the $N$th particle position of $q$-TASEP with the initial condition 
(\ref{h-st-ic}) (resp. (\ref{h-st-ic0})) with (\ref{para1}). 
We define
\begin{align}
\label{Gdef}
G(\z)
&=
\left\langle\frac{1}{(\z q^{X_N(t)+N};q)_{\infty}}\right\rangle
=
\sum_{l\in\Z} \frac{1}{(\zeta q^l;q)_\i} \P[X_N(t)+N = l],
\\
G_0(\z)
&=
\left\langle\frac{1}{(\z q^{X^{(0)}_N(t)+N-1};q)_{\infty}}\right\rangle
=
\sum_{l\in\Z} \frac{1}{(\zeta q^l;q)_\i} \P[X_N^{(0)}(t) +N-1= l].
\label{f11}
\end{align}
It is easy to see that they are finite for $\zeta\notin q^n,n\in\Z$. 
Moreover when $\zeta\notin \R_{\geq 0}:=\{x\in\R|x\geq 0\}$, this decays fast 
as $r\to\i$ for $\zeta=re^{i\th},0<\th<2\pi$. 
Next we discuss a relation between $G(\z)$ and $G_0(\z)$ in the case $\a<a$. 
Later we perform analytic continuation  to a region 
including $a=\a$. Since 
$
X_N(t)=X^{(0)}_N(t)-\chi-1,
$
where $\chi\sim q{\text{Po}(\a/a)}$ and $X_N^{(0)}$ and $\chi$ are
independent with each other, one sees
\begin{align}
G(\z)
=
(\a/a;q)_{\infty}
\sum_{m=0}^{\infty}\frac{(\a/a)^m}{(q;q)_m}
G_0(\z q^{-m}), ~ \zeta\notin \R_{\geq 0}. 
\label{f12}
\end{align}
The inversion formula can also be obtained.
\begin{lemma}
\label{G0G}
For $\zeta\notin \R_{\geq 0}$, 
\begin{align}
G_0(\z)=\frac{1}{(\a/a;q)_{\infty}}
\sum_{k=0}^{\infty}
\frac
{(-1)^kq^{k(k-1)/2}(\a/a)^k}
{(q;q)_k}
G(\z q^{-k}).
\label{fl11}
\end{align}
\end{lemma}

\noindent
{\bf Proof.}
Substituting~\eqref{f12} into rhs of~\eqref{fl11}, one has
\begin{align}
\sum_{k,m=0}^{\i}
\frac
{(-1)^kq^{k(k-1)/2}(\a/a)^{k+m}}
{(q;q)_k(q;q)_{m}}
G_0(\z q^{-(k+m)})
=
\sum_{\ell=0}^{\infty}G_0(\z q^{-\ell})
\left(\frac{\a}{a}\right)^{\ell}
\sum_{k=0}^{\ell}
\frac
{(-1)^kq^{k(k-1)/2}}
{(q;q)_k(q;q)_{\ell-k}}.
\label{fl13}
\end{align}
Thus it is sufficient to prove
\begin{align}
\sum_{k=0}^{\ell}
\frac
{(-1)^kq^{k(k-1)/2}}
{(q;q)_k(q;q)_{\ell-k}}
=
\delta_{\ell,0},
\label{fl14}
\end{align}
and this is a direct consequence of a version of the  $q$-binomial theorem
(\ref{qbinom3}), 
\begin{align}
\sum_{k=0}^{\ell}
\frac
{(-1)^kq^{k(k-1)/2}(q;q)_{\ell}}{(q;q)_k(q;q)_{\ell-k}}x^k
=
(1-x)(1-xq)\cdots(1-xq^{\ell-1}).
\label{fl15}
\end{align}
Setting $x=1$ in~\eqref{fl15}, we obtain~\eqref{fl14}.
\qed

\smallskip
\noindent
By using the $q$-shift operator $T_q$ defined by
$T_qf(z)=f(qz)$ and the $q$-binomial theorem (\ref{qbinom}), 
\eqref{f12} is rewritten as
\begin{align}
G(\z)
=
\frac{(\a/a;q)_{\infty}}
{(\a T_{q^{-1}}/a;q)_{\infty}}
G_0(\z).
\label{fl12}
\end{align}
Formally this can be inverted as 
\begin{align}
G_0(\z)
=
\frac{(\a T_{q^{-1}}/a;q)_{\infty}}
{(\a/a;q)_{\infty}}
G(\z),
\label{G0G}
\end{align}
from which one readily obtains~\eqref{fl11}  by another 
$q$-binomial formula (\ref{qbinom2}), 
\begin{equation}
 \sum_{n=0}^\i \frac{(-1)^n q^{n(n-1)/2}}{(q;q)_n} z^n = (z;q)_\i. 
\end{equation} 

\bigskip
So far our discussions have been for the $q$-Laplace transform. 
Though this is enough for considering the long time limit of the distribution
as we will see in section 5.3, we here give 
a formula for the distribution for finite time $t$. 
Let us consider the generating functions, 
\begin{align}
Q_0(x)
&=
\sum_{l\in\Z}\P(X^{(0)}_N(t)+N-1\ge l)x^{l},
\label{lp2-1}
\\
Q(x)
&=
\sum_{l\in\Z}\P(X_N(t)+N\ge l)x^{l}.
\label{lp21}
\end{align}
Noting $X_N(t)=X_N^{(0)}(t)-\chi-1$, where $X_N^{(0)}$ and $\chi$ are independent and 
$\chi\sim q\text{Po}(\a/a)$, one has
\begin{align}
Q(x)
&=
\sum_{l\in\Z}
\left(
\sum_{m=0}^{\infty}
\P(X_N^{(0)}(t)\ge l+m+1)\P(\chi=m)
\right)
x^l
=
\sum_{n\in\Z}
\P(X_N^{(0)}(t)\ge n+1) x^n
\sum_{m=0}^{\infty}
\P(\chi=m)x^{-m}
\notag
\\
&=
Q_0(x)
\frac
{(\a/a;q)_{\infty}}
{(\a/ax;q)_{\infty}}.
\label{lp22}
\end{align} 
Applying (\ref{invqLapF0}),  we have the following. 
\begin{proposition}
\label{propF0Gq}
The distribution of $X_N^{(0)}(t)$ can be written in terms of $G(z)$, 
the $q$-Laplace transform of $X_N(t)$, as follows. 
\begin{align}
\P (X_N^{(0)}(t)\ge n-N+1)
=
\frac{1}{2\pi i}
\int_{C_0}\frac{dx}{x^{n}}
\frac{(\a/ax;q)_{\infty}}
{(\a/a;q)_{\infty}}
 \frac{(q;q)_{\infty}}{(x;q)_{\infty}}
\sum_{k\in\Z} (qx)^k R_k(G).
\label{F0Gq}
\end{align}
where $C_0$ is a small contour around the origin
and $R_k(G)$ means the residue at the pole $z=q^{-k}$ of the function $G$. 
\end{proposition}

\subsection{Stationary limit}
From the discussions in the last subsection,  we can study the $q$-TASEP in the 
stationary regime using (\ref{fl11}) from the result for the $q$-TASEP with the 
initial condition (\ref{h-st-ic}) with (\ref{para1}) and then taking the limit $a\to\a$. 
Here note that, as we have seen in section 3, the information about the distribution of the 
position of the $N$th particle $X_N(t)$ is contained in the two-sided $q$-Whittaker 
measure and the $q$-Whittaker functions $P_\l(a), Q_\l(\a,t)$ 
are symmetric functions. Therefore instead of (\ref{para1}) we can and will set the parameters 
in the formulas to 
\begin{align}
\a_1=\cdots=\a_{N-1}=0, \a_N=\a, \quad
a_1=\cdots=a_{N-1}=1,a_N=a, \quad 0<\a<a<1, 
\label{paraN}
\end{align}
and take the limit $a\to\a$. 

To discuss the stationary limit, it is useful to rewrite the kernel (\ref{Kernel})  in the form, 
\begin{equation}
 K = K_{a,\a} + (a-\a) B_1\otimes B_2, 
\end{equation}
where
\begin{align}
K_{a,\a}(m,n)&=\sum_{l=0}^{N-2} \phi_l(m) \psi_l(n).
\label{f27}
\end{align}
Here the functions $\phi_n,\psi_n,B_1,B_2$ in the kernel are given by  
\begin{align}
\phi_l(n) 
 &=
\tau(n) 
\int_D 
\frac{dv}{2\pi i} \frac{e^{-vt}}{v^{n+N-l}}\left(\frac{1}{v-1}\right)^{l+1}
 \frac{1}{(qv;q)_{\infty}^{N-1}} 
\frac{(q\a/v;q)_{\infty}}{(qv/a;q)_{\infty}},
\label{f23}
\\
B_1(n)
&:=\phi_{N-1}(n)=
\tau(n)
\int_D 
\frac{dv}{2\pi i} \frac{e^{-vt}}{v^{n+1}}\left(\frac{1}{v-1}\right)^{N-1}
\frac{1}{v-a}
 \frac{1}{(qv;q)_{\infty}^{N-1}} 
\frac{(q\a/v;q)_{\infty}}{(qv/a;q)_{\infty}},
\label{f220}
\\
\psi_l(n)
&=
\frac{1}{\tau(n)} 
\oint_{C_r} 
\frac{dz}{2\pi i z} 
e^{zt} z^{n+N-l-1}
(z-1)^l (q z;q)_{\infty}^{N-1}
\frac{(q z/a;q)_\i}{(q \a/z;q)_\i},
\label{f24}
\\
B_2(n)&:=\frac{\psi_{N-1}(n)}{a-\a}=
\frac{1}{\tau(n)}
\oint_{C_r}
\frac{dz}{2\pi i z} \frac{e^{zt} z^{n+1}}{z-\a}
(z-1)^{N-1} (q z;q)_{\infty}^{N-1}
\frac{(q z/a;q)_\i}{(q \a/z;q)_\i},
\label{f25}
\end{align}
where $l=0,1,\cdots,N-2$, the contour $C_r$ is around 0 and $\a q^j,j\in\N$,  
and the contour $D$ is around 1 and $a$. $\tau(n)$ is defined in (\ref{tau}). 
The bounds in Lemma \ref{fl2-2} still hold
but note that 
since the integrand of $\psi_l,0\leq l\leq N-2$ does not have a pole at $\a$,
the bound for this function can now be replaced by 
\begin{equation}
 |\psi_l(n)| \leq (\a q /b)^n, \quad n\geq 0, ~0\leq l\leq N-2. 
 \label{phi_bd2}
\end{equation} 

Let us  set 
\begin{equation}
A=f K_{a,\a}. 
\label{defA}
\end{equation}
where $f$ is an operator with kernel $f(m,n)=f(n)\d_{m,n}$. 
By the same arguments as in Proposition \ref{trcl},  
the operator $A$ is a trace-class operator on $\ell^2(\Z)$. 
Because of the conditions in Lemma \ref{fl2-2} and (\ref{phi_bd2}),  this is so in the region 
$0<\a q< b <a <1$. The operator $fK$ is also a trace-class 
operator on  $\ell^2(\Z)$ but it is so in the smaller 
region $0<\a<b<a<1$ (see  Lemma \ref{fl2-2}) due to the 
extra term in terms of $B_1,B_2$ in the kernel. 

Furthermore we have
\begin{lemma}
\label{fll2} For $\z<0$ and $0< q \a<b<a<1$, we have 
\begin{align}
||A||<1,
\label{fll24}
\end{align}
where $||\cdot ||$ denotes the operator norm in $\ell^2(\Z)$. This implies that
$1-A$ is invertible.
\end{lemma}

\noindent
{\bf Proof.}
First we note that by the biorthonormality  (\ref{on}) of $\phi_l$ and $\psi_l$, we have
$K_{a,\a}^2=K_{a,\a}$ leading to $||K_{a,\a}||=1$. 
Next, since $f(n)=1/(1-q^n/\z)<1$ (when $\zeta<0$) for any $n\in\Z$, one finds $||f||\leq 1.$
Now observe that $f$ can be bounded as 
\begin{equation}
 f \leq \frac12(f_1 + f_2), 
\end{equation}
where $||f_1||\leq 1$ and $f_2$ is a step function. For example one can take 
$f_1(n)=f(n+n_0), f_2(n)= 1_{n+n_0\geq 0}$ for large enough $n_0\in\N$. 
By the argument in Appendix B.3 of \cite{FS2006}(also Lemma D.3 of \cite{BCFV2015}), 
we know $||K_{a,\a} f_2|| <1$. Hence we have 
$||A|| \leq \frac12 ||(f_1+f_2) K_{a,\a}|| \leq \frac12(||f_1|| ||K_{a,\a}|| + ||f_2 K_{a,\a}||) < 1$. 
\qed

From Lemma~\ref{fll2} we see that in the parameter region $0<\a q< a<1$,
$G(\z)$~\eqref{Gdef} can be rewritten as
\begin{align}
G(\z) 
&=
 \det(1-A-(a-\a)fB_1\otimes B_2)_{\ell^2(\Z)}
\notag
\\
&=
(a-\a)\det(1-A)\left(\frac{1}{a-\a}-\sum_{n\in\Z} (\rho f B_1)(n)B_2(n)\right) 
\notag\\
&=
(a-\a)\det(1-A)\left(L_{a,\a}-\sum_{n\in\Z} (A\rho f  B_1)(n)B_2(n)\right),
\label{f29}
\end{align}
where
\begin{align}
&\rho(m,n)=(1-A)^{-1}(m,n),
\label{f210}
\\
&L_{a,\a}=\frac{1}{a-\a}-\sum_{n\in\Z} f(n)B_1(n)B_2(n).
\label{f211}
\end{align}
To perform the analytic continuation, we separate the contributions from the 
poles in $B_1,B_2$ as 
\begin{align}
B_1(n)&=B_1^{(1)}(n)+B_1^{(2)}(n),~B_2(n)=B_2^{(1)}(n)+B_2^{(2)}(n),
\label{f212-2}
\end{align}
where $B_1^{(1)}(n)$ (resp. $B_2^{(1)}(n)$) denotes
the residue at $v=a$ (resp. $z=\a$) in (\ref{f220}) (resp. (\ref{f25})) 
and $B_1^{(2)}(n)$  (resp. $B_2^{(2)}(n)$) denotes the
remaining contribution:
\begin{align}
B_1^{(1)}(n)
&=
\frac
{(-1)^{N-1}\tau(n)e^{-a t}(q \a/a;q)_{\infty}}
{a^{n+1}(a;q)_{\infty}^{N-1}(q;q)_{\infty}},
\label{f213}
\\
B_1^{(2)}(n)
&=
\tau(n)
\int_{D_1} \frac{dv}{2\pi i} \frac{e^{-vt}}{v^{n+1}}\left(\frac{1}{v-1}\right)^{N-1}
\frac{1}{v-a}
 \frac{1}{(qv;q)_{\infty}^{N-1}} 
\frac{(q\a/v;q)_{\infty}}{(qv/a;q)_{\infty}},
\label{f214}
\\
B_2^{(1)}(n)
&=
\frac
{(-1)^{N-1}\a^{n+1}e^{\a t}(\a;q)_{\infty}^{N-1}(q \a/a;q)_{\infty}}
{\tau(n)(q;q)_{\infty}},
\label{f215}
\\
B_2^{(2)}(n)
&=
\frac{1}{\tau(n)}
\oint_{q\a<|z|<\a} \frac{dz}{2\pi i z} \frac{e^{zt} z^{n+1}}{z-\a}
(z-1)^{N-1} (q z;q)_{\infty}^{N-1}
\frac{(q z/a;q)_\i}{(q \a/z;q)_\i},
\label{f216}
\end{align}
where $D_1$ denotes the contour enclosing 1 only.
Note that, to separate the pole contribution in $B_2$,
we should take the radius of the contour in (\ref{f216}) to be 
smaller than $\a$ and one can not take $d$ in (\ref{psi_bd}) to be  
arbitrary as in Lemma \ref{fl2-2}. 
The bound for $B_2^{(2)}$ reads $|B_2^{(2)}(n)| \leq (\a  q/b)^n, n\geq 0, (\a /c)^n, n<0$.
Substituting~\eqref{f213}--\eqref{f216} into (\ref{f211}), we write
\begin{align}
L_{a,\a}=\frac{1}{a-\a}-\sum_{n\in\Z} f(n)B_1^{(1)}(n)B_2^{(1)}(n)
-\sum_{\substack{i,j\in\{1,2\}\\(i,j)\neq (1,1)}}
\sum_{n\in\Z}f(n)B_1^{(i)}(n)B_2^{(j)}(n).
\label{LBB}
\end{align}
From the above remark, for the last sum for $i=1,j=2$ on rhs to make sense we should take $\a>a q$.
Combining this with the condition in Lemma \ref{fll2}, we will consider the problem in the region 
\begin{equation}
\label{aa_region}
0<qa<\a<1, 0<q\a<a<1. 
\end{equation}
For the moment we should put the condition $a\neq \a$ because of the apparent singularity at $a=\a$
in (\ref{LBB}). Below we remove this limitation $a\neq \a$ by analytic continuation. 
 
For the first two terms in (\ref{LBB}), we have
\begin{align}
&\quad
\frac{1}{a-\a}-\sum_{n\in\Z}f(n)B_1^{(1)}(n)B_2^{(1)}(n)
\notag\\
&=
\frac{1}{a-\a}
-
\frac{e^{(\a-a)t}}{a}
\left(
\frac
{(q\a/a;q)_{\infty}}
{(q;q)_{\infty}}
\right)^2
\left(
\frac
{(\a;q)_{\infty}}
{(a;q)_{\infty}}
\right)^{N-1}
\sum_{n\in\Z}
\frac{(\a/a)^n}
{1-q^n/\z}.
\label{fl22}
\end{align}
Now we recall the Ramanujan\rq{}s summation formula (\ref{ram}). 
Setting $a=1/\z,~b=q/\z, z=\a/a$, we have, for $\zeta\neq q^n, n\in\Z$, 
\begin{align}
\sum_{n\in\Z}
\frac{(\a/a)^n}{1-q^n/\z}
=
\frac
{(\a/a\z;q)_{\infty}(q;q)_{\infty}^2(a\z q/\a ;q)_{\infty}}
{(\a/a;q)_{\infty}(1/\z;q)_{\infty}(q a/\a;q)_{\infty}(q\z;q)_{\infty}}.
\label{fl24}
\end{align}
Note that, the assumption in (\ref{ram}), $|b/a|<|z|$, gives $qa<\a$ 
in our case which is satisfied in our discussions. 
Substituting this into~\eqref{fl22}, 
we arrive at the following expression of $L_{\a,a}$~\eqref{f211}:
\begin{align}
L_{\a,a}
=
&
\frac{1}{a-\a}
\left(
1
-
\frac
{(\a/a\z;q)_{\infty}(a\z q/\a ;q)_{\infty}(\a q/a ;q)_{\infty}}
{(1/\z;q)_{\infty}(q\z;q)_{\infty}(q a/\a;q)_{\infty}}
\left(
\frac
{(\a;q)_{\infty}}
{(a;q)_{\infty}}
\right)^{N-1}
e^{(\a-a)t}
\right)
\notag
\\
&-\sum_{\substack{i,j\in\{1,2\}\\(i,j)\neq (1,1)}}
\sum_{n\in\Z}f(n)B_1^{(i)}(n)B_2^{(j)}(n)
.
\label{f2100}
\end{align}
Combining~\eqref{f29} with~\eqref{f2100}, we have
\begin{align}
G(\zeta)
&=
(a-\a)\det(1-A) \notag\\
&\quad \times
\left(
\frac{1}{a-\a}
\left(
1
-
\frac
{(\a/a\z;q)_{\infty}(a\z q/\a ;q)_{\infty}(\a q/a ;q)_{\infty}}
{(1/\z;q)_{\infty}(q\z;q)_{\infty}(q a/\a;q)_{\infty}}
\left(
\frac
{(\a;q)_{\infty}}
{(a;q)_{\infty}}
\right)^{N-1}
e^{(\a-a)t}
\right)
\right.
\notag
\\
&\quad \left.
-\sum_{\substack{i,j\in\{1,2\}\\(i,j)\neq (1,1)}}
\sum_{n\in\Z}f(n)B_1^{(i)}(n)B_2^{(j)}(n)
-\sum_{n\in\Z} (A\rho f  B_1)(n)B_2(n)\right).
\label{f217}
\end{align}

We have the following. 
\begin{lemma}
\label{fll4}
$G (\z)$~\eqref{f217} is analytic in the region (\ref{aa_region}). 
\end{lemma}

\noindent
{\bf Remark.}
By "analytic for $\a,a$ in (\ref{aa_region})" we actually mean $G(\zeta)$ is an 
analytic function in both $\a$ and $a$ in some complex domains containing 
the region (\ref{aa_region}). From the explicit formulas we have 
obtained so far, our main concern about a possible non-analyticity is at $a=\a$.

\noindent
{\bf Proof.}
We divide~\eqref{f217} into several parts and consider the analyticity of each part. 

\smallskip
\noindent
(i)The factor $(a-\a)\det(1-A)$ is analytic for $0<q\a<a<1$
since, as mentioned before Lemma \ref{fll2},  
$A$ is a trace-class operator on $\ell^2(\Z)$ in this region. 

\smallskip
\noindent
(ii) Next we consider the second line in (\ref{f217}).
Note that the term
\begin{align}
\frac
{(\a/a\z;q)_{\infty}(a\z q/\a ;q)_{\infty}(\a q/a ;q)_{\infty}}
{(1/\z;q)_{\infty}(q\z;q)_{\infty}(q a/\a;q)_{\infty}}
\left(
\frac
{(\a;q)_{\infty}}
{(a;q)_{\infty}}
\right)^{N-1}
e^{(\a-a)t}
\label{f219}
\end{align}
is analytic when $q a/\a<1$, which is included in (\ref{aa_region}). 
Moreover,~\eqref{f219} is of order $1+O(a-\a)$ when $a$ and $\a$ are close.
Thus the second line in (\ref{f217}) 
is analytic in the region (\ref{aa_region}).

\smallskip
\noindent
(iii) We consider the first term in the third line of (\ref{f217}), i.e., 
\begin{align}
\sum_{\substack{i,j\in\{1,2\}\\(i,j)\neq (1,1)}}
\sum_{n\in\Z}f(n)B_1^{(i)}(n)B_2^{(j)}(n).
\label{fll42}
\end{align}
First for  the case $i=1,j=2$,
examining the large $|n|$ behaviors of~\eqref{f216}, (\ref{f})
and recalling the bounded for $B_2^{(2)}$ mentioned below (\ref{f216}),  
one sees that the sum over $n$ converges when $\a q/a<1$ for $n\to +\i$ and  when $a q/\a<1$ 
for $n\to -\i$.  One easily finds that both conditions are satisfied in (\ref{aa_region}). 
Thus \eqref{fll33} is analytic for (\ref{aa_region}). 

For the cases  with $i=2,~j=1$ and $i=j=2$, we take the sum over $j=1,2$ and consider 
$\sum_{n\in\Z} f(n) B_1^{(2)} B_2$.  Since the contour $D_1$ in (\ref{f214}) is only around 1, 
$B_1^{(2)}(n)$ is $\tau(n)$ times an $l$th order polynomial in $n$. On the other hand, for $B_2(n)$,
one has the bounds (\ref{psi_bd}) from Lemma \ref{fl2-2}. Combining these, 
one sees that the sum over $n$ is finite whenever $\a<1$. 

\smallskip
\noindent
(iv) At last we consider the second term in the third line of (\ref{f217}), i.e., 
\begin{align}
\sum_{n\in\Z} (A\rho f  B_1)(n)B_2(n).
\label{fll41}
\end{align}
By using the definition of $A$ (\ref{defA}), we rewrite it as
\begin{align}
\sum_{l=0}^{N-2}
\sum_{n\in\Z}f(n)\phi_{l}(n)B_2(n)
\sum_{n_1,n_2\in\Z}\psi_{l}(n_1)\rho (n_1,n_2)f(n_2)B_1(n_2).
\label{2factors}
\end{align}
For the first factor $\sum_{n\in\Z}f(n)\phi_{l}(n)B_2(n)$, 
since the contour $D$ goes around 1 in (\ref{f23}),
$\phi_l(n)$ is an $l$th order polynomial in $n$
and one can apply the same argument as for (iii) above and 
sees that the summation over $n$ converge whenever $\a<1$.
We also find that from~\eqref{phi_bd2} and Lemma~\ref{fl2-2}, $\psi_l\in\ell^2(\Z)$ 
and $fB_1\in\ell^2(\Z)$ when $q\a<b<a$. Recalling $\rho=(1-A)^{-1}$ is a bounded 
operator by Lemma \ref{fll2}, 
one sees that the second factor in (\ref{2factors}) is also analytic in $q\a < b<a$. 

Therefore combining the above (i)-(iv), we see that ~\eqref{fll41} is analytic (at least)
in the region (\ref{aa_region}). 
\qed

Next we consider the stationary limit $a\to \a$. If we restore the dependence 
on $a,\a$ in $B_j^{(i)},1\leq i,j\leq 2$, (\ref{f213})-(\ref{f216})  as 
$B_j^{(i)}(n;a,\a), 1\leq i,j\leq 2$, it is easy to see that one can simply take 
the limit $a\to\a$ of them: we set 
$B_j^{(i)}(n;\a)=\lim_{a\rightarrow \a}B_j^{(i)}(n;a,\a), 1\leq i,j\leq 2$. 
For $L_{a,\a}$ in (\ref{f211}) we have 
\begin{lemma}
\label{fl2}
\begin{align}
L:=\lim_{a\rightarrow \a}L_{a,\a}
&=-\frac{1}{\a}\sum_{n=0}^{\infty}
\left(
\frac{q^n/\z}{1-q^n/\z}
-\frac{\z q^{n+1}}{1-\z q^{n+1}}
+
\frac{2q^{n+1}}{1-q^{n+1}}
+
\frac{(N-1)\a q^{n}}{1-\a q^{n}}
\right)
+t
\notag
\\
&~~-\sum_{\substack{i,j=1,2\\(i,j)\neq (1,1)}}\sum_{n\in\Z}f(n)B_1^{(i)}(n;\a)B_2^{(j)}(n;\a).
\label{fl21}
\end{align}
\end{lemma}

\noindent
{\bf Proof.}
Using
\begin{align}
((1+c)x;q)_{\infty}=\prod_{n=0}^{\infty}(1-(1+c)xq^n)=(x;q)_{\infty}
-c (x;q)_{\infty}\sum_{n=0}^{\infty}
\frac
{xq^n}
{1-xq^n}
+O(c^2)
\label{fl26}
\end{align}
we find that when $a$ and $\a$ are close each factor 
of (\ref{f219}) can be expanded as
\begin{align}
&
\frac
{(\a/a\z;q)_{\infty}}
{(1/\z;q)_{\infty}}
=
1-\frac{\a-a}{a}\sum_{n=0}^{\infty}
\frac{q^n/\z}{1-q^n/\z}
+O((a-\a)^2),
\label{fl27}
\\
&
\frac
{(a q\z/\a;q)_{\infty}}
{(q\z;q)_{\infty}}
=
1-\frac{a-\a}{\a}\sum_{n=0}^{\infty}
\frac{\z q^{n+1}}{1-\z q^{n+1}}
+O((a-\a)^2),
\label{fl28}
\\
&
\frac
{(\a q/a;q)_{\infty}}
{(q a/\a;q)_{\infty}}
=
1-
\left(
\frac{\a-a}{a}
+
\frac{\a-a}{\a}
\right)
\sum_{n=0}^{\infty}
\frac{q^{n+1}}{1-q^{n+1}}
+O((a-\a)^2),
\label{fl29}
\\
&
\frac
{(\a;q)_{\infty}}
{(a;q)_{\infty}}
=
1-
\frac{\a-a}{a}
\sum_{n=0}^{\infty}
\frac{a q^{n}}{1-a q^{n}}
+O((a-\a)^2).
\label{fl210}
\end{align}
Thus from~\eqref{fl27}--\eqref{fl210}, we see that~\eqref{f219} is written as
\begin{align}
&\quad
\frac{1}{a-\a}-\sum_{n\in\Z}f(n)B_1^{(1)}(n)B_2^{(1)}(n)
\notag
\\
&=
-\frac{1}{a}\sum_{n=0}^{\infty}
\frac{q^n/\z}{1-q^n/\z}
+\frac{1}{\a}\sum_{n=0}^{\infty}
\frac{\z q^{n+1}}{1-\z q^{n+1}}
-
\frac{a+\a}{a \a}
\sum_{n=0}^{\infty}
\frac{q^{n+1}}{1-q^{n+1}}
-
\frac{N-1}{a}
\sum_{n=0}^{\infty}
\frac{a q^{n}}{1-a q^{n}}
+t.
\label{fl211}
\end{align}
From~\eqref{f211} and~\eqref{fl211} and using the fact that the term 
$\sum_{n\in\Z} B_1^{(i)}B_2^{(j)}$ remain finite in the stationary limit except
$i=j=1$ by the same arguments as for (iii) in Lemma \ref{fll4},
we arrive at~\eqref{fl21}.
\qed

Using Lemma \ref{G0G}, \ref{fll4} and \ref{fl2}, we have an expression of $G_0(\z)$ in the stationary 
limit.
\begin{proposition}
\label{fl3}
For $q$-TASEP with parameters (\ref{para1}) with $a=\a$, we have, for $\zeta\notin \R_{\geq 0}$, 
\begin{equation}
\left\langle \frac{1}{(\zeta q^{X^{(0)}_N(t)+N-1};q)_\i} \right\rangle 
=
G_0(\z)
=\frac{\a}{(q;q)_{\infty}}
\sum_{k=0}^{\infty}
\frac
{(-1)^kq^{k(k+1)/2}}
{(q;q)_k}
\left(V_N(\z q^{-k})-V_N(\z q^{-k-1})\right)
\label{fl31}
\end{equation}
where
\begin{align}
V_N(\z)=
\det(1-A)
\left(
L
-\sum_{n\in\Z}(\rho A f B_1)(n;\a)B_2(n;\a)\right).
\label{fl34}
\end{align}
Here $A$ is given by (\ref{defA}),(\ref{f27}),(\ref{f}) with $a=\a$, $L$ by (\ref{fl21}), $f(n)$ by (\ref{f}), $B$ by 
(\ref{f212-2})-(\ref{f216}) with $a=\a$ and $\rho$ by (\ref{f210}) with $a=\a$. 
Note that $A$ and $f$ also have $\zeta$ dependence. 
\end{proposition}

\noindent
{\bf Proof.}
First we consider the region (\ref{aa_region}).  
Note that in this region $G(\z)$ is well-defined due to Lemma~\ref{fll4}. 

Substituting $G(\z q^{-k})=q^kG(\z q^{-k})+(1-q^k)G(\z q^{-k})$ into~\eqref{fl11},
we have
\begin{align}
G_0(\z)
&
=\frac{1}{1-\a/a}\frac{1}{(\a q/a;q)_{\infty}}
\sum_{k=0}^{\infty}
\frac
{(-1)^kq^{k(k-1)/2}(\a q/a)^k}
{(q;q)_k}
\left(
G(\z q^{-k})
-
\frac{\a}{a}
G(\z q^{-k-1})
\right)
\notag
\\
&
=
\frac{1}{1-\a/a}\frac{1}{(\a q/a;q)_{\infty}}
\sum_{k=0}^{\infty}
\frac
{(-1)^kq^{k(k-1)/2}(\a q/a)^k}
{(q;q)_k}
\sum_{m=0}^1
\left(-
\frac{\a}{a}
\right)^m
G(\z q^{-k-m}).
\label{fl32}
\end{align}
Substituting~\eqref{f29} into~\eqref{fl32}, we get
\begin{multline}
G_0(\z)
=
\frac{a}{(\a q/a;q)_{\infty}}
\sum_{k=0}^{\infty}
\frac
{(-1)^kq^{k(k-1)/2}(\a q/a)^k}
{(q;q)_k}
\\
\times\sum_{m=0}^1
\left(-
\frac{\a}{a}
\right)^m
\left.
\det(1-A)\left(L_{\a,a}-\sum_{n\in\Z}f(n)(\rho A B_1)(n)B_2(n)\right)
\right|_{\z\rightarrow \z q^{-k-m}}.
\label{fl33}
\end{multline}
We take the stationary limit $a\rightarrow \a$ in this expression
using Lemma \ref{fl2}. 
To show the convergence of the last term, we notice that the 
resolvent converges in the operator norm in $\ell^2(\R)$, 
\begin{equation}
 \lim_{a\to \a} || (1-K_{a,\a})^{-1} -  (1-K_{a,a})^{-1} || = 0,
\end{equation}
which follows from the convergence of the kernel as in Lem 7.6 of \cite{BCFV2015}. 
Therefore we arrive at~\eqref{fl31}.
 \qed
 
 By taking $a\to\a$ limit in Proposition \ref{F0Gq}, one gets a formula for the 
 distribution of the particle position. 
 \begin{proposition}
 For the stationary $q$-TASEP with parameter $\a$, the distribution of $X_N^{(0)}(t)$
 is written as 
 \begin{align}
\P (X_N^{(0)}(t)\ge n-N+1)
=
\frac{1}{2\pi i}
\int_{C_0}\frac{dx}{x^{n}}
 \frac{(q;q)_{\infty}(x;q)_{\infty}}{(x;q)_{\infty}}
\sum_{k\in\Z} (qx)^k R_k(\hat{G}).
\label{F0Gq}
\end{align}
where $C_0$ is a small contour around the origin
and $R_k(\hat{G})$ means the residue at the pole $z=q^{-k}$ of the function 
$\hat{G}(\zeta)=\lim_{a\to\a} G(\zeta)/(a-\a)$. 
 \end{proposition} 

\subsection {Long time limit}
\label{LTL}
For the step initial condition, the long time limit of the $q$-TASEP was already discussed 
in \cite{FerrariVeto2015,Barraquand2015}. 
In this subsection we discuss the large time limit for the stationary case with the scaling 
discussed in section 2 based on the formulas obtained in the previous subsection. 
We consider the scaling (\ref{scalingQT}) with 
(\ref{f32}), (\ref{f33}),(\ref{f34}),(\ref{f31}) and  
\begin{align}
\z=-q^{-\eta N+\gamma N^{1/3}s},
\label{zeta_sc}
\end{align}
in Proposition \ref{fl3}. 
We will show that rhs of~\eqref{fl31} tends to the Baik-Rains distribution in the 
long time limit. By the same reasoning as for the step initial condition in sec. 5 of 
\cite{FerrariVeto2015}, for which the GUE Tracy-Widom distribution appears, 
this implies that the limiting distribution of the particle position, 
$\lim_{N\rightarrow\infty}\P(X^{(0)}_N(\kappa N)>(\eta-1) N-\g N^{1/3}s)$, is also 
the Baik-Rains distribution. 
Hereafter we focus on the limiting behavior of rhs of~\eqref{fl31}. 
First we show that $\det(1-A)$ tends to the GUE Tracy-Widom distribution. 
In Appendix \ref{TWlimit}, we will provide rather general lemmas to establish 
the GUE Tracy-Widom limit for a kernel of a specific form and we apply them to our case. 
Set 
\begin{align}
C_{N,n,l}
=
\frac{v_c^{l-N-n+1}}{(v_c-1)^{l+1}}\frac{1}{(v_c ;q)_{\infty}^{N-1}} , \quad v_c=q^{\th}, 
\end{align}
and define
\begin{align}
 \tilde{\phi}_l(n) 
 &= 
 \frac{\g N^{1/3}}{C_{N,n,l}}\phi_{l}(n),
\label{fl41}
\\
\tilde{\psi}_l(n)
&=
\gamma N^{1/3}(1-v_c)C_{N,n,l}\psi_{l}(n).
\label{fl42}
\end{align}
The reason of considering the factor $C_{N,n,l}$ becomes clear in the 
following lemma. 
\begin{lemma}
\label{phipsiasy}
The functions $\tilde{\phi}_l,\tilde{\psi}_l,0\leq l\leq N-2$ satisfy the assumptions of 
Lemma \ref{AiryK} with 
$(x,a,\g,c)\to (n, -\eta,\g,(1-v_c)\gamma)$.  
\end{lemma} 

\noindent
{\bf Proof.} 
First we show that the assumption (a) in Lemma \ref{AiryK} is satisfied. 
We provide a proof for only ~\eqref{fl41} since the one for ~\eqref{fl42} is similar. 
Let us write $\phi_l(n)$~\eqref{f23}
as 
\begin{align}
 \phi_l(n) 
 =
 \frac{(-1)^N}{2\pi i}\int_D dv \frac{e^{-Ng(v)}}{v^{\g N^{1/3}\xi}}
\left(
\frac{v}{v-1}
\right)^{-\g (1-v_c)N^{1/3}\l}
\frac
{(qv;q)_{\infty}(qa/v;q)_{\infty}}
{(v-1)(qv/a;q)_{\infty}},
\label{fl46}
\end{align}
where 
\begin{equation}
 g(v) = \k v -\eta \log v + \log (v;q)_\i,
\end{equation}
and set the scaling 
\begin{align}
n=-\eta N+\g N^{1/3}\xi,~l=N+\g (v_c-1)N^{1/3}\l.
\label{f311}
\end{align}
We find that $v_c=q^{\theta}$ is 
a solution of the saddle point equation, 
\begin{align}
g\rq{}(v_c)
=
\k-\frac{\eta}{v_c}-\sum_{n=0}^{\infty}\frac{q^n}{1-q^n v_c}=0.
\label{fl47}
\end{align}
This can be checked by simple calculations,
\begin{align}
g\rq{}(v_c)=\sum_{n=0}^{\infty}
\frac
{q^n-q^{\theta+2n}-(1-q^{\theta+n})q^n}
{(1-q^{\theta+n})^2}=0.
\label{fl49}
\end{align}
Furthermore the second and the third order derivatives of $g(v)$
can be calculated as
\begin{align}
g\rq{}\rq{}(v_c)=0,~g\rq{}\rq{}\rq{}(v_c)
=-2v_c^{-3}\g^3,
\label{fl411}
\end{align}
where $\g$ is given by~\eqref{f34}.
Thus $g(v)$ is expanded as
\begin{align}
g(v)=g(v_c)-\frac{\g^3}{v_c^{3}}\frac{(v-v_c)^3}{3}+O((v-v_c)^4).
\label{fl413}
\end{align}
If we take the contour $D$ to be the circle around 1 which passes through 
$v=v_c$, we see that the main contribution comes from around the saddle 
point $v=v_c$. Hence we focus on the contribution of the integral in~\eqref{fl46}
around the saddle point $v_c$. Scaling $v$ as
\begin{align}
v=
v_c
\left(1-
\frac{\bv}{\g N^{1/3}}
\right),
\label{fl414}
\end{align}
we find that each factor in~\eqref{fl46} behaves asymptotically as
\begin{align}
-Ng(v)
&\sim 
-N g(v_c)-\frac{\bv^3}{3},
\label{fl422}
\\
\frac{1}{v^{\g N^{1/3}\xi}}
&\sim
\frac{e^{\bv\xi}}{v_c^{\g N^{1/3}\xi}},
\label{fl423}
\\
\left(\frac{v}{v-1}\right)^{\g (v_c-1)N^{1/3}\l}
&\sim
\left(\frac{v_c}{v_c-1}\right)^{\g (v_c-1)N^{1/3}\l}
e^{\bv\l},
\label{fl424}
\\
\frac
{(qv;q)_{\infty}(qa/v;q)_{\infty}}
{(v-1)(qv/a;q)_{\infty}}
&\sim
\frac
{(qv_c;q)_{\infty}}
{v_c-1}
\label{fl415}
\end{align}
with the errors of order $O(N^{-1/3})$. 
Here we also applied the scaling for $\a$ given in~\eqref{f31}.
From~\eqref{fl422}-\eqref{fl415}, we see that in the scaling limit,
\begin{align}
\lim_{N\to\infty}\tilde{\phi}_l(n)
= 
\lim_{N\to\infty} \frac{\phi_l(n)}{C_{N,n,l}}
=
\int_{i\R} \frac{d\bv}{2\pi}
e^{-\frac{\bv^3}{3}+\bv(\xi+\l)}
= \Ai(\xi+\l).
\label{fl416}
\end{align}
This is uniform for $\xi,\l$ in a bounded domain. 

Next we show the assumptions (b),(c) in the Lemma \ref{AiryK}. Basic idea is 
taken from section 5  of \cite{Johansson2000}. We first observe $g(v)-g(v_c)$ can be written as
\begin{equation}
 g(v)-g(v_c) = -\frac{\g^3}{v_c^{3}}\frac{(v-v_c)^3}{3}+ \frac{\g^3}{v_c^4} (v-v_c)^4 \tilde{g}(v),
\end{equation}
where $\tilde{g}(v)$ satisfies $|\tilde{g}(v)|\leq 1/(3\delta_0)$ if $|v-v_c|/v_c < \delta_0$
for some $\delta_0>0$. 
By taking the contour $D$ to be a circle around 1
which passes $v_c(1+\delta)$, one sees that integrand is bounded by the value at 
$v_c(1+\delta)(<1)$ and hence one has 
\begin{equation}
 |\phi_n(x)| \leq C e^{N(g(v_c)-g(v_c(1+\delta)))-(y_1+y_2/(1-v_c))\d}. 
\end{equation} 
Then we see, due to the property of $\tilde{g}$ above and 
since $y_1 =\g N^{1/3} \xi,y_2=(1-v_c) \g N^{1/3}\l$,  
\begin{equation}
 |\phi_n(x) | \leq  C e^{\frac{2\g^3N}{3}\delta^3 -(\xi+\l) \g N^{1/3} \d}.
\end{equation}
Choose $\d = \sqrt{\xi+\l}/(\g N^{1/3})$ if $\xi+\l< \g^2 N^{2/3}\d_0^2$ and 
 $\d = \d_0$ if $\xi+\l \geq \g^2 N^{2/3}\d_0^2$. Then we have 
\begin{equation}
  |\phi_n(x) | \leq  c e^{-\frac13 \min(\sqrt{\xi+\l},\g N^{1/3}\d_0)(\xi+\l)}. 
\end{equation}
From this it is easy to see that the assumptions (b),(c) are satisfied for 
the function $\phi_n$. The same argument applies to the function $\psi_n$. 
\qed

\begin{lemma}
\label{Afcheck}
 The kernel (\ref{defA}) and the function $f$ (\ref{f}) with (\ref{zeta_sc}) and (\ref{f311}) 
 satisfy the assumptions of Lemma \ref{KernelTW}. 
\end{lemma}
\smallskip
\noindent
{\bf Proof.}
(i)(ii) follows from Lemma \ref{AiryK} and Lemma \ref{phipsiasy}
with (\ref{fl41}),(\ref{fl42}). 
For (iii), 
one sees in the scaling limit~\eqref{zeta_sc} and~\eqref{f311},
\begin{align}
\lim_{N\rightarrow\i}f(n)
=
\lim_{N\rightarrow\i}
\frac{1}{1+q^{\g N^{1/3}(\xi-s)}}=1_{\ge s}(\xi)
\end{align}
where $1_{\ge s}(\xi)=1 (\text{resp.}\, 0)$ for $\xi\ge s$ (resp. $\xi<s$). 
In fact it is easy to see that this limit holds as a convergence in $L^1(\R)$ norm. 
The first condition of (iv) is included in Lemma \ref{fll2} and the second one 
can also be checked in a similar manner. 
\qed

Combining Lemma \ref{KernelTW} and Lemma \ref{Afcheck}, we have
\begin{lemma}
 \begin{equation}
  \lim_{N\to\i} \det(1-A) = F_2(s). 
 \end{equation}
\end{lemma}

Next we consider the remaining factors in (\ref{fl31}). 
First we consider the asymptotics of $B_1(n)$ and $B_2(n)$. Set
\begin{equation}
D_{N,n}
=
-\frac{(v_c;q)_{\infty}e^{-Ng(v_c)}}{v_c^{\eta N+n+1}}
\end{equation} 
and 
\begin{align}
\B_{\o}^{(1)}(\xi)=e^{\o^3/3-\o\xi},
~~
\B_{\o}^{(2)}(\xi)=-\int_{0}^{\i}dz 
e^{\o z}\Ai(\xi+z),
\label{fl45}
\end{align}
\begin{equation}
\label{Bo}
\B_{\o}(\xi)=\B_{\o}^{(1)}(\xi)+\B_{\o}^{(2)}(\xi).
\end{equation}

\begin{lemma}\label{fl4}
With the scaling~\eqref{f31}, (\ref{f311}), 
we have 
\begin{align}
\lim_{N\to\i} B^{(i)}_1(n)/D_{N,n}
&= \B_{\o}^{(i)}(\xi),
~~i=1,2,
\label{fl43}
\\
\lim_{N\to\i} v_cD_{N,n}B^{(i)}_2(n)
&=\B_{-\o}^{(i)}(\xi),
~~i=1,2.
\label{fl44}
\end{align}
\end{lemma}
\smallskip
\noindent
{\bf Proof.} 
Since \eqref{fl44} can be shown in a similar way to~\eqref{fl43}, 
we provide the proof for only ~\eqref{fl43}.
To prove~\eqref{fl43} with $i=1$, we rewrite $B^{(1)}_1(n)$~\eqref{f213}
as
\begin{align}
B_1^{(1)}
=
e^{-Ng(a)}
\frac
{(aq;q)_{\infty}(a-1)}
{a^{\g N^{1/3}+1}}.
\label{fl420}
\end{align}
Noting that when $a$ is scaled as~\eqref{f31}, each factor 
in~\eqref{fl420} behaves as
\begin{align}
-Ng(a)
\sim 
-N g(v_c)+\frac{\o^3}{3},
~~
\frac{1}{a^{\g N^{1/3}\xi}}
\sim
\frac{e^{-\o\xi}}{v_c^{\g N^{1/3}\xi}},
~~
(aq;q)_{\infty}(a-1)
\sim
(v_cq;q)_{\infty}(v_c-1),
\label{fl425}
\end{align}
we get~\eqref{fl43} with $i=1$.

To prove~\eqref{fl43} with $i=2$, we rewrite~\eqref{f214} as
\begin{align}
B_1^{(2)}(n)=\int_{D_1}\frac{dv}{2\pi i v} 
\frac
{e^{-Ng(v)}}
{v^{\g N^{1/3}}}
\frac
{v-1}
{v-a}
\frac
{(qv;q)_{\infty}(qa/v;q)_{\infty}}
{(qv/a;q)_{\infty}}.
\label{fl418}
\end{align}
Considering the scaling behaviors~\eqref{fl422}-\eqref{fl415} and~\eqref{fl425},
we obtain
\begin{align}
B_1^{(2)}(n)
\sim D_{N,n}
\int_{i\R+\eta} \frac{d\bv}{2\pi}
\frac
{e^{-\frac{\bv^3}{3}+\bv(\xi+\l)}}
{\o+\bv}.
\label{fl419}
\end{align}
Here $\eta<-\o$. This comes from the fact that in~\eqref{f214}, $a<|v|$ is satisfied
since the contour $D_1$ in~\eqref{f214} encloses only $1(>a)$.
\eqref{fl43} with $i=2$ follows immediately from~\eqref{fl419}. 
\qed

One can see that the convergence in the above lemma is uniform in a bounded domain 
and the remainder becomes small as $N\to\i$ as in the case of $\phi_n,\psi_n$ in Lemma \ref{phipsiasy}. 
Thus we find that in the limit~\eqref{f31}, the ingredients in $V_N(\z)$~\eqref{fl34}
besides $\det(1-A)$ goes to
\begin{align}
&\lim_{N\rightarrow\i}\frac{v_c}{\g N^{1/3}}\sum_{\substack{i,j=1,2\\(i,j)\neq (1,1)}}\sum_{n\in\Z}f(n)B_1^{(i)}(n;\a)B_2^{(j)}(n;\a)
=
\sum_{\substack{i,j=1\\(i,j)\neq (1,1)}}^2\int_{s}^{\i}d\xi\B_\o^{(i)}(\xi)\B_{-\o}^{(j)}(\xi)
\label{f38},
\\
&
\lim_{N\rightarrow\i}
\frac{v_c}{\g N^{1/3}}
\sum_{n\in\Z} f(n)(\rho A B_1)(n;\a)B_2(n;\a)
=
\int_{s}^{\infty}d\xi
(\rho_{\A}\A\B_{\o})(\xi)\B_{-\o}(\xi),
\label{f39}
\end{align}
where 
$\B_{\o}(\xi)$ is defined in (\ref{Bo}). 
$\A$ is the operator which has the kernel $K(\xi,\zeta)1_{\geq s}$ with (\ref{AiryK}) and 
$\rho_\A = (1-\A)^{-1}$. 

\smallskip
Furthermore we have the following asymptotic behavior for the remaining part in~\eqref{fl33}:
\begin{lemma}
\label{fl5}
Under the scaling limit~\eqref{f31}, we have
\begin{align}
\lim_{N\rightarrow\infty}\frac{v_c}{\g N^{1/3}}\left[t-\frac{1}{\a}\sum_{n=0}^{\infty}
\left(
\frac{q^n/\z}{1-q^n/\z}
-
\frac{\z q^{n+1}}{1-\z q^{n+1}}
+
\frac{2q^{n+1}}{1-q^{n+1}}
+
\frac{(N-1)\a q^{n}}{1-\a q^{n}}
\right)\right]
=
s-\o^2
\label{f310}
\end{align}
where $v_c=q^{\th}$ and $\g$ is defined by~\eqref{f34} respectively.
\end{lemma}

\noindent
{\bf Proof.}
Note that since $\eta>0$,
\begin{align}
\lim_{N\rightarrow\infty}\frac{1}{\a}\sum_{n=0}^{\infty}
\frac{q^n/\z}{1-q^n/\z}
=
\lim_{N\rightarrow\infty}
\frac{1}{\a}\sum_{n=0}^{\infty}
\frac{q^{n+\eta N-\g N^{1/3}s}}{1+q^{n+\eta N-\g N^{1/3}s}}
=
0
\label{fl51}
\end{align}
and clearly one sees
\begin{align}
\lim_{N\rightarrow\infty}
\frac{v_c}{\g N^{1/3}}\sum_{n=0}^{\infty}
\frac{2q^{n+1}}{1-q^{n+1}}
=
0.
\label{fl52}
\end{align}
Thus nontrivial contributions come from the second and forth terms in 
the summation in~\eqref{f310}. 
For the second term,  we would like to show 
\begin{align}
\sum_{n=0}^{\infty}
\frac{\z q^{n+1}}{1-\z q^{n+1}}
=
-\sum_{n=0}^{\infty}
\frac{q^{n+1-\eta N+\g N^{1/3} s}}{1+q^{n+1-\eta N+\g N^{1/3} s}}
=
-\eta N+\g N^{1/3} s +O(N^0). 
\label{fl53}
\end{align}
Notice that, for a special case $\z=-q^{-m-1},m\in\N$, minus the second expression is 
\begin{equation}
 \sum_{n=0}^{\i} \frac{1}{1+q^{m-n}} 
 = 
 \sum_{n=0}^{2m} \frac{1}{1+q^{m-n}} +\sum_{n=2m+1}^{\i} \frac{1}{1+q^{m-n}}
 =
 m+\frac12 + \sum_{n=2m+1}^{\i} \frac{1}{1+q^{m-n}}
\end{equation} 
and (\ref{fl53}) holdes. 
For general $\z=-q^{-\mu},\mu\in\R$, the value of the second term is between those 
for $\zeta=-q^{-m}$ and $\zeta=-q^{-m-1}$ where $m=[\mu]$ is the largest integer which 
is smaller than or equal to $\mu$. Since the difference of these two 
values are of $O(m^0)$ for large $m$, one sees that (\ref{fl53}) holds generally. 
Considering also the scaling for $\a$ in~\eqref{f31}, we get
\begin{align}
\frac{1}{\a}\sum_{n=0}^{\infty}
\frac{\z q^{n+1}}{1-\z q^{n+1}}
=
-\frac{\eta}{v_c}N+\frac{\eta \o}{v_c\g} N^{2/3} 
+
\frac{\g^3s-\eta\o^2}{v_c\g^2}N^{1/3} +O(N^0). 
\label{fl54}
\end{align}

For the forth term in the sum in~\eqref{f310}, expanding in $\o$, we have
\begin{align}
&\quad -\frac{N-1}{\a}\sum_{n=0}^\i 
\frac{\a q^n}{1-\a q^n}
\notag\\
&=
-\sum_{n=0}^\i
\frac{q^n}{1-v_c q^n} 
N
-
\frac{\o}{\g}\sum_{n=0}^{\i}\frac{v_cq^{2n}}{(1-v_c q^n)^2}
N^{2/3}
-
\frac{\o^2}{\g^2}\sum_{n=0}^{\i}\frac{v_c^2q^{3n}}{(1-v_c q^n)^3}
N^{1/3}+O(N^0)
\notag
\\
&=
\left(-\k+\frac{\eta}{v_c}\right)N
-\frac{\eta\o}{v_c\g}N^{2/3}
+\left(\frac{\eta \o^2}{v_c\g^2}-\frac{\g \o^2}{v_c}\right)N^{1/3}
+O(N^0),
\label{fl55}
\end{align}
where in the last step we used (\ref{f32}),(\ref{f33}),(\ref{f34}). 
Considering~\eqref{fl54} and~\eqref{fl55} with $t=\k N$, we obtain
~\eqref{f310}.
\qed

Using Lemmas~\ref{fl3}-\ref{fl5}, we arrive at the following result:
\begin{theorem}
\label{main2}
For the stationary $q$-TASEP, when the parameter $\a$ determining the 
average density through (\ref{rhoal}) scales as 
$\a = q^{\th}(1+\o/(\g N^{1/3})),\th>0,\omega\in\R$,    
we have, for $\forall s\in\R$, 
\begin{align}
\lim_{N\rightarrow\infty}\P(X^{(0)}_N(\kappa N) >(\eta-1) N-\g N^{1/3}s)
=
F_\o(s):=
\frac{\partial}{\partial s}
\nu_{\o}(s),
\label{fp51}
\end{align}
where $\k,\eta,\g$ are given by (\ref{f32}),(\ref{f33}),(\ref{f34}). 
$F_\o(s)$ is a distribution function with $\nu_{\o}(s)$ expressed as
\begin{align}
\nu_{\o}(s)
=
F_2(s) 
\left(
s-\o^2-\sum_{\substack{i,j=1\\(i,j)\neq (1,1)}}^2\int_{s}^{\i}d\xi\B_\o^{(i)}(\xi)\B_{-\o}^{(j)}(\xi)
-\int_{s}^{\infty}d\xi
(\rho_{\A}\A\B_{\o})(\xi)\B_{-\o}(\xi)
\right)
\label{fp52}
\end{align}
where $F_2$ is the GUE Tracy-Widom distribution, $\A,\rho_{\A}$ are given below (\ref{f39})
and $\B_{\o}, \B_{\o}^{(i)},i=1,2$ are defined in (\ref{fl45}),(\ref{Bo}). 
\end{theorem}
\noindent
{\bf Proof.}
As we have mentioned at the beginning of this subsection, it is enough to show that 
rhs of ~\eqref{fl31} goes to $F_\o(s)$ as $t\to\i$. 
For taking the limit on rhs, we want to consider 
\begin{align}
~\lim_{N\rightarrow\i}
\frac{\a}{(q;q)_{\infty}}
\sum_{k=0}^{\infty}
\frac
{(-1)^kq^{k(k+1)/2}}
{(q;q)_k}
\left(V_N(\z q^{-k})-V_N(\z q^{-k-1})\right). 
\end{align}
Combining Proposition \ref{fl3}, Lemmas~\ref{fl4},\ref{fl5} and (\ref{f38}),(\ref{f39}), 
we see, for large $N$, 
\begin{equation}
 V_N(\zeta) = \nu_{\o}(s) + O(N^{-1/3}).
\end{equation}
From this it is easy to see, for a given $k$, 
\begin{align}
\lim_{N\to\i} \a (V_N(\z q^{-k})-V_N(\z q^{-k-1}))
=
\frac{\partial}{\partial s} \nu_{\o}(s).
\label{fp53}
\end{align}
On the other hand, we see that $\det(1-A)$ in (\ref{fl31}) and hence $V_N(\zeta q^{-k})$ 
goes to zero as $k\to\i$ for $|\zeta |\to\i$ for $\zeta\notin \R_{\geq 0}$ uniformly in $N$. 
Thus we find 
\begin{align}
&~\lim_{N\rightarrow\i}
\frac{\a}{(q;q)_{\infty}}
\sum_{k=0}^{\infty}
\frac
{(-1)^kq^{k(k+1)/2}}
{(q;q)_k}
\left(V_N(\z q^{-k})-V_N(\z q^{-k-1})\right)
\notag
\\
&=
\frac{1}{(q;q)_{\infty}}
\sum_{k=0}^{\infty}
\frac
{(-1)^kq^{k(k+1)/2}}
{(q;q)_k}
\frac{\partial}{\partial s} \nu_{\o}(s)
\label{fp55}
\end{align}
Using the relation (a special case of (\ref{qbinom2}))
\begin{align}
\sum_{k=0}^{\infty}
\frac
{(-1)^kq^{k(k-1)/2}q^k}
{(q;q)_k}=(q;q)_{\infty},
\end{align}
we arrive at our desired expression.
\qed

\medskip
\noindent
{\bf Remark.}
The distribution function $F_\o$ was 
first introduced in \cite{BR2000}. The representation in ~\eqref{fp51} with 
(\ref{fp52}) was given in \cite{IS2013}. 
In ~\cite{FS2006,BFP2010}, somewhat different representation was discussed. 
In the context of the $q$-TASEP,  this comes from the difference in the decomposition 
of the third factor in~\eqref{f29} for the $q$-TASEP.
We decompose the factor in~\eqref{f29} as
\begin{align}
&~\frac{1}{a-\a}-\sum_{n\in\Z}(\rho f B_1)(n)B_2(n)
\notag
\\
&=
\frac{1}{a-\a}-\sum_{n\in\Z} f(n)B_1^{(1)}(n)B_2^{(1)}(n)
-\sum_{\substack{i,j\in\{1,2\}\\(i,j)\neq (1,1)}}
\sum_{n\in\Z}f(n)B_1^{(i)}(n)B_2^{(j)}(n)
-\sum_{n\in\Z} (A\rho f  B_1)(n)B_2(n).
\label{l134}
\end{align}
On the other hand, in the case of~\cite{FS2006,BFP2010},
the corresponding decomposition is 
\begin{align}
&~\frac{1}{a-\a}-\sum_{n\in\Z}(\rho f B_1)(n)B_2(n)
\notag
\\
&=
\frac{1}{a-\a}-\sum_{n\in\Z} f(n)B_1^{(1)}(n)B_2(n)
-\sum_{n\in\Z}
\left(
(A \rho f B_1^{(1)})(n)+(\rho fB_1^{(2)})(n)\right)
B_2(n).
\label{l134}
\end{align}
Considering in the $N\rightarrow\infty$ limit with
$a\rightarrow\a$, each function goes to
\begin{align}
&f(n)\rightarrow 1_{\ge s}(\xi),
~B_1^{(1)}(n)\rightarrow e^{\o^3/3-\o\xi},
~B_2^{(1)}(n)\rightarrow e^{-\o^3/3+\o\xi},
\notag
\\
&~B_1^{(2)}(n)\rightarrow -\int_{0}^{\infty}d\l e^{\o\l}\Ai(\xi+\l),
~B_2^{(2)}(n)\rightarrow -\int_{0}^{\infty}d\l e^{-\o\l}\Ai(\xi+\l),
\label{l140}
\end{align}
we easily find the representation in~\cite{FS2006,BFP2010}.

\section{TASEP} 
In this section, we consider the stationary TASEP by taking $q\to 0$ limit 
in the previous sections. 
The stationary TASEP was already studied and the limiting distribution was 
established in \cite{FS2006}.  But there the relation 
between the stationary TASEP and the Schur process is not exact microscopically. 
Our approach provides a formula for the position of a tagged particle in the stationary 
case which is true even for finite time. Note that in the limit of reflecting Brownian 
motions, such formulas have been discussed \cite{FerrariSpohnWeiss2015}.

When $q\to 0$, our skew and ordinary $q$-Whittaker functions~\eqref{skewWh} and~\eqref{Wh} 
become a two-sided version of the skew and the ordinary Schur functions ~\cite{Borodin2011}. 
We denote them as $s_{\l/\mu}(a)$ and 
$s_{\l}(a_1,\cdots,a_N)$ respectively. On the other hand, the $q\rightarrow 0$
limit of $Q_{\l}(\a,t)$ in (\ref{defQ}) becomes
\begin{align}
s_\l\left(\alpha;t\right):=\lim_{q\rightarrow 0} Q_{\l}(\a;t)
 =
\int_{\T^N}\prod_{i=1}^N\frac{dz_i}{z_i}\cdot s_\l \left(1/z\right)
 \Pi^0\left(z;\a,t\right)m_N^0\left(z\right), 
\label{l113}
\end{align}
where
\begin{align}
&m_N^0(z)=\frac{1}{(2\pi i)^N N!}\prod_{1\le i<j\le N}
\left(1-\frac{z_i}{z_j}\right)\left(1-\frac{z_j}{z_i}\right),
~~\Pi^0\left(z;\a,t\right)=\prod_{i,j=1}^N\frac{1}{1-\a_i/z_j}
\cdot
\prod_{j=1}^Ne^{z_j t}.
\label{l114}
\end{align}
As a special case of the remark below Definition~\ref{d2}, 
we find that $s_\l\left(0;t\right)$ becomes the Schur function
with the Plancherel specialization on the ordinary Gelfand-Tsetlin
cone~$\G_N^{(0)}$~\eqref{GN0} while  $s_\l\left(\a;0\right)=s_{-\l}(\a),~-\l\in\G_N^{(0)}$. Hence we see that 
the $q\rightarrow 0$ limit of $P_{t}(\underline{\l}_N)$~\eqref{qWh-meas}, 
which is written as
\begin{align}
P_t^{0}(\underline{\l}_N)=\lim_{q\rightarrow 0} P_t(\underline{\l}_N)
=
 \frac{\prod_{j=1}^N s_{\l^{(j)}/\l^{(j-1)}} \left( a_j\right) \cdot s_{\l^{(N)}} \left(\a;t\right)}{\Pi^0(a;\a,t)} 
\label{l115}
\end{align}
is a type of the two-sided Schur process introduced by~\cite{Borodin2011}. 
In addition, as $q\rightarrow 0$, the generator $L(\underline{\mu}_N,\underline{\l}_N)$~\eqref{genL} becomes a simple form
\begin{align}
L^0(\underline{\mu}_N,\underline{\l}_N)
=\lim_{q\rightarrow 0}L(\underline{\mu}_N,\underline{\l}_N)
=
\sum_{1\le j\le k\le N}
a_k1_{\underline{\mu}_N^{jk+}\in\G_N}(\underline{\mu}_N)
\left(
\delta_{\underline{\l}_N,\underline{\mu}^{jk+}_N}
-\delta_{\underline{\l}_N,\underline{\mu}_N}
\right).
\label{l116}
\end{align}
The dynamics defined by the generator $L^0(\underline{\mu}_N,\underline{\l}_N)$~\eqref{l116} has been introduced in~\cite{BF2014, Borodin2011}.
In particular, the marginal $(\l^{(1)}_1,\l^{(2)}_2,\cdots,\l^{(N)}_N)$
describes (the ordinary) TASEP where the rate of the $k$th particle
is $a_k$. (So we call the $q\rightarrow 0$ limit the TASEP limit).

We readily obtain the TASEP limit of Theorem \ref{ql-det}:
\begin{corollary}
\label{lc1}
For the two-sided Schur process~\eqref{l115}, we have 
\begin{align}
\P\left(\l^{(N)}_N\ge m\right)
=
\det\left(1-K_0\right)_{L^2(\Z_{\ge -m+1})}, ~m\in\Z,
\label{lc11}
\end{align}
where the kernel $K_0(m,n)$ is written as
\begin{align}
K_0(m,n)&=\sum_{l=0}^{N-1}\phi_l^{(0)}(m)\psi_l^{(0)}(n)
\label{lc12}
\\
 \phi_l^{(0)}(n) &=\lim_{q\rightarrow 0}\phi_l(n)=\int_D dv \frac{e^{-vt}}{v^{n+N}}\frac{1}{v-a_{l+1}}
 \prod_{j=1}^l \frac{v-\a_j}{v-a_j},
 \label{l45} 
\\
 \psi_l^{(0)} (n)&=\lim_{q\rightarrow 0}\psi_l(n)=
(a_{l+1}-\a_{l+1})
 \int_{C_r} dz \frac{e^{zt} z^{n+N}}{z-\a_{l+1}}
 \prod_{j=1}^l \frac{z-a_j}{z-\a_j}
 \label{l46}
\end{align}
for $l=0,1,\cdots,N-2$. The contour $C_r$ is around 0 and $\a_j,1\leq j\leq N$,  
and the contour $D$ is around 1 and $a_j,1\leq j\leq N$. 
\end{corollary}

\smallskip
\noindent
{\bf Proof.}
In~\eqref{det_formula} we set 
\begin{align}
\zeta=-(1-q)q^{-m+\delta}
\label{l44}
\end{align}
where $m\in\Z_{\ge 0}$ and take the $\delta\downarrow 0$ limit.  
Noting that under~\eqref{l44}
\begin{align}
\lim_{\d\downarrow 0}
\lim_{q\rightarrow 0}\frac{1}{(\zeta q^x;q)_{\infty}}=1_{\ge 0}(x-m)
\label{l105}
\end{align}
where $1_{\ge 0}(x)=1$ for $x\ge 0$ and $=0$ for $x<0$,
we find that the lhs of~\eqref{det_formula} goes to 
\begin{align}
\lim_{q\rightarrow 0}
\left\langle
\frac{1}{(\zeta q^{\l_N^{(N)}};q)_{\infty}}
\right\rangle
=\P\left(\l^{(N)}_N\ge m\right).
\label{l103}
\end{align}

On the other hand we can also easily take the $q\to 0$ limit of the 
rhs of~\eqref{det_formula}. As in~\eqref{l105}, under~\eqref{l44} we have
\begin{align}
\lim_{\d\downarrow 0}
\lim_{q\rightarrow 0}
\frac{1}{1-q^{n}/\zeta}=1_{\ge 0}(n-m+1).
\label{l104}
\end{align}
In addition, 
$\phi_l^{(0)}(n)~\eqref{l45}$ and $\psi_l^{(0)}(n)$~\eqref{l46}
are easily obtained by taking $q\rightarrow 0$ limit of
$\phi_l(n)$~\eqref{phi} and $\psi_l(n)$~\eqref{psi} respectively.
From these, we see that the 
rhs of~\eqref{det_formula} goes to that of~\eqref{lc11}.
\qed
\smallskip

The relation~\eqref{lc11} can be interpreted as 
that for the $X_N(t)$, the position of $N$th particle of TASEP 
since $\l^{(N)}_N(t)=X_N(t)+N-1$. One finds
\begin{align}
\P\left(\l^{(N)}_N\ge m\right)
=\P_{\text{TASEP}}\left(X_N(t)\ge m-N+1\right),
\label{l117}
\end{align}
where in the rhs $\P_{\text{TASEP}}$
represents the probability measure with respect to both
the TASEP dyanmics and initial configurations. For general
parameters $a_j,~\a_j,~j=1,\cdots,N$, the pdf of the initial configurations
is given as the marginal density of~\eqref{l115}
on $\l^{(1)}_1,\l^{(2)}_2,\cdots,\l^{(N)}_N$. For more
special case where one of $\a_j$ remain finite (and it is set to be $\a$)
while all the others are zero, the initial configuration becomes
the half stationary initial data, i.e. the gap between $j-1$th and $j$th 
particle obeys the geometric distribution, 
$X_j(0)-X_{j-1}(0)\sim {\rm Ge} (\a/a_j), 1\leq j\leq N$  
with $X_0(0)=0$. 

\begin{corollary}
For TASEP with the parameters $a,\a$ described above, 
$\P_{\text{TASEP}}\left(X_N(t)\ge m-N+1\right)$ is written as the Fredholm 
detereminant in rhs of (\ref{lc11}) where in the functions $\phi^{(0)}_l,\psi^{(0)}_l$
the parameters are specialized as described above. 
\end{corollary}

When we further specialize the paramters as
$\a_1=\cdots=\a_N=0,~a_1=\cdots=a_N=1$, the initial configuration
reduces to the step initial condition, $X_j(0)=-j$. In this case, 
the kernel $K_0(m,n)$of the Fredholm determinant can be written as
\begin{align}
K_0(m,n)=\sum_{l=0}^{N-1}\frac{t^{2l-n-N+1}e^{-t}}{l!(l-n-N+1)!}
C_{l}(l-n-N;t)C_{l}(l-n-N+1;t)
\label{l118}
\end{align}
Here $C_{l}(x;t)$ is the $l$th order Charlier polynomial, and the 
orthogonal relation is given by
\begin{align}
\sum_{x=0}^{\infty} \frac{t^x e^{-t}}{x!}C_{m}(x;t)C_n(y;t)=t^{-n}n!\delta_{m,n}.
\label{l108}
\end{align}
This expression follows from the relation: in the case
$\a_1=\cdots=\a_N=0,~a_1=\cdots=a_N=1$, 
\begin{align}
\phi_l^{(0)}(n)&=\int_D dv \frac{e^{-vt}}{v^{n-l+N}}\frac{1}{(v-1)^{l+1}}
=\frac{t^le^{-t}}{l!}C_l(l-n-N;t),
 \label{l119} 
\\
 \psi_l^{(0)} (n)&=
 \int_{C_r} dz e^{zt} z^{n+N-l-1} (z-1)^l
=\frac{t^{l-n-N+1}}{(l-n-N+1)!}C_l(l-n-N+1;t).
\label{l120}
\end{align}
Though these functions are not exactly the same as the ones in 
\cite{Johansson2000,Johansson2002}, one could perform the 
asymptotics with this kernel and establish the GUE Tracy-Widom 
distribution.  

Next we take the TASEP limit of~Proposition~\ref{fl3} for 
the stationary $q$-TASEP. Let us set
\begin{align}
B_{10}^{(1)}(n)
&=
\lim_{q\rightarrow0}
B^{(1)}_1(n)
=
\frac
{e^{-\a t}}
{\a^{n+1}(\a-1)^{N-1}},
\label{lc23}
\\
B_{10}^{(2)}(n)
&=
\lim_{q\rightarrow0}
B^{(2)}_1(n)
=
\int_{D_1} \frac{dv}{2\pi i} \frac{e^{-vt}}{v^{n+1}}\left(\frac{1}{v-1}\right)^{N-1}
\frac{1}{v-\a},
\label{lc24}
\\
B_{20}^{(1)}(n)
&=
\lim_{q\rightarrow0}
B^{(1)}_2(n)
=
\a^{n+1}e^{\a t}(\a-1)^{N-1},
\label{lc25}
\\
B_{20}^{(2)}(n)
&=
\lim_{q\rightarrow0}
B^{(2)}_2(n)
=
\oint_{0<|z|<\a} \frac{dz}{2\pi i z} \frac{e^{zt} z^{n+1}}{z-\a}
(z-1)^{N-1},
\label{lc26}
\end{align}
where $B^{(i)}_j(n),i,j=1,2$ are defined by~\eqref{f213}--\eqref{f216} respectively
and we set $B_{10}(n):=B_{10}^{(1)}(n)+B_{10}^{(2)}(n)$ and 
$B_{20}(n):=B_{20}^{(1)}(n)+B_{20}^{(2)}(n)$.

We also define the kernel $A_0(m,n)$ as
the $q\rightarrow 0$ limit of $A(m,n)$~\eqref{defA}
with~\eqref{l44},
\begin{align}
A_0(m,n)=\lim_{q\rightarrow 0}A(n_1,n_2)=1_{\ge 0}(n_1-m+1)
\sum_{l=0}^{N-2}
\phi_{l}^{(0)}(n_1)\psi_{l}^{(0)}(n_2),
\label{lc27}
\end{align}
where $\phi_{l}^{(0)}(n_1)$ and $\psi_{l}^{(0)}(n_2)$ are 
given by~\eqref{l45} and~\eqref{l46} respectively.

Then taking the $q\rightarrow 0$ limit of~\eqref{fl11} with~\eqref{l44}, 
we have the following. 
\begin{corollary}
Let $X_N^{(0)}(t)$  
be the position of the $N$th particle in the stationary TASEP, with the conditioning 
that there is a particle at the origin initially, with parameter $\alpha$ at time $t$. 
We have
\label{lc2}
\begin{align}
\P(X_{N}^{(0)}(t)\ge m-N)=\a(V_0(m)-V_0(m+1)),
\label{lc21}
\end{align}
where for $m\in\Z$ 
\begin{align}
V_0(m)
&=\det\left(1-A_0\right)
\left(
-\frac{1}{\a}(m-1)+\frac{(N-1)\a}{1-\a}+t
\right.
\notag\\
&-\sum_{\substack{i,j=1\\i,j\neq (1,1)}}
\sum_{n\in\Z}1_{\ge 0}(n+m-1)B^{(i)}_{10}(n;\a)B^{(i)}_{20}(n;\a)
\notag\\
&\left.-\sum_{n\in\Z}1_{\ge 0}(n+m-1)
(\rho_{A_0}A_0B_{1,0})(a;\a)B_2(n;\a)
\right)
\label{lc22}
\end{align}
and $\rho_{A_0}=(1-A_0)^{-1}$. 
\end{corollary}

\smallskip
\noindent
{\bf Proof.}
We can show that $G_0(\zeta)$ with $a=\a$ 
goes to the lhs of~\eqref{lc21} in the same way as~\eqref{l103}.
For the rhs, we note under~\eqref{l44}
\begin{align}
&\lim_{\d\downarrow 0}\lim_{q\rightarrow0}\frac{q^n/\zeta}{1-q^n/\zeta}
=
\lim_{\d\downarrow 0}
\frac
{-q^{n+m-\delta}/(1-q)}
{1+q^{n+m-\delta}/(1-q)}
=
\begin{cases}
0,& n\ge -m+1,\\
-1, &n \le -m
\end{cases}
\label{lc28}
\\
&\lim_{\d\downarrow 0}\lim_{q\rightarrow0}
\frac{-\zeta q^{n+1}}{1-\zeta q^{n+1}}
=
\lim_{\d\downarrow 0}
\frac
{(1-q)q^{n+m+1+\delta}}
{1+(1-q)q^{n-m+\delta}}
=
\begin{cases}
0,& n\ge m-1,\\
-1, & n\le m-2.
\end{cases}
\label{lc29}
\end{align}
Combining these, we have the simple relation
\begin{align}
\lim_{\d\downarrow 0}\lim_{q\rightarrow0}
\left(\frac{q^n/\zeta}{1-q^n/\zeta}
-\frac{\zeta q^{n+1}}{1-\zeta q^{n+1}}
\right)=m-1.
\label{lc30}
\end{align}
From~\eqref{lc23}--\eqref{lc26} with the simple fact
\begin{align}
\lim_{q\rightarrow 0}
\left(2\frac{q^{n+1}}{1-q^{n+1}}+\frac{(N-1)\a q^n}{1-\a q^n}\right)
=
\frac{(N-1)\a}{1-\a},
\end{align}
we see 
\begin{align}
\lim_{q\rightarrow 0} V(\zeta)=V_0(m).
\label{lc26}
\end{align}
Furthermore noting in the rhs of~\eqref{fl31}, only the term with $k=0$
contributes to the $q\rightarrow 0$ limit, we find that the rhs goes to
the one in~\eqref{lc21}.
\qed

\smallskip
By considering the $q=0$ case of our analysis for $q$-TASEP in the 
previous section, one can prove that the fluctuations of  a tagged particle 
in stationary TASEP is described by the Baik-Rains distribution. 

So far the fluctuation properties of the stationary TASEP
have been studied in~\cite{FS2006,BFP2010}. 
There the authors introduced a last passage percolation (LPP) model
whose last passage time approximates a particle position of the 
stationary TASEP. The LPP model has a nice property that
the distribution function of the last passage time is represented 
in the language of the Schur measure.
However it has been known that
there is a discrepancy between the last passage time and the position
of the stationary TASEP for finite time $t$, although Proposition 3 in \cite{FS2006} shows 
that the discrepancy vanishes in the large time limit.
Note that our formula~\eqref{lc21} provides the exact distribution function
of the particle position since we directly deal with the particle position of the
($q$-)TASEP without using the approximations above.

\appendix
\section{Some $q$-functions and $q$-formulas}
\label{q}
In this appendix, we summarize a few $q$-notations, $q$-functions and $q$-formulas. 
The first is the $q$-Pochhammer symbol, or the $q$ shifted factorial.  
For $|q|<1$ and $n\in\N$,
\begin{equation}
 (a;q)_\i = \prod_{n=0}^\i (1-a q^n), ~~ (a;q)_n = \frac{(a;q)_\i}{(a q^n;q)_\i}.
\end{equation}
The $q$-binomial theorem will be useful in various places in the discussions,
\begin{equation}
\sum_{n=0}^\i \frac{(a;q)_n}{(q;q)_n} z^n = \frac{(az;q)_\i}{(z;q)_\i}, ~ |z|<1.
\label{qbinom}
\end{equation}
In particular the $a=0$ case appears in many applications. 
Another $q$-binomial formula reads (see e.g. Cor.10.2.2.b in \cite{AAR1999})
\begin{equation}
 \sum_{n=0}^\i \frac{(-1)^n q^{n(n-1)/2}}{(q;q)_n} z^n = (z;q)_\i. 
 \label{qbinom2}
\end{equation} 
There is yet another version of the  $q$-binomial theorem
(see e.g. Cor.10.2.2.c in \cite{AAR1999}),
\begin{align}
\sum_{k=0}^{\ell}
\frac
{(-1)^kq^{k(k-1)/2}(q;q)_{\ell}}{(q;q)_k(q;q)_{\ell-k}}x^k
=
(1-x)(1-xq)\cdots(1-xq^{\ell-1}).
\label{qbinom3}
\end{align}
The $q$-exponential function, denoted as $e_q(z)$ is defined to be 
\begin{equation}
 e_q(z) := \frac{1}{((1-q)z;q)_\i} = \sum_{n=0}^\i \frac{(1-q)^n}{(q;q)_n} z^n. 
\label{qexp}
\end{equation}
The second equality is by the above $q$-binomial theorem (\ref{qbinom}). 
From the series expansion expression, it is easy to see that this tends to the 
usual exponential function in the $q\to 1$ limit. 

The $q$-Gamma function $\Gamma_q(x)$ is defined by
\begin{align}
\Gamma_q(x)
=
(1-q)^{1-x}
\frac
{(q;q)_{\infty}}
{(q^x;q)_{\i}}.
\label{qgamma}
\end{align}
The $q$-digamma function is defined by $\Phi_q(z)=\partial_z\log\Gamma_q(z)$.
In the $q\to 1$ limit, they tends to the usual $\Gamma$ function and the 
digamma function respectively.  

Ramanujan's summation formula
(cf \cite{AAR1999} p502, \cite{GR2004} p138) is a two-sided generalization 
of the above $q$-binomial theorem (\ref{qbinom}).  
For $|q|<1, |b/a|<|z|<1$, 
\begin{equation}
\sum_{n\in\Z} \frac{(a;q)_n}{(b;q)_n} z^n 
=
\frac{(az;q)_\i (\frac{q}{az};q)_\i (q;q)_\i (\frac{b}{a};q)_\i}
       {(z;q)_\i  (\frac{q}{a};q)_\i (b;q)_\i (\frac{b}{az};q)_\i} . 
\label{Ram}
\end{equation}


\section{$q$-Whittaker functions and $q$-Whittaker process}
\label{stdqWh}
In this appendix, we introduce the ordinary (skew) $q$-Whittaker functions 
labeled by partitions and the $q$-Whittaker process and then discuss some 
of their properties. Most of them are standard \cite{Mac1995} 
but some of them are reformulated for the applications in the main text.  

First a partition of length $n\in\N$ is an $n$-tuple $\l=(\l_1,\ldots ,\l_n)$ with 
$\l_j\in\Z_+, 1\leq j\leq n$ s.t. $\l_1\geq \ldots \geq \l_n$. The set of all partitions
of length $n$ is denoted by $\Pe_n$. A partition $\l$ can also be represented 
(and identified ) as a Young diagram with $n$ rows of length $\l_1,\ldots ,\l_n$. 
For two partitions $\l\in\Pe_n,\mu\in\Pe_m$ s.t.
$m\leq n$ and $\l_i-\mu_i\geq 0, 1\leq i\leq n$ (with the understanding $\mu_i\equiv 0,
m<i\leq n)$, a pair $(\l,\mu)$ is called a skew diagram  and is denoted by $\l/\mu$. 
The transpose $\l'$ is the partition of length $\l_1$ defined as 
$\l_i'=\#\{j\in\Z_+ | \l_j\geq i\},1\leq i\leq \l_1$.

The Gelfand-Tsetlin (GT) cone 
for partitions, denoted by 
$\G_N^{(0)},N\in\N$ here, is defined by  
\begin{align}
\mathbb{G}^{(0)}_N
:=
&\{ (\l^{(1)},\l^{(2)},\cdots,\l^{(N)}), \l^{(n)}\in\Pe_n,1\leq n\leq N
| \notag\\
&\quad 0\le \l^{(m+1)}_{\ell+1}\le \l^{(m)}_{\ell}\le \l^{(m+1)}_{\ell}, 
1\le \ell \le m \le N-1
\}.
\label{GN0}
\end{align}
An element $\underline{\l}_N\in\G_N^{(0)}$ is called a Gelfand-Tsetlin pattern. 
Next we define the (skew) $q$-Whittaker functions. 

\begin{definition}{\label{dqWh}}
Let 
$\l\in\Pe_n,\mu\in\Pe_{n-1}$ be two partitions of order $n$ and $n-1$ respectively 
and $a$ an indeterminate. 
The skew $q$-Whittaker function (with 1 variable) is defined as 
(\cite{Mac1995}VI.7, (7.14) and Ex. 2.)
\begin{align}
 P_{\l/\mu}\left(a \right)
 =\prod_{i=1}^n a^{\l_i}\cdot
	\prod_{i=1}^{n-1}\frac{ a^{-\mu_i} (q;q)_{\l_i-\l_{i+1}}}
        {(q;q)_{\l_i-\mu_i}(q;q)_{\mu_i-\l_{i+1}}} .
\label{skewWhp}
\end{align} 
Using this, for a partition $\l\in\Pe_N$ and $a=(a_1,\cdots,a_N)$ being $N$ indeterminates, 
we define the $q$-Whittaker function with $N$ variables  as  
(cf. \cite{Mac1995}VI.7, (7.9') )
\begin{equation}
P_\l\left(a\right)
=
\sum_{\substack{\l_i^{(k)}, 1\leq i\leq k\leq N-1\\
            \l_{i+1}^{(k+1)} \leq \l_i^{(k)} \leq \l_i^{(k+1)}}}
	\prod_{j=1}^N P_{\l^{(j)}/\l^{(j-1)}}\left(a_j\right).
	\label{Whp}
\end{equation}
Here the sum is over the Gelfand-Tsetlin cone $\G_N^{(0)}$ with the condition 
$\l^{(N)}=\l$ and $\l^{(0)}=\phi$.
\end{definition}
It is known that the $q$-Whittaker function $P_\l(a)$ forms a basis of $\Lambda_N$, 
the space of $N$-variable symmetric polynomials with coefficients being rational 
functions in $q$ (cf \cite{Mac1995}VI).  
There is an inner product $\langle ~,~ \rangle$ in this space for which $P_\l$'s are 
orthogonal (\cite{Mac1995}VI (6.19)):
\begin{align}
\langle P_\lambda,P_\mu\rangle=(q;q)_{\lambda_N}\prod_{i=1}^{N-1}(q;q)_{\l_{i}-\l_{i+1}}
\delta_{\lambda,\mu}. 
\label{Pnorm}
\end{align}
Using this, we also introduce $Q_\l(x)$.
\begin{definition}
For $\l\in\Pe_N$ and for $x=(x_1,\cdots,x_N)$, we define
(\cite{Mac1995}VI (4.11),(4.12))
\begin{align}
Q_\l(x)
&=
\frac{P_\l(x)}{\langle P_\lambda,P_\lambda\rangle}.
\label{Q}
\end{align}
\end{definition}

\smallskip
\noindent
Note that $P_\l$ and $Q_\l$ are orthonormal: $\langle P_\l, Q_\mu\rangle = \delta_{\l,\mu}$. 

If we take a sum over all partitions of length $N$ of a product of $P_\l$ and $Q_\l$, 
the following Cauchy identity for the $q$-Whittaker functions holds
(\cite{Mac1995}VI(4.13),(2.5)),
\begin{align}
\sum_{\l\in\Pe_N} P_{\l}(x) Q_{\l}(y)=\prod_{ij=1}^N\frac{1}{(x_i y_j;q)_{\infty}}
=:\Pi(x;y). 
\label{CI}
\end{align}
Here we rewrite this for applications in the main text.  
First one notices that one can express $P_{\lambda}(x)$ in a form, 
\begin{align}
P_\l(x)=X^{\l_N}R_\ell(x),
\label{p81}
\end{align}
where $X=x_1 \cdots x_N$
and $\ell=(\ell_1,\cdots,\ell_{N-1})$ with $\ell_j=\l_{j}-\l_{j+1}, 1\leq j\leq N-1$
and $R_\ell \in \Lambda_N$.  This is seen as follows. Since 
$(\l^{(1)},\ldots, \l^{(N)})$ in (\ref{Whp}) is an element of $\G_N$,
the order of each term in the sum in (\ref{Whp}) is bigger than or equal to 
$\l^{(N)}_N$ and one can factor out this lowest order common factor. 
Then a partition $\l$ is uniquely determined by $\l^{(N)}_N$ and $\ell_j,j=1,...,N-1$
and the coefficient of each term depends only on $\ell_j,j=1,\ldots ,N-1$.
Then we have (\ref{p81}). 
We do not write explicitly the form of the function $R_\ell(x)$ since it is not 
necessary in the following discussion. In terms of $R_\ell$, the Cauchy 
identity~\eqref{CI} is equivalent to 
\begin{align}
\sum_{\ell_1,\cdots,\ell_{N-1}=0}^{\infty}
R_\ell(x)R_\ell(y)
\prod_{j=1}^{N-1}
\frac{1}{(q;q)_{\ell_j}}
=
\frac{\left(XY ;q\right)_{\infty}}{\prod_{i,j=1}^N( x_i y_j;q)_{\infty}}
\label{hci}
\end{align}
with $Y=y_1 \cdots y_N$. 
In fact if one substitutes (\ref{p81}) into lhs of (\ref{CI}) and uses (\ref{hci})
and the $q$-binomial theorem (\ref{qbinom}) with $a=0$, we get rhs of (\ref{CI}).

There is another inner product~$\langle \cdot,\cdot\rangle'$
in $\Lambda_N$ called the torus scaler product:
For the $N$-variable functions, $f(z), g(z)\in\Lambda_N$, 
the torus scalar product is defined by (\cite{Mac1995}VI (9.10))
\begin{align}
\langle f,g\rangle'_N
&=
\int_{\T^N}\prod_{j=1}^N
\frac{dz_j}{z_j}
\cdot
f(z)\overline{g(z)} m_N^q(z),
\label{torus}
\end{align}
where 
\begin{equation}
m_N^q(z)=\frac{1}{(2\pi i)^NN!}\prod_{1\le i<j\le N}(z_i/z_j;q)_{\infty}(z_j/z_i;q)_{\infty}
\label{Sk2}
\end{equation} 
is the $q$-Sklyanin measure. The $q$-Whittaker functions of $N$ variables 
are known to satisfy the following orthogonality relations with respect 
to this inner product
~\cite{Mac1995},
\begin{align}
\langle P_{\l},P_{\mu}\rangle'_N
=
\prod_{i=1}^N
\frac{1}{(q^{\l_i-\l_{i+1}+1};q)_{\infty}}
\cdot
\delta_{\l,\mu}. 
\label{torus-ortho}
\end{align}
Here we rewrite the orthogonality relation (\ref{torus-ortho}).  
From the definition~\eqref{Whp}, we find that 
$P_{\l}(e^{i\th})=P_{\l}(e^{i\th_1},\cdots,e^{i\th_N})$ 
can be expressed as
\begin{align}
P_{\l}(e^{i\th})=e^{iN\bar{\th}\bar{\l}}\tilde{P}_{\ell}(\tilde{\th}),
\label{nl46}
\end{align}
where $\bar{\th}=(\th_1+\cdots+\th_N)/N$ and $\bar{\l}=(\l_1+\cdots+\l_N)/N$ represent the barycentric coordinates
while relative coordinates are denoted by 
$\tilde{\th}=(\tilde{\th}_1,\cdots,\tilde{\th}_{N-1})$ and
$\ell=(\ell_1,\cdots,\ell_{N-1})$
with $\tilde{\th}_j=\th_j-\th_{j+1}$ and
$\ell_j=\l_j-\l_{j+1}, 1\leq j\leq N-1$.
Considering this we see that \eqref{torus-ortho} can be rewritten as,
with $m_j=\mu_j-\mu_{j+1}, 1\leq j\leq N-1$
\begin{align}
\langle P_{\l},P_{\mu}\rangle'_N
&=
\int_{0}^{2\pi}d\bar{\th} e^{iN\bar{\th}(\bar{\l}-\bar{\mu})}
\int_{(-\pi,\pi]^{N-1}}d\tilde{\th}
\tilde{P}_{\ell}(\tilde{\th})\tilde{P}_{m}(-\tilde{\th})
\prod_{1\le j<k\le N-1}
\left|
(e^{i(\tilde{\th}_j+\cdots+\tilde{\th}_k)};q)_{\infty})
\right|^2
\notag
\\
&=
\prod_{i=1}^N
\frac{1}{(q^{\ell_i+1};q)_{\infty}}
\cdot
\delta_{\l,\mu}.
\label{nl47}
\end{align}
Hence for the difference variables $\ell,m\in\Z_{\ge 0}^{N-1}$,
the orthogonality relation (\ref{torus-ortho}) can be restated as  
\begin{align}
\int_{(-\pi,\pi]^{N-1}}d\tilde{\th}
\tilde{P}_{l}(\tilde{\th})\tilde{P}_{m}(-\tilde{\th})
\prod_{1\le j<k\le N-1}
\left|
(e^{i(\tilde{\th}_j+\cdots+\tilde{\th}_k)};q)_{\infty})
\right|^2
=
\prod_{i=1}^N
\frac{1}{(q^{l_i+1};q)_{\infty}}
\cdot
\delta_{l,m}.
\label{nl48}
\end{align}

One finds a representation of $Q_\l$ using torus scalar product,
\begin{lemma}
For $y\in\R^N,\l\in\Pe_N$,
\begin{align}
Q_\l\left(y\right)
&=\frac{1}{\langle P_\l,P_\l\rangle'_N }
\langle \Pi(\cdot,y),P_\l(\cdot)\rangle_N'
\notag\\
&=\prod_{i=1}^{N-1}(q^{\l_i-\l_{i+1}+1};q)_{\infty}
\int_{\T^N}\prod_{i=1}^N\frac{dz_i}{z_i}\cdot P_\l \left(1/z\right)
 \Pi\left(z;y\right)m_N^q\left(z\right), 
\label{Qtorus}
\end{align}
where $1/z$ in $P_\l$ is a shorthand notation for $(1/z_1,\cdots,1/z_N)$.  
\end{lemma}

\smallskip
\noindent
{\bf Proof.}
It immediately follows from the Cauchy identity (\ref{CI}) 
and the orthogonality (\ref{torus-ortho}).
\qed

\smallskip

We can consider a generalization of the function $Q_\l$ by modifying 
$\Pi(z;y)$ in  (\ref{Qtorus}) to 
\begin{equation}
 \Pi(z;\{\a,\beta,\gamma\}) 
 = 
 \prod_{i=1}^N\prod_{j=1}^M  \frac{1+z_i \beta_j}{(z_i \a_j;q)_\i} e^{\gamma z_i}
\label{nnsp}
\end{equation}
with $\a_j,\b_j, \g\ge 0$ for $j=1,\cdots,M$. 
We call this $Q_{\l} \left(\{\a,\b,\g \}\right)$.
The case where only some of $\a_j$'s
(resp. $\b_j$'s or $\g$) are positive is called the $\a$-specialization 
(resp. $\b$-specialization or Plancherel specialization) in \cite{BC2014}. 
Note that the (\ref{nnsp}) for the $\a$-specialization is nothing but $\Pi(z;\a)$ in (\ref{CI}).  
The (ascending) $q$-Whittaker process corresponding to this generalization is defined by
\begin{align}
P (\underline{\l}_N)
 := 
 \frac{\prod_{j=1}^N P_{\l^{(j)}/\l^{(j-1)}} \left( a_j\right) \cdot Q_{\l^{(N)}} \left(\{\a,\b,\g \} \right)}
        {\Pi(z;\{\a,\beta,\gamma\})}.
\label{qWh-pr}
\end{align}

For instance we define $Q^{(\b)}_{\l}(y),y\in\R^M,M\in\N$ by~\eqref{Qtorus} with $\Pi(z;y)$ replaced
by $\Pi_M^{(\b)}(z;y):=\Pi(z;{\a=0,\b=y,\g=0})=\prod_{i=1}^N\prod_{j=1}^M(1+z_i y_j)$. 
We state two properties of the $q$-Whittaker functions with this 
$\beta$-specialization, which will be also useful to discuss our 
$q$-TASEP with a random initial condition (cf \cite{Mac1995}VI 7).
\begin{align}
&\sum_{\k}P_{\k/\l}(a)Q^{(\beta)}_{\k/\mu}(r)
=
\sum_{\t}P_{\mu/\t}(a)Q^{(\beta)}_{\l/\t}(r)\cdot
(1+ra)
\label{l413},
\\
&\sum_{\mu}Q_{\l/\mu}^{(\beta)}(r_{M+1})
Q_{\mu}^{(\beta)}(\{r\}_M)
=
Q_{\l}^{(\beta)}(\{r\}_{M+1}),
\label{l415}
\end{align}
where in~\eqref{l415}, $Q^{(\b)}_{\l/\mu}(r)$ with one variable is defined as 
\begin{align}
Q^{(\beta)}_{\l/\mu}(r)=\prod_{i\ge 1: \l_i=\mu_i,\l_{i+1}=\mu_{i+1}+1}
(1-q^{\mu_i-\mu_{i+1}})r^{|\l|-|\mu|}
\label{a1}
\end{align}
and we introduced a notation $\{r\}_M=(r_1,\cdots,r_M)$ to consider a change of $M$ below. 

In the rest of this appendix we consider some properties of the $q$-Whittaker 
process corresponding to this pure $\b$ specialization. 
\begin{definition}
For $M\in\N$, 
\begin{align}
P_M(\underline{\l}_N)
=
\frac{1}{\Pi^{(\beta)}(a;r)}\prod_{j=1}^NP_{\l^{(j)}/\l^{(j-1)}}(a_j)\cdot Q_{\l^{(N)}}^{(\beta)}(\{r\}_M). 
\label{a2}
\end{align}
\end{definition}
\smallskip 
As mentioned above, the positivity of~\eqref{a2} is known. 
This process describes a distribution function at time $M$ 
of a discrete time Markov process on $\G_N^{(0)}$ described by  
the following Kolmogorov forward equation. 
\begin{proposition}
\label{p26}
\begin{align}
P_{M+1}(\underline{\l}_N)
=
\sum_{\underline{\mu}_N}
P_{M}(\underline{\mu}_N)G_{M}(\underline{\mu}_N,\underline{\l}_N)
\label{a3}
\end{align}
where the transition matrix $G_M(\underline{\mu}_N,\underline{\l}_N)$ is
\begin{align}
&G_M(\underline{\mu}_N,\underline{\l}_N)
=
\prod_{j=1}^N
\frac
{P_{\l^{(j)}/\l^{(j-1)}}(a_j)Q^{(\beta)}_{\l^{(j)}/\mu^{(j)}}(r_{M+1})}
{\Delta(\l^{(j-1)},\mu^{(j)})},
\label{a4}
\end{align}
with $\Delta(\l,\mu)=\sum_{\k}P_{\k/\l}(a)Q^{(\beta)}_{\k/\mu}(r_M)$.
\end{proposition}

\noindent
{\bf Remark.} It is easy to see that each factor in (\ref{a4}) is a transition probability 
matrix from the definition of $\Delta(\l,\mu)$. 

{\smallskip}
\noindent
{\bf Proof.}
We rewrite $G_{M}(\underline{\mu}_N,\underline{\l}_N)$ as
\begin{align}
G_{M}(\underline{\mu}_N,\underline{\l}_N)
=
\prod_{r=1}^N A^{(\beta)}_r(r_{M+1})\cdot B^{(\beta)}(r_{M+1})\prod_{j=1}^N 
P_{\l^{(j)}/\l^{(j-1)}}(a_j),
\label{a5}
\end{align}
where
\begin{align}
A^{(\beta)}_r(r_{M+1})
=
\frac
{Q^{(\beta)}_{\l^{(r-1)}/\mu^{(r-1)}}(r_{M+1})}
{\Delta(\l^{(r-1)},\mu^{(r)})},~
B^{(\beta)}(r_{M+1})=Q^{(\beta)}_{\l^{(N)}/\mu^{(N)}}(r_{M+1}).
\label{a6}
\end{align}
Using~\eqref{l413} and~\eqref{l415}, we have
\begin{align}
&\sum_{\mu^{(r-1)}}P_{\mu^{(r)}/\mu^{(r-1)}}(a_r)
A_{r}^{(\b)}(r_{M+1})=\frac{1}{1+a_r r_{M+1}},~1\leq r\leq N,
\label{a7}
\\
&
\sum_{\mu^{(N)}}Q^{(\beta)}_{\l^{(N)}}(\{r\}_{M})B^{(\beta)}(r_{M+1})
=Q_{\l^{(N)}}^{(\b)}(\{r\}_{M+1}).
\label{a8}
\end{align}
Thus we have
\begin{align}
&\quad 
\sum_{\underline{\mu}_N}
P_{M}(\underline{\mu}_N)G_M(\underline{\mu}_N,\underline{\l}_N)
\notag
\\
&=
\frac{1}{\Pi_M^{(\beta)}(a;r)}
\prod_{r=1}^N
\left(\sum_{\mu^{(r-1)}}P_{\mu^{(r)}/\mu^{(r-1)}}(a_r)
A_{r}^{(\b)}(r_{M+1})\right)
\notag\\
&\quad\times 
\sum_{\mu^{(N)}}Q^{(\beta)}_{\l^{(N)}/\mu^{(N)}}(r_{M+1})B^{(\beta)}(r_{M+1})
\cdot
\prod_{j=1}^N 
P_{\l^{(j)}/\l^{(j-1)}}(a_j)
\notag
\\
&=
\frac{1}{\Pi_M^{(\beta)}(a;r)}
\prod_{r=1}^N
\frac{1}
{1+a_r r_{M+1}}
\cdot
Q_{\l^{(N)}}^{(\b)}(\{r\}_{M+1})
\cdot
\prod_{j=1}^N 
P_{\l^{(j)}/\l^{(j-1)}}(a_j)
=
P_{M+1}(\underline{\l}_N).
\label{a9}
\end{align}
\qed

\section{Two lemmas regarding the Airy kernel and the GUE Tracy-Widom limit} 
\label{TWlimit}
Here we provide two lemmas for establishing the GUE Tracy-Widon limit when 
a kernel of a specific form is given. The Hermite kernel for the GUE is a simplest 
example. For GUE, one can also use a bound due to Ledoux, which holds for GUE and 
simplifies the proof for the case(cf \cite{Ledoux2003, AGZ2009}), but in other applications
such a bound is not available. Our lemmas do not rely on such extra information 
but focus only on properties of the kernel. 
The essential part of the arguments are given in \cite{Johansson2000} and 
\cite{BFP2007} but we reformulate and generalize them in a way which would be suited 
for various applications. 
The Airy kernel $K(\xi,\zeta)$ is defined as 
\begin{equation}
 K(\xi,\zeta)  = \int_0^\i d\l \Ai(\xi+\l) \Ai(\zeta+\l). 
 \label{AiryK}
\end{equation}
Our discussions below are given in a continuous setting but can be 
also applied to a discrete setting as well. 

\begin{lemma}
\label{AiryK}
Suppose that we have a kernel of the form, 
\begin{equation}
 K_N^{(0)}(x,y) = \sum_{n=0}^{N-1} \vp_n(x) \psi_n(y), 
 \label{KN0}
\end{equation}
where $\vp_n,\psi_n,n\in\N$ are complex functions on $\R$. 
Assume that the functions $\vp_n(x)$ satisfy the followings for 
some $a\in\R, \g>0, c>0$. 

\smallskip
\noindent
\noindent
(a)For $\forall M>0$ and $\forall L>0$, 
\begin{equation}
\lim_{N\to\i} 
\vp_{N-cN^{1/3}\l}(aN+\g N^{1/3}\xi) = \Ai(\xi+\l),
\end{equation}
uniformly for $|\xi|<L$ and for $\l\in[0,M]$. 

\smallskip
\noindent
(b)
For $\forall L>0, \forall \e>0$ and $N$ large enough, 
\begin{equation}
 \vp_{N-cN^{1/3}\l}(aN+\g N^{1/3}\xi) \leq c_b e^{-\xi-\l},  
\end{equation}
for $\xi,\l$ satisfying $\l>0, L\leq |\xi  +\l | \leq \e N^{2/3}$ and for some constant $c_b$. 

\smallskip
\noindent
(c)
For $\forall\e >0$ and $N$ large enough, 
\begin{equation}
 \vp_{N-cN^{1/3}\l}(aN+\g N^{1/3}\xi) \leq c_c e^{-\xi-\l} , 
\end{equation}
for $\xi,\l$ satisfying $\l>0, |\xi  +\l |> \e N^{2/3}$ and for some constant $c_c$. 

The functions $\psi_n$ are also assumed to satisfy the same conditions 
(a),(b),(c) with the same parameters $a\in\R, \g>0, c>0$.
Define the rescaled kernel, 
\begin{equation}
 K_N(\xi,\zeta) = cN^{1/3} K_N^{(0)}(aN+\g N^{1/3}\xi,aN+\g N^{1/3}\zeta). 
\end{equation}
Then we have 

\smallskip
\noindent
(i)For $\forall L>0$,
\begin{equation}
\label{K2limit}
 \lim_{N\to\i} K_N(\xi,\zeta ) = K(\xi,\zeta), \quad \text{uniformly (in $N$) on} \quad  [-L,L]^2.
\end{equation}

\smallskip
\noindent
(ii)
For $\forall L>0$, 
\begin{equation}
 | K_N(\xi,\zeta)| \leq c_1 e^{-\max(0,\xi)-\max(0,\zeta)} ,
 \quad \text{uniformly (in $N$) for} \quad \xi,\zeta  \geq -L 
\end{equation} 
for some constant $c_1$. 
\end{lemma}

\noindent
{\bf Proof.} 
First we divide the sum over $n$ in (\ref{KN0}) as 
$\sum_{n=0}^{N-1} = \sum_{n=0}^{[N-cN^{1/3}M]} + \sum_{n=[N-cN^{1/3}M]+1}^{N-1}$,
where $[x]$ is the maximum integer which is less than or equal to $x\in\R$. 
For the second sum, we have, due to the uniform convergence in (a),
\begin{equation}
  \lim_{N\to\infty} cN^{1/3}\sum_{n=N-cN^{1/3}M}^{N-1} \vp_n(aN+c N^{1/3}\xi) \psi_n(aN+c N^{1/3}\zeta) 
  = 
   \int_0^M d\l \Ai(\xi+\l) \Ai(\zeta+\l)
\end{equation}
for $\xi,\zeta\in [-L,L]$ for $\forall L>0$. The first sum is, due to (c),(d), bounded by 
$\int_M^\i e^{-2\l} d\l = e^{-2M}/2$. By taking the $M\to\i$ limit, we have (\ref{K2limit}) in (i). 
The bound in (ii) easily follows from the uniform convergence in (a) 
(Note $\Ai(\xi)$ is bounded by a constant as  $\xi\to-\i$ and by $e^{-\xi}$ as $\xi\to \i$) 
combined with the bounds in  (b),(c). 
\qed

\begin{lemma}
\label{KernelTW}
Suppose that a (sequence of ) kernel $K_N: \R\times\R \to \C, N\in\N$ satisfies

\smallskip
\noindent
(i) For $\forall L>0$, 
\begin{equation}
 \lim_{N\to\infty} K_N(\xi,\zeta) = K(\xi,\zeta), \quad \text{uniformly in} \quad [-L,L]^2.
\end{equation}

\smallskip
\noindent
(ii)For $\forall L>0$, 
\begin{equation}
 | K_N(\xi,\zeta)| \leq c_1 e^{-\max(0,\xi)-\max(0,\zeta)} ,
 \quad \text{uniformly (in $N$) for} \quad \xi,\zeta \geq -L, 
\end{equation}
for some constant $c_1$. 

\smallskip
\noindent
In addition suppose that a (sequence of ) function $f_N: \R\to\R,N\in\N$ satisfy  

\smallskip
\noindent
(iii) The functions $f_N,N\in\N$ are uniformly bounded and converges to $1_{(s,\i)}$ 
for a $s\in\R$ in $L^1(\R)$ norm, 
\begin{equation}
 \lim_{N\to\infty} \int_\R |f_N (\xi) - 1_{(s,\i)} (\xi)|d\xi=0.
\end{equation}

\smallskip
\noindent
(iv) For $\forall L(>|s|)$,
\begin{equation}
\sqrt{|f_N(\xi)f_N(\zeta)|} |K_N(\xi,\zeta)| \leq e^{\xi+\zeta} 
\end{equation}
and 
\begin{equation}
\lim_{N\to\i}  \sqrt{|f_N(\xi)f_N(\zeta)|} |K_N(\xi,\zeta)| =0
\end{equation}
uniformly (in $N$) for $\xi,\zeta$ satisfying $\xi\leq -L$ or $\zeta\leq -L$. 

\noindent
Then for $\forall s\in\R$, 
\begin{equation}
 \lim_{N\to\infty} \det(1-f_N K_N)_{L^2(\R)} = F_2(s). 
\end{equation} 
\end{lemma}

\noindent
{\bf Remark.} 
The condition (i),(ii) are the same as (i)(ii) in Proposition \ref{AiryK}. 
For $f_N= 1_{(s,\i)},\forall N\in\N$, (iii)(iv) are trivially satisfied.
Hence the kernel $K_N$ from Proposition \ref{AiryK} with this special 
$f_N$ automatically satisfies the above four conditions. 
This special case appears in many applications, for example in GUE and TASEP. 

\noindent
{\bf Proof.} 
In this proof $c_i,i=1,2,\ldots$ are some constants. 
We first prove the $f_N= 1_{(s,\i)},\forall N\in\N$ case.  By the remark above, we will 
show that for a kernel safisfying (i)(ii) in Propsosition \ref{AiryK} we have 
\begin{equation}
 \lim_{N\to\infty} \det(1-K_N)_{L^2(s,\i)} = F_2(s). 
\end{equation} 
The general case will be treated by considering 
the difference to this special case later in the proof.   
For a given $s\in\R$, take $L(>|s|)$. 
By the Hadamard's inequality, 
\begin{equation}
 |\det A | \leq \prod_{i=1}^n (\sum_{j=1}^n |a_{i,j}|^2|)^{1/2},
\end{equation}
which holds for general $n\times n$ matrix \cite{HornJohnson2012}, 
and (ii), we have for $\xi,\zeta \geq -L$, 
\begin{align}
 |\det (K_N(\xi_i,\xi_j))_{1\leq i,j\leq k}|
 &\leq 
 \prod_{i=1}^k \left( \sum_{j=1}^k |K_N(\xi_i,\xi_j)|^2 \right)^{1/2} \notag\\
 &\leq
 \prod_{i=1}^k  (\sum_{j=1}^k c_1^2( e^{-\max(0,\xi_i)-\max(0,\xi_j)} )^2)^{1/2} \notag\\
 &\leq 
 c_1^k k^{k/2} \prod_{i=1}^k e^{-\max(0,\xi_i)},
 \label{bnd_detK}
\end{align}
and hence 
\begin{equation}
 \left| \int_{(s,\i)^k} \det (K_N(\xi_i,\xi_j))_{1\leq i,j\leq k} \prod_{i=1}^k d\xi_i \right|
 \leq
 c_1^k k^{k/2} \prod_{i=1}^k \int_s^{\i} e^{-\max(0,\xi_i)} d\xi_i
 \leq 
 c_2^k k^{k/2}. 
 \label{k_bound}
\end{equation}
Take $\e>0$. By (\ref{k_bound}), the Fredholm expansion,
\begin{equation}
 \det(1-K_N)_{L^2(s,\i)}
 =
 \sum_{k=0}^\i \frac{(-1)^k}{k!} \int_{(s,\i)^k} \det (K_N(\xi_i,\xi_j))_{1\leq i,j\leq k} \prod_{i=1}^k d\xi_i, 
\end{equation} 
converges and there exists $l\in\N$ s.t. 
\begin{equation}
\left| \det(1-K_N)_{L^2(s,\i)} 
- \sum_{k=0}^l \frac{(-1)^k}{k!} \int_{(s,\i)^k} \det (K_N(\xi_i,\xi_j))_{1\leq i,j\leq k} \prod_{i=1}^k d\xi_i \right|
\leq 
\sum_{k=l+1}^\i \frac{c_2^k k^{k/2}}{k!} 
\leq 
\frac{\e}{6}. 
\label{Kl}
\end{equation}
By (ii), we have
\begin{align}
 &\left| \left(\int_{(s,\i)^k}-\int_{(s,L]^k}\right) \det (K_N(\xi_i,\xi_j))_{1\leq i,j\leq k} \prod_{i=1}^k d\xi_i \right|
 \notag\\
 &\leq
 \int_{\substack{(s,\i)^k \\ \text{some}~ \xi_j>L}} \left |\det (K_N(\xi_i,\xi_j))_{1\leq i,j\leq k} \right| \prod_{i=1}^k d\xi_i 
 \notag\\
 &\leq
 \sum_{j=1}^k \int_{\substack{\xi_j>L\\ \xi_l>s, l\neq j}} 
 c_1^k k^{k/2} \prod_{i=1}^k e^{-\max(0,\xi_i)} \prod_{i=1}^k d\xi_i 
 \notag\\
 &=
 c_1^k k^{k/2+1} \int_L^{\i} e^{-\max(0,\xi_j)}d\xi_j \left(\int_s^{\i} e^{-\max(0,\xi)} d\xi \right)^{k-1},
 \label{KNiL}
\end{align}
where in the second inequality we used (\ref{bnd_detK}). 
Hence, for $L$ large enough, 
\begin{align}
 &\quad \left|\sum_{k=0}^l \frac{(-1)^k}{k!} 
 \left(\int_{(s,\i)^k}-\int_{(s,L]^k}\right) \det (K_N(\xi_i,\xi_j))_{1\leq i,j\leq k} \prod_{i=1}^k d\xi_i \right|
 \notag\\
 &\leq \sum_{k=0}^{\i} \frac{c_1^k k^{k/2+1}}{k!} \int_L^{\i} e^{-\max(0,\xi_j)}d\xi_j 
 \leq c_2 \int_L^{\i} e^{-\max(0,\xi_j)}d\xi_j 
 \leq 
 \frac{\e}{6},
 \label{KlL}
\end{align}
uniformly in $N$. 
Combining (\ref{Kl}) and (\ref{KlL}), we have
\begin{equation}
\left| \det(1-K_N)_{L^2(s,\i)} 
- 
\sum_{k=0}^l \frac{(-1)^k}{k!} \int_{(s,L]^k} \det (K_N(\xi_i,\xi_j))_{1\leq i,j\leq k} \prod_{i=1}^k d\xi_i \right|
\leq 
\frac{\e}{3}. 
\label{KNK}
\end{equation} 
By the uniform convergence (i), for $N$ large enough,  
\begin{align}
&\left| \sum_{k=0}^l \frac{(-1)^k}{k!} \int_{(s,L]^k} \det (K_N(\xi_i,\xi_j))_{1\leq i,j\leq k} \prod_{i=1}^k d\xi_i \right.
\notag\\
&\quad \left.
-
\sum_{k=0}^l \frac{(-1)^k}{k!} \int_{(s,L]^k} \det  (K(\xi_i,\xi_j))_{1\leq i,j\leq k} \prod_{i=1}^k d\xi_i 
\right|
\leq
\frac{\e}{3}. 
\label{KNKL}
\end{align}
By the same argument as to get (\ref{KNK}), we have 
\begin{equation}
\left| \det(1-K)_{L^2{(s,\i)} }
- 
\sum_{k=0}^l \frac{(-1)^k}{k!} \int_{(-\i,L]^k} \det (K(\xi_i,\xi_j))_{1\leq i,j\leq k} \prod_{i=1}^k d\xi_i \right|
\leq 
\frac{\e}{3}. 
\label{KL}
\end{equation}
Combining (\ref{KNK}),(\ref{KNKL}),(\ref{KL}), we have 
\begin{equation}
|\det(1-K_N)_{L^2(s,\i)} - \det(1-K)_{L^2{(s,\i)}} | \leq \e. 
\end{equation}
This completes the proof for the $f_N=1_{(s,\i)},\forall N\in\N$ case. 

Next we consider the general $f_N$ case. 
It is enough to show 
\begin{equation}
 \left| \int_{\R^k} \det (K_N(\xi_i,\xi_j))_{1\leq i,j\leq k} \prod_{i=1}^k f_N(\xi_i) d\xi_i - \int_{(s,\i)^k} \det (K_N(\xi_i,\xi_j))_{1\leq i,j\leq k} \prod_{i=1}^k d\xi_i \right| 
\leq 
c^k k^{k/2+1}\e 
\label{KNfs}
\end{equation} 
for large enough $N$. In the second term on lhs, the $\i$ as the upper limit of the integrals 
can be replaced by a large $L$ with an error of the form $c_1^k k^{k/2+1}\e$ as in (\ref{KNiL}). 
Similarly, in the first term on lhs of (\ref{KNfs}),  the $\i$ as the upper limit of the integrals 
can be replaced by $L$ with an error of the form $c_2^k k^{k/2+1}\e$ and the $-\i$ as the 
lower limit by $-L$ with an error of the form $c_3^k k^{k/2+1}\e$ due to (iv). 
Combining these, we can replace the integrals in (\ref{KNfs}) within $(-L,L]$ for large $L$
with an error $c_4^k k^{k/2+1}\e$. 
By a similar argument as to get (\ref{KNiL}),  we have
\begin{align}
&\quad \left| \int_{(-L,L]^k} \det (K_N(\xi_i,\xi_j))_{1\leq i,j\leq k} \prod_{i=1}^k f_N(\xi_i) d\xi_i - 
\int_{(s,L]^k} \det (K_N(\xi_i,\xi_j))_{1\leq i,j\leq k} \prod_{i=1}^k d\xi_i \right| 
\notag\\
&\leq
\sum_{j=1}^k \int_{(-L,s]}d\xi_j \int_{(-L,L]^{k-1}} \prod_{i=1,i\neq j}^k d\xi_i 
\prod_{i=1}^k  |f_N(\xi_i)| \left| \det (K_N(\xi_i,\xi_j))_{1\leq i,j\leq k} \right| 
\notag\\
&\quad +
\int_{(s,L]^k} | \prod_{i=1}^k f_N(\xi_i)-1 | 
\left| \det (K_N(\xi_i,\xi_j))_{1\leq i,j\leq k} \right| \prod_{i=1}^k d\xi_i .
\label{Kf_bd} 
\end{align}
Using (\ref{bnd_detK}) and (iii), one observes that the integral over $\xi_i, i\neq j$ in the 
first sum is finite. By using (iii), we see that the integral over $\xi_j$ in the first sum  
is bounded for large $N$ by 
$ c_5^k k^{k/2+1} \e$.  
Similarly for the second term in (\ref{Kf_bd}), one first sees 
$| \prod_{i=1}^k f_N(\xi_i)-1 | \leq c_6 \sum_{i=1}^k |f_N(\xi_i)-1|$ since $f_N$ is bounded. Then 
using (iii) and (\ref{bnd_detK}) one sees that it is bounded by 
$ c_7^k k^{k/2+1} \e $
for large $N$. Combining all of the above, we get (\ref{KNfs}). 
Taking the sum over $k$ and taking $N\to\i$ limit, we find 
\begin{equation}
\lim_{N\to\i} \det(1-f_N K_N)_{L^2(\R)} 
=
\lim_{N\to\i} \det(1-K_N)_{L^2{(s,\i)}}.
\end{equation}
This completes the proof for general $f_N$ case. 
\qed

\section{Inverse $q$-Laplace transforms}
\label{invqLap}
In this appendix we discuss inversion formulas of the $q$-Laplace transform. 
In Appendix B of \cite{IS2017p2}, we discussed a few inversion formulas for the 
ordinary Laplace transform, which is summarized as follows. First the 
Laplace transform is defined as 
\begin{equation}
 \hat{\vp}(u) = \int_0^{\i} e^{-ux} \vp(x) dx, ~ u\in\C. 
 \label{Lap}
\end{equation}
(In \cite{IS2017p2} we used the notation $\tilde{\vp}$. Here we use $\hat{\vp}$
to keep the former for the transformed one, see below.) 
When $\vp(x)$ is a (probability) distribution function on $(0,\i)$, 
the Laplace transform is analytic for ${\rm Re}\,u>0$. 
The usual inversion formula is 
\begin{equation}
 \vp(x) = \frac{1}{2\pi i} \int_{\d + i\R} du e^{ux} \hat{\vp}(u),~x>0,
 \label{invLap}
\end{equation}
where $\d$ should be taken so that the singularities of $\hat{\vp}$ are to the 
left of the integration contour. If $\vp$ is associated with a random variable $X$, 
we have $G(u):=\langle e^{-uX} \rangle = u\hat{\vp}(u)$. The formula can be 
restated for the random variable $Y=\log X$. For instance for the distribution 
$F(y)=\P[Y\leq y] = \P[X\leq e^y] = \vp(e^y)$, (\ref{Lap}) is rewritten as 
\begin{equation}
 \hat{\vp}(u) = \int_{\R} e^{-ue^y+y} F(y) dy =:\tilde{F}(u) .
\end{equation}
For discussing the distribution of the O'Connell-Yor polymer model, the following 
inversion formula was useful. 
\begin{proposition}
\label{fR}
For a random variable $Y$, set $G(u) = \langle e^{-ue^Y}\rangle$. 
The distribution function of $Y$ is recovered from $G(u)$ as
\begin{equation}
 F(y) 
 = 
 \frac{1}{2\pi i} \int_{\d+i\R} d\xi \frac{e^{y \xi}}{\Gamma(\xi+1)}
 \int_0^{\i} u^{\xi-1} G(u)du,
 \label{GinvLap2F}
\end{equation}
where $\d>0$. 
The corresponding density function $f(y)=F'(y)$, if it exists, is given by 
\begin{equation}
 f(y) 
 = 
 \frac{1}{2\pi i} \int_{\d+i\R} d\xi \frac{e^{y \xi}}{\Gamma(\xi)}
 \int_0^{\i} u^{\xi-1} G(u) du. 
 \label{invLap2f}
\end{equation}
\end{proposition} 
\noindent
The formulas discussed in this appendix are $q$-analogues of the above. 

\begin{figure}[h]
\begin{center}
\includegraphics[scale=0.6]{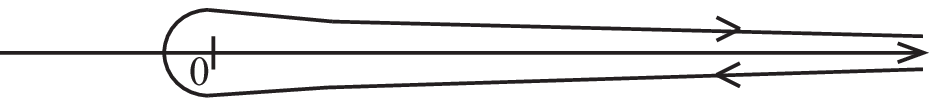}
\caption{\label{figgcontour}
The contour $\g$. }
\end{center}
\end{figure}

Suppose we have a function $f(n),n\in\Z$ 
and denote by $\tilde{f}_q(z)$ its $q$-Laplace transform, 
\begin{equation}
 \tilde{f}_q(z) := \sum_{n\in\Z} \frac{f(n)}{(zq^n;q)_{\i}}, 
 \label{qLap}
\end{equation}
In this appendix, we mainly consider the case where $f(n),n\in\Z$ is a 
discrete probability density function, i.e., $f(n)\geq 0, \sum_{n\in\Z} f(n)=1$, 
for which the $q$-Laplace transform converges and analytic except 
$z\neq q^n, n\in\Z$. 
By using the fact that the $q$-exponential function tends to the usual 
exponential function in a $q\to 1$ limit (see the comment after (\ref{qexp})), 
one sees that this formula 
goes to (\ref{Lap}) as 
\begin{align}
 \tilde{f}_q(-(1-q)u)
 = 
 \sum_{n\in\Z} \frac{f(n)}{(-(1-q)q^nu;q)_\i} \to \int_\R dy e^{y-ue^y} f(y) = \tilde{f}(u),
\end{align}
where in the limit ($\to$) we set $n=y/\log q$ and took $q\to 1$. 
An inverse formula for the $q$-Laplace transform is given by 
\begin{equation}
 f(n) = \int_\gamma \frac{dz}{2\pi i} q^n(q^{n+1}z;q)_\i \tilde{f}_q(z),
 \label{invqLap}
\end{equation}
where $\gamma$ is a contour enclosing $\R_+$ clockwise, see Fig. \ref{figgcontour}. 
Because of the factor $(q^{n+1}z;q)_\i$, the poles at $z=q^{-k},k=n+1,n+2,\cdots$ vanish
and the contour can be taken to be around the poles at $z=1,2,\ldots, n$, but the infinite 
contour in Fig. 6 has an advantage that it can be used for any $n$. 

In a certain $q\to 1$ limit, this, with the change of variable $z=-(1-q)u$, tends to (\ref{invLap}) as
\begin{align}
 f(n) &= q^n(1-q)\int_{\g}\frac{du}{2\pi i} (-(1-q)q^{n+1}u;q)_\i \tilde{f}_q(-(1-q)u)
 \to \int_{\d+i\R} \frac{du}{2\pi i}e^{u e^y}\tilde{f}(u). 
\end{align}
There is also an inversion formula corresponding to (\ref{invLap2f}).
\begin{proposition}
When $f(n),n\in\Z$ is a discrete probability density function,  
\begin{equation}
 f(n) 
 = 
 -\int_{C_0} \frac{dx}{2\pi i x^{n+1}} 
 \frac{(q;q)_{\infty}}{(qx;q)_{\infty}}
\sum_{k\in\Z} (qx)^k R_k(\tilde{f}_q).
\label{lp20}
\end{equation}
Here $R_k(\tilde{f}_q)$ is the residue of $\tilde{f}_q(z)$ at the pole $z=q^{-k}$ and $C_0$ is a 
small contour around the origin. 
\end{proposition}

\smallskip
\noindent
{\bf Proof.} 
It is equivalent to showing 
\begin{equation}
 \sum_{n\in\Z} f(n) x^n 
 =
 -\frac
{(q;q)_{\infty}}
{(qx;q)_{\infty}}
\sum_{k\in\Z} (qx)^k R_k(\tilde{f}_q),
\label{qLapinv1}
\end{equation}
for $x=e^{i\th},\th\in (-\pi,\pi]$. 
In fact considering the contour integral around the origin 
of both sides of (\ref{qLapinv1}) divided by $x^{n+1}$, we get (\ref{lp20}). 
By (\ref{invqLap}), this is further equivalent to showing 
\begin{align}
-\sum_{n\in\Z}x^n
\int_{\gamma}
\frac{dz}{2\pi i}
\tilde{f}_q(z)
q^n(q^{n+1}z;q)_{\infty}
=
\frac
{(q;q)_{\infty}}
{(qx;q)_{\infty}}
\sum_{k\in\Z} (qx)^k R_k(\tilde{f}_q).
\label{ll11}
\end{align}
Taking the poles at $z=q^{-k},k\in\Z$, one can rewrite lhs of~\eqref{ll11} as
\begin{align}
\sum_{k\in\Z}
R_k(\tilde{f}_q)
\sum_{n\in\Z}
(qx)^n(q^{n-k+1};q)_{\infty}
&=
\sum_{k\in\Z}
(qx)^k
R_k(\tilde{f}_q)
\sum_{n\in\Z}
(qx)^{n-k}(q^{n-k+1};q)_{\infty}.
\label{ll12}
\end{align}
The last sum in (\ref{ll12}) can be taken owing to 
the $q$-binomial theorem (\ref{qbinom}), 
\begin{align}
\sum_{n\in\Z}
(qx)^{n-k}(q^{n-k+1};q)_{\infty}
&=
\sum_{n=k}^{\infty}
(qx)^{n-k}(q^{n-k+1};q)_{\infty}
=
\frac
{(q;q)_{\infty}}
{(qx;q)_{\infty}}.
\label{ll13}
\end{align}
Substituting this into~\eqref{ll12}, we get~\eqref{ll11}.
\qed

\noindent 
The last sum in (\ref{lp20}) formally seems to be a consequence of 
taking poles at $z=q^{-k},k\in\Z$ of the integral,  
$\int_{\g} \frac{dz}{2\pi i z} z^{\xi} \tilde{f}_q(z)$ (though this is not 
true since $z^{\xi}$ has a cut along $\R_+$). 
Setting $x=q^{-\xi}$ and taking the $q\to 1$ 
limit of this and recalling the factor $(q;q)_\i/(qx;q)_\i$ can be written 
in terms of the $q$-Gamma function (\ref{qgamma}) which tends 
to the Gamma function as $q\to 1$, one would observe that 
(\ref{lp20}) may be regarded as a $q$-analogue of (\ref{invLap2f}). 

As a corollary of proposition \ref{fR}, we get a formula for the distribution function 
$F(n)=\P [X\leq n], n\in\N$. It reads 
\begin{align}
 F(n) 
 &= 
 -\int_{C_0} \frac{dx}{2\pi i x^{n+1}} 
 \frac{(q;q)_{\infty}}{(x;q)_{\infty}} 
\sum_{k\in\Z} (qx)^k R_k(\tilde{f}_q).
\label{invqLapF}
\end{align}
For the density function, $f(n):=F(n)-F(n-1)$, this gives (\ref{lp20}). 
(\ref{invqLapF}) is an analogue of (\ref{GinvLap2F}). 
If we introduce the generating function $Q(x)=\sum_{n\in\Z} F(n) x^n$, 
clearly 
\begin{equation}
F(n) = \int_{C_0} \frac{dx}{2\pi i x^{n+1}} Q(x)
\label{FQ}
\end{equation}
and (\ref{invqLapF}) is equivalent to 
\begin{equation}
 Q(x) = \frac{(q;q)_{\infty}}{(x;q)_{\infty}} \sum_{k\in\Z} (q x)^k R_k(\tilde{f}_q). 
 \label{Qx}
\end{equation} 
Suppose further that a random variable $X$ having the density $f(n),n\in\Z$
is written as $X=X_0+\chi$ where $X_0$ and $\chi$ are independent random variables on $\Z$.  
By the independence, the generating functions of them, 
$Q(x)$, $Q_0(x)=\sum_{n\in\Z} \P[X_0 \leq n] x^n, g(x)=\sum_{n\in\Z} \P[\chi = n] x^n$, 
are related as $Q(x) = g(x)Q_0(x)$. Combining this and (\ref{FQ}),(\ref{Qx}), we find a 
formula for the distribution of $X_0$, 
\begin{align}
 \P [X_0\leq n] 
 &= 
 \int_{C_0} \frac{dx}{2\pi i x^{n+1}} 
 \frac{(q;q)_{\infty}}{(x;q)_{\infty}g(x)}
\sum_{k\in\Z} (qx)^k R_k(\tilde{f}_q).
\label{invqLapF0}
\end{align}
Noting $\tilde{f}_q(\zeta)= \langle \frac{1}{(\zeta q^X;q)}\rangle$, this is a formula 
which gives the distribution function of $X_0$ in terms of the $q$-Laplace transform 
of $X$.

\providecommand{\bysame}{\leavevmode\hbox to3em{\hrulefill}\thinspace}
\providecommand{\MR}{\relax\ifhmode\unskip\space\fi MR }
\providecommand{\MRhref}[2]{%
  \href{http://www.ams.org/mathscinet-getitem?mr=#1}{#2}
}
\providecommand{\href}[2]{#2}


\begin{thebibliography}{10}

\bibitem{Aggarwal2016p}
A.~Aggarwal, \emph{{Current Fluctuations of the Stationary ASEP and Six-Vertex
  Model. arXiv:1608.04726}}.

\bibitem{AB2016p}
A.~Aggarwal and A.~Borodin, \emph{{Phase Transitions in the ASEP and Stochastic
  Six-Vertex Model. arXiv:1607.08684}}.

\bibitem{ACQ2011}
G.~Amir, I.~Corwin, and J.~Quastel, \emph{{Probability distribution of the free
  energy of the continuum directed random polymer in $1+1$ dimensions}}, Comm.
  Pure Appl. Math. \textbf{64} (2011), 466--537.

\bibitem{AGZ2009}
G.W. Anderson, A.~Guionnet, and O.~Zeitouni, \emph{{An Introduction to Random
  Matrices}}, Cambridge University Press, 2009.

\bibitem{Andjel1982}
E.~D. Andjel, \emph{{Invariant Measures for the Zero Range Process}}, Ann.
  Prob. \textbf{10} (1982), 525--547.

\bibitem{AAR1999}
G.~E. Andrews, R.~Askey, and R.~Roy, \emph{{Special Functions. Volume 71 of
  Encyclopedia of Mathematics and its Applications}}, Cambridge University
  Press, 1999.

\bibitem{BFP2010}
J.~Baik, P.~L. Ferrari, and S.~P\'ech\'e, \emph{{Limit process of stationary
  TASEP near the characteristic line}}, Comm. Pure Applied Math. \textbf{63}
  (2010), 1017--1070.

\bibitem{BR2000}
J.~Baik and E.~M. Rains, \emph{Limiting distributions for a polynuclear growth
  model with external sources}, J. Stat. Phys. \textbf{100} (2000), 523--541.

\bibitem{BS1995}
A.L. Barab{\'a}si and H.E. Stanley, \emph{Fractal concepts in surface growth},
  Cambridge University Press, 1995.

\bibitem{Barraquand2015}
G.~Barraquand, \emph{{A phase transition for $q$-TASEP with a few slower
  particles}}, Stoch. Process Their Appl. \textbf{125} (2015), 2674--2699.

\bibitem{BG1997}
L.~Bertini and G.~Giacomin, \emph{{Stochastic Burgers and KPZ Equations from
  Particle Systems}}, Comm. Math. Phys \textbf{183} (1997), 571--607.

\bibitem{Borodin2009}
A.~Borodin, \emph{{Determinantal point processes, arXiv:0911.1153}}.

\bibitem{Borodin2016p}
\bysame, \emph{Stochastic higher spin six vertex model and Madconald measures,
  {arXiv:1608.01553}}.

\bibitem{Borodin2011}
\bysame, \emph{{Schur dynamics of the Schur processes}}, Adv. Math.
  \textbf{228} (2011), 2268--2291.

\bibitem{BC2014}
A.~Borodin and I.~Corwin, \emph{{Macdonald processes}}, Prob. Th. Rel. Fields
  \textbf{158} (2014), 225--400.

\bibitem{BCFV2015}
A.~Borodin, I.~Corwin, P.L. Ferrari, and B.~Veto, \emph{{Height fluctuations
  for the stationary KPZ equation}}, Math. Phys. Anal. Geom. \textbf{18}
  (2015), 1--95.

\bibitem{BCR2013}
A.~Borodin, I.~Corwin, and D.~Remenik, \emph{{Log-gamma polymer free energy
  fluctuations via a Fredholm determinant identity}}, Comm. Math. Phys.
  \textbf{324} (2013), 215--232.

\bibitem{BCS2014}
A.~Borodin, I.~Corwin, and T.~Sasamoto, \emph{{From duality to determinants for
  $q$-TASEP and ASEP}}, Ann. Prob. \textbf{42} (2014), 2314--2382.

\bibitem{BF2014}
A.~Borodin and P.~L. Ferrari, \emph{{Anisotropic growth of random surfaces in
  2+1 dimensions}}, Comm. Math. Phys. \textbf{325} (2014), 603--684.

\bibitem{BFP2007}
A.~Borodin, P.~L. Ferrari, and M.~Pr{\"a}hofer, \emph{{Fluctuations in the
  discrete TASEP with periodic initial configurations and the Airy$_1$ process
  }}, Int. Math. Res. Papers (2007), rpm002.

\bibitem{BFPS2007}
A.~Borodin, P.~L. Ferrari, M.~Pr{\"a}hofer, and T.~Sasamoto, \emph{{Fluctuation
  properties of the TASEP with periodic initial configuration}}, J. Stat. Phys.
  \textbf{129} (2007), 1055--1080.

\bibitem{Bump1989}
D.~Bump, \emph{{The Rankin Selberg Method: A Survey, Number Theory, Trace
  Formulas and Discrete Groups}}, Academic Press (1989).

\bibitem{Burke1956}
P.~J. Burke, \emph{{The Output of a Queuing System}}, Oper. Res. \textbf{4}
  (1956), 699--704.

\bibitem{CLDR2010}
P.~Calabrese, P.~Le Doussal, and A.~Rosso, \emph{Free-energy distribution of
  the directed polymer at high temperature}, Euro. Phys. Lett. \textbf{90}
  (2010), 200002.

\bibitem{CGRS2016}
G.~Carinci, C.~Giardina, F.~Redig, and T.~Sasamoto, \emph{{ A generalized
  asymmetric exclusion process with $U_q(sl_2)$ stochastic duality}}, Prob. Th.
  Rel. Fields. \textbf{166} (2016), 887.

\bibitem{CFP2010}
I.~Corwin, P.~L. Ferrari, and S.~P\'ech\'e, \emph{{Universality of slow
  decorrelation in KPZ growth}}, Ann. Inst. H. Poincar\'e B \textbf{48} (2012),
  134--150.

\bibitem{CP2015}
I.~Corwin and L.~Petrov, \emph{{The q-PushASEP: A New Integrable Model for
  Traffic in 1+1 Dimension}}, J. Stat. Phys. \textbf{160} (2015), 1005--1026.

\bibitem{Dotsenko2010a}
V.~Dotsenko, \emph{{Replica Bethe ansatz derivation of the Tracy-Widom
  distribution of the free energy fluctuations in one-dimensional directed
  polymers}}, J. Stat. Mech. \textbf{P07010} (2010).

\bibitem{Ferrari2016p}
P.~Ferrari, \emph{{TASEP hydrodynamics using microscopic characteristics}},
  arXiv:1601.05346.

\bibitem{FerrariFontes1994}
P.~A. Ferrari and L.R.G. Fontes, \emph{{The net output process of a system with
  infinitely many queues}}, Ann. Appl. Prob. \textbf{4} (1994), 1129--1144.

\bibitem{FS2006}
P.~L. Ferrari and H.~Spohn, \emph{Scaling limit for the space-time covariance
  of the stationary totally asymmetric simple exclusion process}, Comm. Math.
  Phys. \textbf{265} (2006), 1--44.

\bibitem{FerrariSpohnWeiss2015}
P.~L. Ferrari, H.~Spohn, and T.~Weiss, \emph{{Brownian motions with one-sided
  collisions: the stationary case}}, Electron. J. Probab. \textbf{20} (2015).

\bibitem{FerrariVeto2015}
P.~L. Ferrari and B.~Vet\H{o}, \emph{{Tracy-Widom asymptotics for q-TASEP}},
  Ann. Inst. H. Poincar\'e Probab. Statist. \textbf{4} (2015), 1465--1485.

\bibitem{Forrester2010}
P.~J. Forrester, \emph{Log gases and random matrices}, Princeton University
  Press, 2010.

\bibitem{FNS1977}
D.~Forster, D.~R. Nelson, and M.~J. Stephen, \emph{Large-distance and long-time
  properties of a randomly stirred fluid}, Phys. Rev. A \textbf{16} (1977),
  732--749.

\bibitem{GR2004}
{G. Gasper, M. Rahman}, \emph{Basic hypergeometric series}, Cambridge, 2004.

\bibitem{Gon2014}
P.~Goncalves, \emph{{On the Asymmetric Zero-Range in the Rarefaction Fan}}, J.
  Stat. Phys. \textbf{154} (2014), 1974--1095.

\bibitem{HornJohnson2012}
R.~A. Horn and C.~R. Johnson, \emph{Matrix analysis, 2nd edition}, Cambridge,
  2012.

\bibitem{IS2017p2}
T.~Imamura and T.~Sasamoto, \emph{{Free energy distribution of the stationary O'Connell-Yor polymer model,
  arXiv:1701.06904}}.

\bibitem{IS2004}
\bysame, \emph{Fluctuations of the one-dimensional polynuclear growth model
  with external sources}, Nucl. Phys. B \textbf{699} (2004), 503--544.

\bibitem{IS2011}
\bysame, \emph{{Current moments of 1D ASEP by duality}}, J. Stat. Phys.
  \textbf{142} (2011), 919--930.

\bibitem{IS2012}
\bysame, \emph{{Exact solution for the stationary KPZ equation}}, Phys. Rev.
  Lett. \textbf{108} (2012), 190603.

\bibitem{IS2013}
\bysame, \emph{Stationary correlations for the {1D KPZ} equation}, J. Stat.
  Phys. \textbf{150} (2013), 908--939.

\bibitem{IS2016}
\bysame, \emph{{Determinantal structures in the O’Connell-Yor directed random
  polymer model}}, J. Stat. Phys. \textbf{163} (2016), 675--713.

\bibitem{Johansson2000}
K.~Johansson, \emph{Shape fluctuations and random matrices}, Comm. Math. Phys.
  \textbf{209} (2000), 437--476.

\bibitem{Johansson2002}
\bysame, \emph{Non-intersecting paths, random tilings and random matrices},
  Probab. Theory Relat. Fields \textbf{123} (2002), 225--280.

\bibitem{Johansson2003}
\bysame, \emph{Discrete polynuclear growth and determinantal processes}, Comm.
  Math. Phys. \textbf{242} (2003), 277--329.

\bibitem{TS2012}
{K.A. Takeuchi and M. Sano}, \emph{{Evidence for geometry-dependent universal
  fluctuations of the Kardar-Parisi-Zhang interfaces in liquid-crystal
  turbulence}}, J. Stat. Phys. \textbf{147} (2010), 853--890.

\bibitem{TS2010}
{K.A. Takeuchi, M. Sano}, \emph{{Growing Interfaces of Liquid Crystal
  Turbulence: Universal Scaling and Fluctuations}}, Phys. Rev. Lett.
  \textbf{104} (2010), 230601.

\bibitem{KN2003}
Y.~Kajihara and M.~Noumi, \emph{{Mutiple elliptic hypergeometric series: An
  approach from the Cauchy determinant}}, Indag. Mathem., N.S. \textbf{14}
  (2003), 395--421.

\bibitem{KPZ1986}
M.~Kardar, G.~Parisi, and Y.~C. Zhang, \emph{Dynamic scaling of growing
  interfaces}, Phys. Rev. Lett. \textbf{56} (1986), 889--892.

\bibitem{Katori2015}
M.~Katori, \emph{Elliptic determinantal process of type A}, Probab. Theory
  Relat. Fields \textbf{162} (2015), 637--677.

\bibitem{Katori2016}
M.~Katori, \emph{Elliptic Bessel processes and elliptic Dyson models realized as temporally 
                    inhomogeneous processes}, J. Math. Phys. \textbf{57} (2016), 103302.

\bibitem{KipnisLandim1999}
C.~Kipnis and C.~Landim, \emph{Scaling limits of interacting particle systems},
  Springer, 1999.

\bibitem{GS1992}
{L.-H. Gwa and H. Spohn}, \emph{{Six-vertex model, roughened surfaces, and an
  asymmetric spin Hamiltonian}}, Phys. Rev. Lett. \textbf{68} (1992), 725--728.

\bibitem{Ledoux2003}
M.~Ledoux, \emph{{A Remark on Hypercontractivity and Tail Inequalities for the
  Largest Eigenvalues of Random Matrices. In S\'eminaire de Pobalilit\'es
  XXXVII, vol. 1832 of Lecture Notes in Mathematics. Springer 2003.}}

\bibitem{Liggett1985}
T.~M. Liggett, \emph{Interacting particle systems}, Springer-Verlag, 1985.

\bibitem{Liggett2010}
\bysame, \emph{Continuous time markov processes: An introduction (graduate
  studies in mathematics}, Amer Mathematical Society, 2010.

\bibitem{Mac1995}
I.G. Macdonald, \emph{{Symmetric Functions and Hall Polynomials. 2nd ed.}},
  Oxford University Press, 1995.

\bibitem{MatveevPetrov2017}
K.~Matveev and L.~Petrov, \emph{$q$-randomized {Robinson-Schensted-Knuth}
  correspondences and random polymers}, Ann. Inst. Henri Poincar\'e D
  \textbf{4} (2017), 1--123.

\bibitem{Mehta2004}
M.~L. Mehta, \emph{Random matrices}, 3rd ed., Elsevier, 2004.

\bibitem{OConnell2012}
N.~O'Connell, \emph{{Directed polymers and the quantum Toda lattice}}, Ann.
  Prob. \textbf{40} (2012), 437--458.

\bibitem{OrrPetrov2016p}
D.~Orr and L.~Petrov, \emph{{Stochastic higher spin six vertex model and
  q-TASEPs, {arXiv:1610.10080}}}.

\bibitem{Po2004}
A.~M. Povolotsky, \emph{{ Bethe ansatz solution of zero-range process with
  nonuniform stationary state}}, Phys. Rev. E \textbf{69} (2004), 061109.

\bibitem{PS2002a}
M.~Pr{\"a}hofer and H.~Spohn, \emph{Current fluctuations for the totally
  asymmetric simple exclusion process}, In and out of equilibrium, vol. 51 of
  {\it Progress in Probability} (V.~Sidoravicius, ed.), 2002, pp.~185--204.

\bibitem{PS2002b}
\bysame, \emph{{Scale Invariance of the PNG Droplet and the Airy Process}}, J.
  Stat. Phys. \textbf{108} (2002), 1071--1106.

\bibitem{PS2004}
\bysame, \emph{{Exact scaling functions for one-dimensional stationary
                   KPZ growth}}, J.
  Stat. Phys. \textbf{115} (2004), 255--279.
  
\bibitem{RS1980}
M.~Reed and B.~Simon, \emph{Methods of modern mathematical physics i:
  Functional analysis}, Academic, 1980.

\bibitem{Sasamoto2005}
T.~Sasamoto, \emph{{Spatial correlations of the 1D KPZ surface on a flat
  substrate}}, J. Phys. A \textbf{38} (2005), L549--L556.

\bibitem{SS2010b}
T.~Sasamoto and H.~Spohn, \emph{{Exact height distributions for the KPZ
  equation with narrow wedge initial condition}}, Nucl. Phys. B \textbf{834}
  (2010), 523--542.

\bibitem{SS2010a}
\bysame, \emph{{The crossover regime for the weakly asymmetric simple exclusion
  process}}, J. Stat. Phys. \textbf{140} (2010), 209--231.

\bibitem{SS2010c}
\bysame, \emph{{Universality of the one-dimensional KPZ equation}}, Phys. Rev.
  Lett. \textbf{834} (2010), 523--542.

\bibitem{SW1998}
T.~Sasamoto and M.~Wadati, \emph{{Exact results for one-dimensional totally
  asymmetric diffusion models}}, J. Phys. A \textbf{31} (1998), 6057--6071.

\bibitem{Schuetz1997a}
G.~M. Sch{\"u}tz, \emph{Duality relations for asymmetric exclusion processes},
  J. Stat. Phys. \textbf{86} (1997), 1265--1287.

\bibitem{Soshnikov2000}
A.~Soshnikov, \emph{{Determinantal Random Point Fields}}, Russian Math. Surveys
  \textbf{55} (2000), 923--975.

\bibitem{Spohn1991}
H.~Spohn, \emph{Large scale dynamics of interacting particles}, Springer, 1991.

\bibitem{Spohn2012}
\bysame, \emph{{KPZ Scaling Theory and the Semi-discrete Directed Polymer
  Model, {\rm MSRI proceedings}}},  (2013).

\bibitem{Stade2002}
E.~Stade, \emph{{Archimedean $L$-factors on GL$(n)\times$GL$(n)$ and
  generalized Barnes integrals}}, Israel J. Math. \textbf{127} (2002),
  201--219.

\bibitem{Stanley1999}
R.~P. Stanley, \emph{Enumerative combinatorics 2}, Springer, 1999.

\bibitem{TW1994}
C.~A. Tracy and H.~Widom, \emph{Level-spacing distributions and the {Airy}
  kernel}, Comm. Math. Phys. \textbf{159} (1994), 151--174.

\bibitem{TW1996}
\bysame, \emph{On orthogonal and symplectic matrix ensembles}, Comm. Math.
  Phys. \textbf{177} (1996), 727--754.

\bibitem{TW1998}
\bysame, \emph{Correlation functions, cluster functions, and spacing
  distributions for random matrices}, J. Stat. Phys. \textbf{92} (1998),
  809--835.

\bibitem{TW2009a}
\bysame, \emph{{Asymptotics in ASEP with step initial condition}}, Commun.
  Math. Phys. \textbf{209} (2009), 129--154.

\bibitem{WW1927}
E.T. Whittaker and G.N. Watson, \emph{A course of modern analysis}, Cambridge,
  1927.

\end{thebibliography}
\end{document}